\documentclass[11pt,a4paper,titlepage,oneside]{memoir}

\usepackage[OT1]{fontenc}

\usepackage[english]{babel}

\usepackage[utf8]{inputenc}

\usepackage[sc]{mathpazo}

\usepackage{amsmath,amssymb,amsfonts,mathrsfs}

\usepackage[amsmath,thmmarks]{ntheorem}

\usepackage{graphicx}

\usepackage{soul}

\usepackage{pdfpages}
\usepackage{tikz}
\usepackage{pgfplots}
\pgfplotsset{compat=1.16}

\usepackage{varioref}

\usepackage{datetime}

\usepackage{mathtools}

\usepackage[h]{esvect}

\usepackage{array}

\usepackage{listings}
\usepackage{microtype}

\usepackage{booktabs}

\usepackage{ETHlogo}

\nonzeroparskip
\defaultlists

\makeatletter

\if@twoside
  \copypagestyle{chapter}{Ruled}
\else
  \copypagestyle{chapter}{ruled}
\fi
\makeoddhead{chapter}{}{}{}
\makeevenhead{chapter}{}{}{}
\makeheadrule{chapter}{\textwidth}{0pt}
\copypagestyle{abstract}{empty}

\makechapterstyle{bianchimod}{%
  \chapterstyle{default}
  \renewcommand*{\chapnamefont}{\normalfont\Large\sffamily}
  
  \renewcommand*{\printchaptername}{%
    \chapnamefont\centering\@chapapp}

  }

\chapterstyle{bianchimod}

\setsecheadstyle{\Large\bfseries\sffamily}
\setsubsecheadstyle{\large\bfseries\sffamily}
\setsubsubsecheadstyle{\bfseries\sffamily}
\setparaheadstyle{\normalsize\bfseries\sffamily}
\setsubparaheadstyle{\normalsize\itshape\sffamily}
\setsubparaindent{0pt}

\captionnamefont{\sffamily\bfseries\footnotesize}
\captiontitlefont{\sffamily\footnotesize}
\setsecnumdepth{subsection}
\settocdepth{subsection}

\pretitle{\vspace{0pt plus 0.7fill}\begin{center}\HUGE\sffamily\bfseries}
\posttitle{\end{center}\par}
\preauthor{\par\begin{center}\let\and\\\Large\sffamily}
\postauthor{\end{center}}
\predate{\par\begin{center}\Large\sffamily}
\postdate{\end{center}}

\def\@advisors{}
\newcommand{\advisors}[1]{\def\@advisors{#1}}
\def\@department{}
\newcommand{\department}[1]{\def\@department{#1}}
\def\@thesistype{}
\newcommand{\thesistype}[1]{\def\@thesistype{#1}}

\renewcommand{\maketitlehookb}{\vspace{1in}%
  \par\begin{center}\Large\sffamily\@thesistype\end{center}}

\renewcommand{\maketitlehookd}{%
  \vfill\par
  \begin{flushright}
    \sffamily
    \@advisors\par
    \@department, ETH Z\"urich
  \end{flushright}
}

\checkandfixthelayout

\makeatother

\theoremstyle{plain}
\numberwithin{equation}{chapter}

\theoremstyle{nonumberplain}
\theorembodyfont{\normalfont}
\theoremsymbol{\ensuremath{\square}}

\newcommand{\R}{\mathbb{R}}

\renewcommand{\epsilon}{\ensuremath\varepsilon}

\renewcommand{\phi}{\ensuremath{\varphi}}

\DeclareMathOperator{\softmax}{softmax}
\DeclareMathOperator{\relu}{ReLU}

\usepackage[linkcolor=black,colorlinks=true,citecolor=black,filecolor=black]{hyperref}

\usepackage[authoryear]{natbib}

\title{On the Mechanistic Interpretability of Neural Networks for Causality in Bio-statistics}
\author{Jean-Baptiste A. Conan\\ 18-814-327}
\thesistype{Master Thesis}
\advisors{Advisors: Prof.\ Dr.\ Mark van der Laan\\ Prof.\ Dr.\ Fadoua Balabdaoui}
\department{Department of Mathematics}

\date{\today}

\begin{document}

\frontmatter

\begin{titlingpage}
  \calccentering{\unitlength}
  \begin{adjustwidth*}{\unitlength-24pt}{-\unitlength-24pt}
    \maketitle
  \end{adjustwidth*}
\end{titlingpage}

\begin{abstract}
Interpretable insights from predictive models remain critical in bio-statistics, particularly when assessing causality, where classical statistical and machine learning methods often provide inherent clarity. While Neural Networks (NNs) offer powerful capabilities for modeling complex biological data, their traditional "black-box" nature presents challenges for validation and trust in high-stakes health applications. Recent advances in Mechanistic Interpretability (MI) aim to decipher the internal computations learned by these networks. This work investigates the application of MI techniques to NNs within the context of causal inference for bio-statistics.

We demonstrate that MI tools can be leveraged to: (1) probe and validate the internal representations learned by NNs, such as those estimating nuisance functions in frameworks like Targeted Minimum Loss-based Estimation (TMLE); (2) discover and visualize the distinct computational pathways employed by the network to process different types of inputs, potentially revealing how confounders and treatments are handled; and (3) provide methodologies for comparing the learned mechanisms and extracted insights across statistical, machine learning, and NN models, fostering a deeper understanding of their respective strengths and weaknesses for causal bio-statistical analysis.
\end{abstract}
\chapter*{Acknowledgement}
I would like to thank Mark for his guidance during this project.

\cleartorecto
\tableofcontents
\mainmatter


\chapter{Introduction}
\begin{figure}[htbp]
    \centering
    \includegraphics[width=0.8\textwidth]{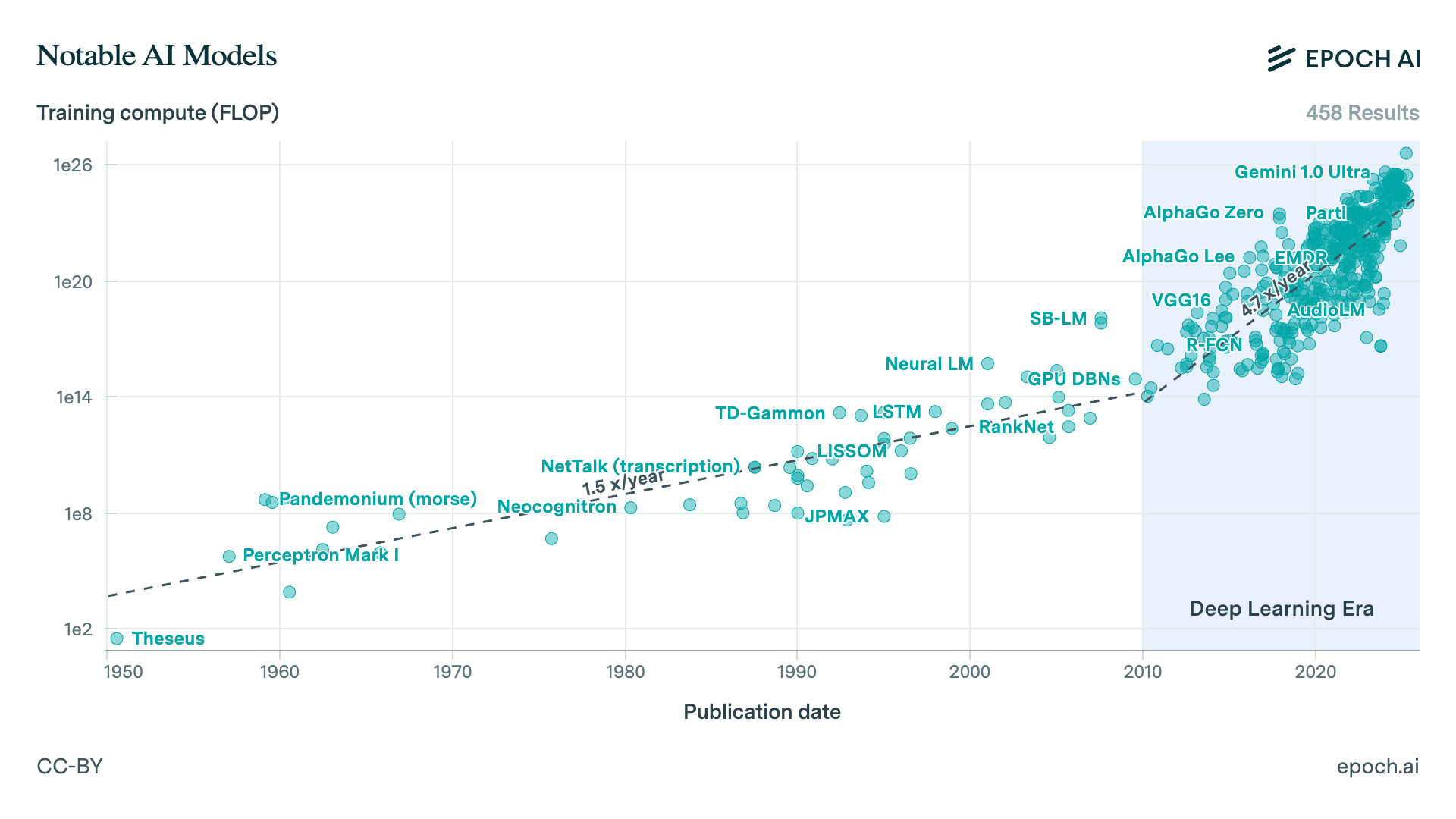}
    \caption{Visualization of AI model scaling trends based on the Epoch AI Notable Models dataset \citep{EpochNotableModels2024}. The figure illustrates key insights including: training compute doubling every five months since 2010 (growing at 4.7x per year), training costs for large models doubling every nine months, dataset sizes for language models doubling every eight months, and power requirements for frontier models doubling annually. This comprehensive dataset tracks over 900 historically significant AI models with over 400 training compute estimates.}
    \label{fig:epoch_ai_models}
\end{figure}

\epigraph{To Explain or to Predict?}{\textit{Galit Shmueli} \citep{Shmueli2010}}
Since the emergence of early statistical frameworks, our comprehension of data has advanced dramatically. From basic correlation analysis of empirical observations \citep{graunt1662natural} and progressing to sophisticated modeling of complete languages \citep{brown2020language} and biological structures \citep{jumper2021highly}, Artificial Intelligence (AI) systems have developed to support virtually all domains of human endeavor. They have also evolved from highly interpretable systems, where we can understand the role of the few variables in the prediction model, to so-called black-box models, where the individual contribution of each datapoint is submerged in the billions of parameters that contribute to the model. \textit{Explain or Predict} indeed, as the prediction became more accurate, the size and complexity of all models grew accordingly, as shown in Figure \ref{fig:epoch_ai_models}, trading off interpretability for predictive capability.

However, recent research regarding neural networks (NNs), the most powerful black-box methods, interpretability has demonstrated interesting capabilities that in no way impute model accuracy. These methods are based on the activation of neurons, in the same way as we observe a human brain. Using techniques and data sets, it is possible to associate certain parts of a NN with a behavior, a computation or a concept. It has thus become possible to know, e.g. when a language model talks about politics \citep{kim2025linear}, without interfering with the capabilities of the model in question.

This Master's thesis therefore bears the following research questions: Is it possible to transpose NN interpretability techniques to bio-statistics, and how effective these new techniques become?

\section{Bio-statistics}

\subsection{Biology, Experiments, and Clinical Trials}
Biology is the broad study of living things, from cells to evolution, disease and the effect of molecules on the human body. The latter is of particular interest to us, as it is one of the main applications of bio-statistics, the sub-branch of statistics that applies statistical methods to the study of living organisms. It's relatively easy to collect information, i.e.  about patients and to record and study the effects of, say, drugs on them. This is called experimentation, and produces a statistically analyzable data set. Today, this process, when optimized, is akin to our clinical trials: a set of patients selected to take part in a study on the effect of a drug. The aim is to discover whether the observed effect is absolutely attributable to the drug, and whether there are no undesirable side-effects.

\subsection{Clinical Trials limitations}
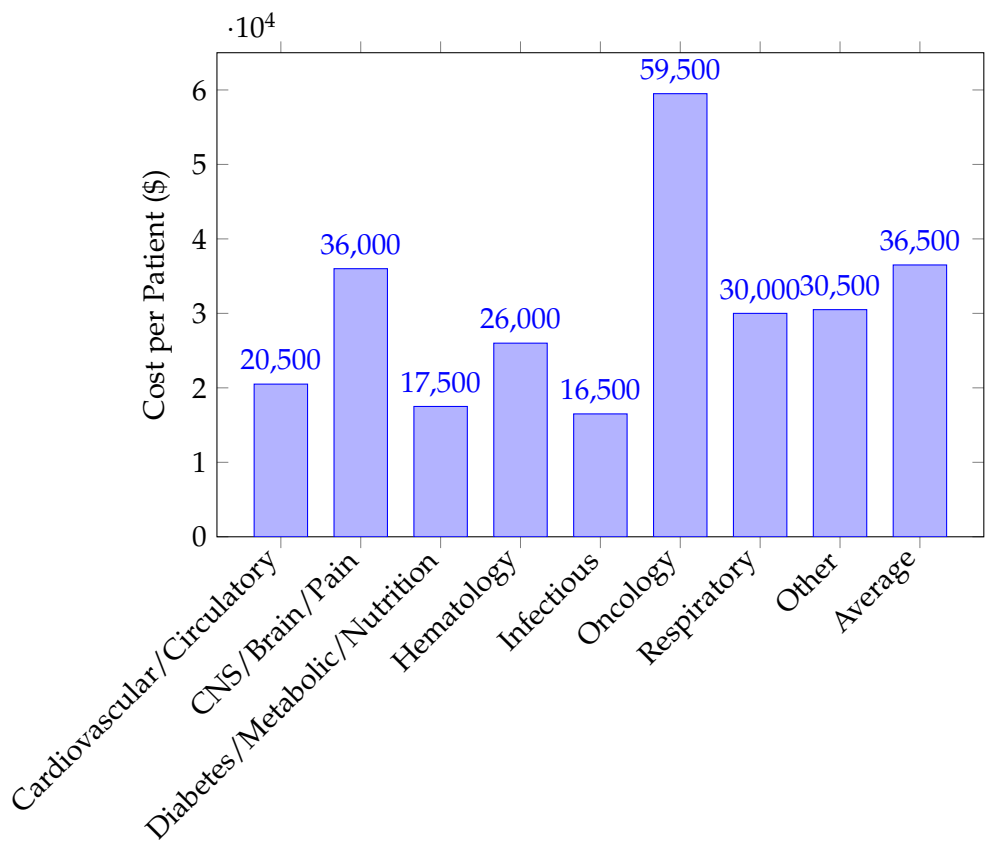
\begin{figure}[htbp]
    \centering
    \begin{tikzpicture}
        \begin{axis}[
            width=0.9\textwidth,
            height=8cm,
            ybar,
            bar width=20pt,
            ylabel={Cost per Patient (\$)},
            symbolic x coords={Cardiovascular/Circulatory, CNS/Brain/Pain, Diabetes/Metabolic/Nutrition, Hematology, Infectious, Oncology, Respiratory, Other, Average},
            xtick=data,
            xticklabel style={rotate=45, anchor=east, align=right},
            nodes near coords,
            nodes near coords align={vertical},
            ymin=0,
            ymax=65000,
            enlarge x limits=0.1,
            ]
            \addplot coordinates {
                (Cardiovascular/Circulatory, 20500)
                (CNS/Brain/Pain, 36000)
                (Diabetes/Metabolic/Nutrition, 17500)
                (Hematology, 26000)
                (Infectious, 16500)
                (Oncology, 59500)
                (Respiratory, 30000)
                (Other, 30500)
                (Average, 36500)
            };
        \end{axis}
    \end{tikzpicture}
    \caption{Estimated Average Per-Patient Clinical Trial Costs by Selected Condition (\cite{Battelle2015})}
    \label{fig:cost_by_condition}
\end{figure}

However, clinical trials face many limitations. Primarily, the costs per patient are enormous, as shown in figure \ref{fig:cost_by_condition}. In the US, costs average \$36.5k per patient, rising to almost \$60k per patient for rarer diseases such as cancer, according to \citep{Battelle2015}. Therefore, reducing the number of patient is a priority. Another component is the number of visit each patient is supposed to make. Indeed, the more interventions, the more expensive the study is per patient. The financial burden intensifies when a patient withdraws from a study, rendering all prior investments in their participation --- such as recruitment, screening, and monitoring --- effectively lost. This underscores the need for optimization, and by refining trial designs, we can mitigate these losses. 

\subsection{Statistics to optimize Clinical Trials}
\subsubsection{Sampling size and statistical significance}
Optimizing clinical trials involves strategically determining the minimal sample size required to achieve statistical significance, thereby reducing costs without compromising result validity. Statistical significance assesses whether observed drug effects are genuinely attributable to the treatment rather than random chance. This evaluation typically employs hypothesis testing, where the null hypothesis (no effect) is compared against an alternative hypothesis (presence of an effect). Trials must carefully balance two critical statistical parameters: \textbf{significance level} \(\alpha\), often set at 0.05, representing the probability of incorrectly rejecting a true null hypothesis (Type I error) ; and \textbf{statistical power}, usually targeted at 80\% or higher, indicating the likelihood of correctly detecting a true effect (avoiding Type II errors).
By selecting appropriate statistical tests and precisely calculating the minimum necessary sample size to meet these criteria, trials can avoid unnecessary patient recruitment, thus significantly lowering costs while maintaining scientific rigor.

\subsubsection{Causality in the statistical model}
Establishing causality is a cornerstone of clinical trials, as the goal is to confirm that the drug directly causes the observed effect, not merely that the two are associated. Randomized controlled trials (RCTs) are the gold standard for this, as randomization ensures that treatment and control groups are comparable, isolating the drug's effect. However, challenges such as confounding variables (e.g., age or lifestyle factors) or patient non-compliance can obscure causal relationships. Optimization strategies include stratified randomization, which balances known confounders across groups, and intention-to-treat analysis, which preserves the benefits of randomization by including all patients as originally assigned, regardless of adherence. These methods enhance the trial's ability to draw valid causal conclusions efficiently.


\subsubsection{Mitigating biases in assignment, outcome and attrition}
Biases pose significant threats to the integrity of clinical trial results. For instance, selection bias occurs when the sample does not represent the target population, performance bias arises from inconsistent treatment administration, detection bias stems from unequal outcome assessment between groups, and attrition bias results from differential dropouts. To optimize trials, we can implement blinding (where patients are unaware of their group assignment) or double-blinding (where both patients and researchers are unaware), reducing performance and detection biases. For attrition bias, techniques like multiple imputation or last observation carried forward can address missing data from withdrawals \citep{madsen2015using}. By minimizing these biases, trials produce more reliable results with fewer resources wasted on flawed designs.

\subsubsection{Right-Censoring Techniques}
Right-censoring is a common issue in survival analysis within clinical trials, occurring when the event of interest (e.g., death, disease progression) has not occurred by the study’s end or when participants drop out. This can lead to biased estimates if not handled properly. In \citep{Prinja2010}, they highlight methods such as Kaplan-Meier for estimating survival functions and Cox models \citep{Cox1972} for assessing hazard ratios, accounting for censored observations.





\section{Targeted Learning}

In bio-statistics, where datasets are often high-dimensional and riddled with complex relationships, traditional statistical methods can stumble due to their dependence on rigid parametric assumptions \citep{erceg-hurn2008modern}. Conversely, machine learning excels at modeling intricate patterns but often prioritizes prediction over the statistical inference \citep{anastasiadiou2023prediction} required for scientific rigor. Targeted Learning (TL), pioneered by Mark van der Laan \citep{vanderLaanRubin2006}, elegantly bridges this divide. By melding the adaptability of machine learning with the precision of causal inference, TL delivers robust, efficient estimates tailored to complex settings like clinical trials and observational studies—cornerstones of bio-statistical research.

\subsection{Methodology}

The TL framework unfolds in two key stages. First, it estimates the data-generating distribution using Super Learning \citep{vanderLaan2007}, an ensemble technique that integrates multiple machine learning algorithms—think random forests, gradient boosting, or neural networks—into a single, optimized predictor. This step sidesteps the pitfalls of relying on a lone, potentially misspecified model, capturing the data’s complexity with flexibility. Second, TL refines this initial estimate through a targeting step, typically via Targeted Maximum Likelihood Estimation (TMLE). TMLE adjusts the estimate to zero in on the parameter of interest—say, the average treatment effect—while ensuring statistical properties like efficiency and minimal bias. The outcome? An estimator that adapts to the data yet yields valid confidence intervals and hypothesis tests, vital for bio-statistical conclusions.

\subsection{Advantages and Applications}

TL shines in bio-statistics for its ability to tackle high-dimensional data—like patient records with dozens of covariates—without choking on restrictive assumptions. It offers efficient estimates, potentially shrinking the sample sizes needed for reliable results, and supports robust inference, making it a game-changer for clinical research. Its applications are broad: TL has been used to estimate treatment effects in randomized trials, dissect longitudinal data with time-varying interventions, and craft individualized treatment rules in precision medicine. Imagine tailoring therapies to patient profiles or decoding the impact of a drug across diverse populations—TL makes it possible with precision and confidence.

\section{Interpretability}
Interpretability in AI refers to the ability to understand and explain how a model arrives at its predictions or decisions. This is essential for building trust in AI systems, debugging potential issues, and ensuring fairness and accountability in their applications. Different types of models exhibit varying levels of interpretability. Simpler models, such as linear regression or decision trees, are inherently more transparent—their decision-making can be traced through coefficients or tree paths. However, more complex models, like neural networks, often function as "black boxes," where the intricate interplay of numerous parameters obscures the reasoning behind their outputs, posing challenges for comprehension across a wide range of AI applications.

In deep learning, interpretability becomes particularly difficult due to the models' complexity, characterized by multiple layers and millions of parameters. To address this, researchers have developed techniques like feature visualization, which reveals patterns the network detects, circuit tracing, which highlight influential input regions, and attention mechanisms, which show where the model focuses, especially in tasks like natural language processing. While these methods offer valuable insights into specific aspects of a deep learning model's behavior, they often fall short of providing a complete picture. The scale of these models and their non-linear nature mean that fully elucidating their decision processes remains a major hurdle, limiting our ability to interpret them comprehensively.

At present, interpretability methods can highlight key features of predictions and provide explanations for individual results, making it easier to validate models or analyse errors. However, they struggle to provide comprehensive explanations that encompass the general behaviour of a model or to ensure that explanations accurately reflect the underlying processes. In the future, advances could lead to more robust techniques and a deeper understanding of complex systems. It is also possible to design intrinsically interpretable models that do not compromise performance, combining transparency and capacity. As AI evolves, improving interpretability will be essential for its ethical and effective use in a variety of domains.

\section{AI Safety}
I wanted to end this introduction with a note about AI safety, the branch of AI research dedicated to ensuring safe outcomes from the training of powerful AI models.

The current understanding of AI models is still extremely limited. The entire field of MI is focused on mitigating the risks associated with the emergence of highly intelligent systems. Recent studies by major LLM research laboratories \citep{Marks2023} found that LLMs can behave in dangerous ways. And with increasingly more capable systems being currently trained and deployed, we have never been so unsure about the security of our future.

This research is thus part of a drive to advance AI Safety, by studying the impact of MI in other adjoining fields. This can have the dual effect of first improving bio-statistics models, as demonstrated in the present paper. But also to improve MI methods, by bringing in new ideas, specific to bio-statistics and the functioning of the human body, but transferable to other fields. I hope this will inspire fellow researchers in this quest to maximize our understandings of AI systems.

\chapter{Neural Networks: Prediction and Generation}

Neural networks are the central concept behind Deep Learning and the automated learning of patterns and features from data. These artificial networks, inspired by the structure and function of biological neural network in animal brains, consist of interconnected nodes or artificial neurons that process and transmit information. 

Neural networks can vary in structure depending on the task. Multi-Layer perceptrons consist of fully connected layers, but other architectures like Convolutional Neural Networks are better for images, capturing spatial patterns, while Recurrent Neural Networks suit sequential data like text. The number of layers and neurons, along with activation functions like ReLU, are chosen to model complexity. For outputs, regression uses linear activations, binary classification uses sigmoid, and multi-class uses softmax.

This chapter builds up knowledge on how to define, train and use this kind of machine learning techniques, exploring different models, neural network architectures and concepts.

\section{Foundations of Machine Learning}

This chapter introduces fundamental concepts and techniques in Machine Learning (ML), laying the groundwork for understanding both classical methods and the more complex Neural Networks (NNs) architectures discussed subsequently, and which correspond to the deep learning subset of ML. ML algorithms enable systems to learn patterns, make predictions, or uncover structures within data without explicit programming for each specific task.

\subsection{Supervised vs. Unsupervised Learning}

A primary distinction within ML is between \textbf{supervised} and \textbf{unsupervised} learning. In supervised learning, the algorithm is trained on a dataset consisting of input-output pairs, denoted as $\{(x_i, y_i)\}_{i=1}^n$, where $x_i$ is the input feature vector and $y_i$ is the corresponding target label or value. The goal is to learn a mapping function $f: \mathcal{X} \to \mathcal{Y}$ that can accurately predict the output $y$ for new, unseen inputs $x$. Common supervised tasks include \textit{regression}, where $y$ is a continuous value, and \textit{classification}, where $y$ belongs to a discrete set of categories.

In contrast, unsupervised learning deals with datasets containing only input data, $\{x_i\}_{i=1}^n$, without corresponding target outputs. The objective here is to discover inherent structures or patterns within the data, such as grouping similar data points (\textit{clustering}) or reducing the dimensionality of the feature space while preserving essential information (\textit{dimensionality reduction}).

\subsection{Data Handling and Evaluation}

Effective ML model development hinges on proper data handling and evaluation strategies to ensure models generalize well to unseen data.

\subsubsection{Datasets: Training, Validation, and Testing} 

Typically, the available data is partitioned into distinct sets. The \textbf{training set} is used to learn the model parameters (e.g., weights in a NN or coefficients in a linear model) by optimizing a specific objective. This optimization usually involves minimizing a \textbf{loss function}, $\mathcal{L}(\hat{y}, y)$, which quantifies the discrepancy between the model's predictions $\hat{y}$ and the true targets $y$.

The \textbf{validation set} plays a crucial role during the training phase. It is used to tune \textbf{hyperparameters} – parameters not learned directly from the training data, such as the learning rate in gradient descent, the depth of a decision tree, or the strength of a regularization penalty. Crucially, the validation set provides an estimate of the model's performance on data it hasn't been trained on, which helps in identifying and preventing \textbf{overfitting}. Overfitting occurs when a model learns the training data too well, capturing noise and specific idiosyncrasies, leading to poor performance on new, unseen data. Monitoring performance on the validation set allows for strategies like early stopping, where training is halted when validation performance begins to degrade.

Finally, the \textbf{test set} is held out completely until the model development process (including training and hyperparameter tuning) is finished. It provides the final, unbiased estimate of the model's generalization performance on completely new data, simulating how the model would perform in a real-world deployment.

\subsubsection{Cross-Validation}

To obtain more robust performance estimates and make better use of limited data, particularly when the initial dataset is not very large, \textbf{cross-validation} (CV) is a widely employed technique. A common approach is $k$\textbf{-fold CV}. In this method, the original training data is randomly partitioned into $k$ equal-sized folds (subsets). The model is then trained and evaluated $k$ times. In each iteration $j$ (from $1$ to $k$), the $j$-th fold is held out as a temporary validation set, and the model is trained on the remaining $k-1$ folds. The performance metric (e.g., accuracy, mean squared error) is calculated on the held-out fold. After $k$ iterations, the performance metrics are averaged across all folds to provide a more stable and reliable estimate of the model's performance than a single train-validation split. This averaged performance is often used to guide hyperparameter selection or compare different model types.

\subsection{Classical Supervised Learning Models}

Before delving into the intricacies of NN, it is instructive to review some fundamental models of supervised learning. These methods are widely used, particularly in fields such as biostatistics, where interpretability and efficiency on smaller, structured datasets are often crucial.

\subsubsection{Linear and Logistic Regression}

\textbf{Linear regression} is the basis for modelling the relationship between a continuous target variable $Y$ and a set of input characteristics $X = (X_1, \dots, X_p)$. It assumes a linear relationship:
\[ Y = \beta_0 + \beta_1 X_1 + \dots + \beta_p X_p + \epsilon = X\beta + \epsilon \]
where $\beta = (\beta_0, \beta_1, \dots, \beta_p)^T$ is the vector of coefficients (including an intercept term $\beta_0$), $X$ is the design matrix (including a column of ones for the intercept), and $\epsilon$ represents the error term, typically assumed to follow a normal distribution $N(0, \sigma^2)$. The coefficients $\beta$ are most commonly estimated using \textbf{Ordinary Least Squares (OLS)}, which minimizes the Residual Sum of Squares (RSS):
\[ \hat{\beta}_{OLS} = \arg\min_{\beta} ||y - X\beta||_2^2 = \arg\min_{\beta} \sum_{i=1}^n (y_i - x_i^T\beta)^2 \]
The OLS solution has a closed form: $\hat{\beta}_{OLS} = (X^T X)^{-1} X^T y$, provided $X^T X$ is invertible.

\textbf{Generalized Linear Models (GLMs)} extend the linear regression framework to account for response variables with non-normal error distribution patterns, and for dependent variables whose relationship with the predictors is non-linear. A GLM is defined by a probability distribution from the exponential family (e.g. Bernoulli, Poisson, Gamma) and a \textbf{link function} $g(\cdot)$ such that $g(E[Y|X]) = X\beta$.

\textbf{Logistic Regression} is a specific, widely used GLM designed for binary classification problems, where the response variable $Y$ takes values in $\{0, 1\}$. It models the probability of the positive class, $p(X) = P(Y=1|X)$, using the \textit{logit} (or log-odds) link function:
\[ g(p(X)) = \log\left(\frac{p(X)}{1-p(X)}\right) = X\beta \]
Equivalently, the probability is modeled via the sigmoid (or logistic/expit) function:
\[ p(X) = \frac{e^{X\beta}}{1 + e^{X\beta}} = \frac{1}{1 + e^{-X\beta}} \]
The parameters $\beta$ are typically estimated using \textbf{Maximum Likelihood Estimation (MLE)}. Given $n$ independent observations $(x_i, y_i)$, the log-likelihood function for Bernoulli outcomes is:
\[ \ell(\beta) = \sum_{i=1}^n [y_i \log(p(x_i)) + (1-y_i)\log(1-p(x_i))] \]
Maximizing $\ell(\beta)$, or minimizing its negative, the binary cross-entropy (BCE) loss, yields the parameter estimates $\hat{\beta}_{MLE}$. Unlike linear regression, there is generally no closed-form solution, and iterative methods like gradient ascent are used.

\subsubsection{Regularized Regression: LASSO}

In scenarios where the number of predictors $p$ is large, potentially even larger than the number of observations $n$, standard OLS or MLE can lead to unstable estimates and poor predictive performance due to high variance or overfitting. \textbf{Regularization} methods address this by adding a penalty term to the objective function, encouraging simpler models by shrinking the coefficient estimates towards zero.

The \textbf{LASSO (Least Absolute Shrinkage and Selection Operator)} \citep{Tibshirani1996} adds an $L_1$ penalty to the OLS objective (for linear regression) or the negative log-likelihood (for logistic regression). For linear regression, the LASSO estimate is:
\[ \hat{\beta}_{LASSO} = \arg\min_{\beta} \left( ||y - X\beta||_2^2 + \lambda ||\beta||_1 \right) \]
where $||\beta||_1 = \sum_{j=1}^p |\beta_j|$ is the $L_1$ norm of the coefficient vector (excluding the intercept), and $\lambda \ge 0$ is a tuning parameter controlling the amount of shrinkage. As $\lambda$ increases, more coefficients are forced to be exactly zero. This property makes LASSO perform simultaneous coefficient shrinkage and automatic feature selection, leading to potentially more interpretable and sparse models. The optimal $\lambda$ is typically chosen using cross-validation. Ridge regression, another popular technique, uses an $L_2$ penalty ($\lambda ||\beta||_2^2 = \lambda \sum \beta_j^2$) which shrinks coefficients but rarely sets them exactly to zero .

\subsection{Ensemble Methods}

While individual models like linear regression or decision trees can be effective, \textbf{ensemble methods} \citep{zhou2012ensemble} often achieve superior performance and robustness by strategically combining the predictions of multiple models, known as base learners. The core principle is that by aggregating the "wisdom" of several diverse models, the weaknesses or biases of individual learners can be averaged out, leading to a stronger, more reliable overall prediction. Different ensemble techniques employ various strategies for creating diversity among the base learners and for combining their outputs. Two highly influential and powerful ensemble paradigms are Random Forests and Gradient Boosting Machines.

\subsubsection{Random Forests}

\textbf{Random Forests} \citep{breiman2001random} represent a prominent ensemble technique primarily utilizing decision trees as base learners. Instead of relying on a single, potentially complex decision tree which might overfit the training data, a random forest constructs a large collection (a "forest") of decision trees during training. The final prediction is then determined by aggregating the outputs of all trees in the forest: for regression tasks, this typically involves averaging the predictions, while for classification, a majority vote among the trees determines the final class label. The key to the effectiveness of random forests lies in the introduction of randomness during the tree-building process, which ensures diversity among the individual trees and reduces correlation between them. This randomness is incorporated through two primary mechanisms. First, \textbf{bagging}, or bootstrap aggregating, is employed, meaning each individual tree is trained on a different bootstrap sample drawn with replacement from the original training dataset. Consequently, each tree learns from a slightly different perspective of the data. Second, when determining the best split at each node within a tree, only a random subset of the total available features is considered as candidates. This \textbf{random feature subspace} method prevents a few dominant features from overly influencing all trees in the forest, further promoting diversity. By averaging the predictions of these numerous, largely uncorrelated trees, random forests significantly reduce the variance compared to single decision trees, leading to improved generalization and robustness against overfitting, even with high-dimensional data.

\subsubsection{Gradient Boosting Machines (XGBoost)}

\textbf{Gradient Boosting Machines (GBMs)} \citep{FreundSchapire1997,friedman2001greedy} constitute another powerful class of ensemble methods, but unlike the parallel construction of trees in random forests, GBMs build the ensemble sequentially. An initial simple model (e.g., predicting the mean) is established, and subsequent models (typically decision trees) are added iteratively. Each new tree is specifically trained to predict and correct the errors, specifically the residuals or, more generally, the negative gradient of a chosen loss function, made by the current ensemble of preceding models. The predictions of all models in the sequence are then combined, usually as a weighted sum, to form the final prediction. This sequential, error-correcting approach allows boosting models to achieve very high accuracy by incrementally focusing on the harder-to-predict instances.

\textbf{XGBoost (Extreme Gradient Boosting)} \citep{xgboost} stands out as a highly optimized and regularized implementation of the gradient boosting concept, addressing some limitations of traditional GBMs and achieving state-of-the-art results on many structured data problems. It enhances the standard gradient boosting framework through several key innovations. Crucially, it incorporates \textbf{regularization} directly into the objective function being optimized during tree construction, including both L1 (LASSO-like) and L2 (Ridge-like) penalties on the leaf weights (the prediction values at the terminal nodes of the trees). This regularization helps to control model complexity and prevent overfitting. Furthermore, XGBoost provides sophisticated \textbf{handling of missing values}, implementing an internal routine that learns the optimal direction (left or right child node) to send instances with missing values during the splitting process, rather than requiring pre-imputation.

\section{Feed Forward Networks: Definitions and properties}\label{sec:ffn}
Feed Forward Networks (FFNs) are the most common category of NNs. They consist of successive (hidden) layers through which the information flows, meaning it cannot go backwards.

The Multi-Layer Perceptron (MLP) represents a foundational architecture in the study of FFN, serving as a fully connected model capable of learning complex, nonlinear relationships from data. Building on the extension of the perceptron, the MLP addresses its predecessor's limitations by introducing multiple layers of interconnected nodes, or neurons, organized into an input layer, one or more hidden layers, and an output layer. This hierarchical structure enables the MLP to approximate arbitrary continuous functions, a property formalized by the Universal Approximation Theorem (\citep{Cybenko1989ApproximationBS}; \citep{Hornik1989Universalapprox}).

From the fully connectedness of MLP, other architecture emerged, allowing for complex data relation and structures. The notion of FFN opposes the notion of \textbf{recurrent connections} within networks, such as Recurrent Neural Networks (RNNs), where information can flow backward through the network, allowing the model to maintain internal states and process sequential or time-dependent data (cf. Section \ref{ssec:rnn}).

\subsection{Multi Layer Perceptron}\label{ssec:mlp}

The perceptron, introduced by \citep{rosenblatt1958perceptron}, serves as the historical and conceptual precursor of modern NNs, embodying a simplified model of a neuron (cf. \ref{ssec:neuron_act}) with a binary output. It operates on the same principle of computing a weighted sum of inputs, followed by a threshold-based decision rule. Formally, given an input vector $\boldsymbol{x}\in \R^n$, weight vector $\boldsymbol{w}\in \R^n$ , and bias $\boldsymbol{b}\in \R$, the perceptron’s output is defined as:
$$y = \begin{cases}
1 & \text{if } \mathbf{w}^T \mathbf{x} + b \geq 0, \\
0 & \text{otherwise},
\end{cases}
$$
where the activation function is implicitly a step function, $\sigma(z)=H(z)$, with $H(z)$ denoting the Heaviside step function ($H(z)=1$ if $z\geq0$, else $0$). Geometrically, the perceptron implements a linear decision boundary in the input space, partitioning it into two regions separated by the hyperplane $w^\top x+b=0$. The weights $w$ determine the orientation of this hyperplane, while the bias $b$ adjusts its position relative to the origin.

The perceptron learning algorithm, a seminal contribution by \citep{rosenblatt1958perceptron}, iteratively updates the weights and bias to minimize classification errors on a training set. For a misclassified sample $(x,t)$, where $t\in\{0,1\}$ is the target label and \(y\) is the predicted output, the update rule is:
$$\mathbf{w} \leftarrow \mathbf{w} + \eta(t - y)\mathbf{x},$$
$$b \leftarrow b + \eta(t - y),$$
where $\eta>0$ is the learning rate. This rule adjusts the weights in the direction of the gradient of a simple error function, converging to a solution when the data is linearly separable \citep{minsky_papert_1969}. However, Minsky and Papert’s critique highlighted the perceptron’s fundamental limitation: its inability to model nonlinear relationships, such as the XOR function, due to its reliance on a single linear boundary.

The MLP overcomes this constraint by stacking multiple layers of perceptron-like units with nonlinear activation functions. For a network with $L$ layers, the output of the $l$-th layer, $\boldsymbol{a}^{(l)}$, is computed recursively from the previous layer’s output 
$$\boldsymbol{z}^{(l)} = \boldsymbol{W}^{(l)}\boldsymbol{a}^{(l-1)} + \boldsymbol{b}^{(l)},$$
$$\boldsymbol{a}^{(l)}=\sigma(\boldsymbol{z}^{(l)}),$$

where $\boldsymbol{W}^{(l)}$ is the weight matrix connecting layer $l-1$ to layer $l$, $\boldsymbol{b}^{(l)}$ is the bias vector, and $\sigma$ is applied element-wise. The input layer is denoted $\boldsymbol{a}^{(0)}=\boldsymbol{x}$, and the output layer $\boldsymbol{a}^{(L)}$ produces the final prediction. This layered composition enables the MLP to construct hierarchical feature representations, a property central to its success in tasks ranging from classification to regression.

\subsection{Neurons and Activations}\label{ssec:neuron_act}
The neuron constitutes the elementary computational unit of any NN, drawing inspiration from biological neural systems while abstracting their functionality into a mathematical framework. A neuron receives a vector of input signals \(\boldsymbol{x} =[x_1, \dots, x_n]^\top\in \R^n\), where \(n\) denotes the dimensionality of the input space. These inputs are modulated by a corresponding weight vector \(\boldsymbol{w} =[w_1, \dots, w_n]^\top\in \R^n\), which quantifies the strength and direction of influence each input exerts on the neuron's output. Additionally, a bias term 
\(b\in\R\) shifts the weighted sum to adjust the neuron's activation threshold. The neuron computes a linear combination of these inputs, expressed as:
\[z=\boldsymbol{w}^\top\boldsymbol{x}+ b = \sum_i w_ix_i + b \]
where \(z\) represents the pre-activation value, or \textit{logit}, capturing the aggregated input signal.

\subsubsection{Activation functions}
To introduce nonlinearity, a critical feature enabling NN to model complex patterns \citep{Hornik1989Universalapprox}, the pre-activation \(z\) is passed through an \textit{activation function} \(\sigma:\R\to\R\), yielding the neuron's output:
\[a=\sigma(z)\]
There are two different use cases for activation functions in a NN: in the output layer, to match the task's output type and shape requirements; and in the hidden layers, to introduce nonlinearity and indicate the `\textit{firing}' of a neuron. Most common activations function are summed up in Table \ref{tab:activation_functions}.

\paragraph{Output Layer}
Activation functions in the output layer are chosen specifically to shape the network's final pre-activation values into predictions appropriate for the given task. For instance, a sipmle \textbf{Linear activation function}, defined as \(\sigma(z) = z\), has a common use case in regression problems where the target is a continuous value, as it does not restrict the output range. For binary classification tasks, the \textbf{Sigmoid function}, \(\sigma(z) = \frac{1}{1+e^{-z}}\), is frequently employed. It maps any real-valued input to the (0, 1) interval, suitable for interpreting the output as a probability of the positive class. Sigmoid can also be applied element-wise in multi-label classification scenarios. In contrast, for multi-class classification problems with mutually exclusive classes, the \textbf{Softmax function} is the standard choice. Applied to the entire vector \(\mathbf{z}\) of the output layer's pre-activations, its formula \(\sigma(\mathbf{z})_i = \frac{e^{z_i}}{\sum_{j=1}^{K} e^{z_j}}\) produces a probability distribution across the \(K\) classes, where outputs are positive and sum to one.

\paragraph{Hidden Layer}
Within hidden layers, activation functions serve the crucial role of introducing nonlinearity, enabling the network to learn complex data representations and hierarchical features. The concept of a neuron `\textit{firing}' refers to its activation \(a = \sigma(z)\) being significantly non-zero, indicating a meaningful response to its input \(z = \mathbf{w} \cdot \mathbf{x}_{\text{prev}} + b\). Several functions are used for this purpose. The \textbf{Sigmoid function} \(\sigma(z) = \frac{1}{1+e^{-z}}\) and the \textbf{Hyperbolic Tangent (Tanh) function} \(\sigma(z) = \tanh(z) = \frac{e^z - e^{-z}}{e^z + e^{-z}}\), mapping to (0, 1) and (-1, 1) respectively, were historically common. However, their tendency to saturate and cause vanishing gradients makes them less favored in deep networks today, although Tanh's zero-centered output is sometimes advantageous. The \textbf{Rectified Linear Unit (ReLU)}, \(\sigma(z) = \max(0, z)\), has become a prevalent choice for hidden layers. Its computational efficiency, non-saturating nature for positive inputs (which alleviates vanishing gradients), and tendency to induce sparse activations make it highly effective. Its clear threshold (\(z>0\) for firing) simplifies the firing interpretation. However, it can suffer from the "dying ReLU" problem where neurons cease to activate. Variants like \textbf{Leaky ReLU}, \(\sigma(z) = \max(\alpha z, z)\) for a small positive constant \(\alpha\) (e.g., 0.01), address this by allowing a small gradient for negative inputs; its common use case is as a direct alternative to ReLU. More recently, functions like the \textbf{Gaussian Error Linear Unit (GELU)}, often approximated as \(\sigma(z) \approx 0.5 z (1 + \tanh[\sqrt{2/\pi} (z + 0.044715 z^3)])\), have gained prominence. GELU offers a smooth, non-monotonic approximation to ReLU and is increasingly used in advanced architectures like Transformers (cf. Section \ref{ssec:transformers}). The choice significantly impacts training dynamics and representation quality.

Interpreting this `\textit{firing}' relative to the input data requires understanding what the neuron has learned. Through training, each hidden neuron adjusts its weights (\(\mathbf{w}\)) and bias (\(b\)) to become sensitive to specific patterns or features within the data it receives from the preceding layer. Therefore, when a neuron `\textit{fires}', it signals the detection of the particular feature or combination of features it is specialized to recognize. For instance, in image recognition tasks, neurons in early hidden layers might learn to fire in response to simple features like edges or specific color gradients in parts of the input image. Neurons in deeper layers receive inputs from these earlier neurons and might fire in response to more complex combinations, such as textures, shapes, or object parts. Common choices like ReLU (\(\sigma(z) = \max(0, z)\)) exhibit a clear firing threshold (firing only when \(z>0\)), while others like Tanh or Sigmoid offer a graded response. Collectively, the activation patterns across a hidden layer form a distributed, abstract representation of the input, which is then processed by subsequent layers. Analysis of this phenomenon is the core discussion of the later Chapter \ref{chap:mech_int}.

\paragraph{Summary Table}

Here is a summary of the common activation functions discussed:

\begin{table}[htbp]
    \caption{Summary of common activation functions.}
    \label{tab:activation_functions}
    \resizebox{\textwidth}{!}{
        \begin{tabular}{@{}lllll@{}}
        \toprule
        \textbf{Function} & \textbf{Formula} & \textbf{Range} & \textbf{Common Use Case(s)} \\
        \midrule
        \textbf{Linear}      & \(z\)                                                                                                  & \((-\infty, \infty)\) & Output (Regression)               \\
        \textbf{Sigmoid}     & \(\frac{1}{1+e^{-z}}\)                                                                                 & \((0, 1)\)       & Output (Binary/Multi-label Class.) \& Hidden layers \\
        \textbf{Tanh}        & \(\tanh(z)\)                                                                                           & \((-1, 1)\)      & Hidden layers                      \\
        \textbf{ReLU}        & \(\max(0, z)\)                                                                                         & \([0, \infty)\)  & Hidden layers (Default choice)    \\
        \textbf{Leaky ReLU}  & \(\max(\alpha z, z)\) (small \(\alpha > 0\))                                                             & \((-\infty, \infty)\) & Hidden layers                      \\
        \textbf{Softmax}     & \(\frac{e^{z_i}}{\sum_{j} e^{z_j}}\) (vector \(\mathbf{z}\))                                              & \((0, 1)\)       & Output (Multi-class Classification)       \\
        \textbf{GELU}        & \(\approx 0.5 z (1 + \tanh[\sqrt{2/\pi} (z + 0.044715 z^3)])\)                                        & \(( \approx -0.17, \infty)\) & Hidden layers (esp. Transformers \ref{ssec:transformers}) \\
        \bottomrule
        \end{tabular}
    }
\end{table}

\subsection{Training a Neural Network}\label{sssec:training}



The process of training a Neural Network involves finding the optimal set of parameters (weights and biases, collectively denoted by \( \theta \)) that enable the network to perform a specific supervised learning task, such as classification or regression, effectively. This is achieved by iteratively adjusting the parameters \( \theta \) to minimize a predefined \textbf{loss function}, \( \mathcal{L}(\hat{y}, y) \), which measures the discrepancy between the network's predictions \( \hat{y} = f_{\theta}(x) \) and the true target values \( y \). The fundamental goal is to learn a parameter set \( \theta \) such that the network generalizes well, making accurate predictions on new, unseen data drawn from the same distribution as the training data.

Training typically proceeds in \textbf{batches}, where the network processes multiple input-output examples simultaneously in each update step. This approach improves computational efficiency, leverages hardware parallelism (especially on GPUs), and often leads to more stable convergence compared to updating parameters after every single example (stochastic gradient descent) or after processing the entire dataset (batch gradient descent). The selection of appropriate batch sizes is itself a hyperparameter that can influence training dynamics and final model performance. As introduced in Section 2.1, the data is typically split into training, validation, and test sets to guide this process and evaluate the final model's generalization capability.

\subsubsection{Loss Functions: Quantifying the Training Objective} 

The specific choice of the \textit{loss function}, \( \mathcal{L}(\hat{y}, y) \), is crucial as it mathematically defines the objective the network aims to minimize. This choice is primarily dictated by the nature of the task (e.g., regression vs. classification).

For \textbf{regression tasks}, where the goal is to predict continuous values, common choices include:
\begin{itemize}
    \item \textbf{Mean Squared Error (MSE):} This loss calculates the average of the squared differences between predictions and targets. It strongly penalizes large errors but can be sensitive to outliers.
    \[
    \mathcal{L}_{\text{MSE}}(\theta) = \frac{1}{n} \sum_{i=1}^{n} (\hat{y}_i - y_i)^2 = \frac{1}{n} \sum_{i=1}^{n} (f_{\theta}(x_i) - y_i)^2
    \]
    \item \textbf{Mean Absolute Error (MAE):} This loss averages the absolute differences, making it generally less sensitive to outliers than MSE.
    \[
    \mathcal{L}_{\text{MAE}}(\theta) = \frac{1}{n} \sum_{i=1}^{n} |\hat{y}_i - y_i| = \frac{1}{n} \sum_{i=1}^{n} |f_{\theta}(x_i) - y_i|
    \]
\end{itemize}
Here, \( n \) represents the number of samples in the current batch.

For \textbf{classification tasks}, where the network predicts class probabilities, \textit{Cross-Entropy Loss} is the standard:
\begin{itemize}
    \item \textbf{Binary Cross-Entropy (BCE):} Used for binary classification (two classes, e.g., 0 and 1), assuming the network outputs a single probability \( \hat{y}_i \in [0, 1] \) (typically via a sigmoid activation) for the positive class.
    \[
    \mathcal{L}_{\text{BCE}}(\theta) = -\frac{1}{n} \sum_{i=1}^{n} [y_i \log(\hat{y}_i) + (1 - y_i) \log(1 - \hat{y}_i)]
    \]
    It encourages \( \hat{y}_i \) towards 1 when \( y_i=1 \) and towards 0 when \( y_i=0 \).
    \item \textbf{Categorical Cross-Entropy (CCE):} Employed for multi-class classification with \( C > 2 \) mutually exclusive classes. It assumes the true labels \( y_i \) are one-hot encoded vectors and the network outputs a probability distribution \( \hat{y}_i \) over the \( C \) classes (typically via a softmax activation).
    \[
    \mathcal{L}_{\text{CCE}}(\theta) = -\frac{1}{n} \sum_{i=1}^{n} \sum_{c=1}^{C} y_{i,c} \log(\hat{y}_{i,c})
    \]
    where \( \hat{y}_{i,c} \) is the predicted probability of the \( i \)-th sample belonging to class \( c \).
\end{itemize}

Beyond these standard choices, more specialized loss functions can be employed. \textit{Hinge Loss} is sometimes used in classification settings, particularly those related to maximum-margin classification ideas from SVMs. \textit{Focal Loss} adapts the cross-entropy loss to address significant class imbalance by down-weighting the loss contribution from well-classified examples, thereby focusing training on difficult, misclassified examples (often from minority classes).

Furthermore, the primary loss function is often augmented with \textbf{regularization terms}. As mentioned in the context of LASSO (Section 2.1.3.2), these terms add a penalty based on the magnitude of the network parameters \( \theta \) to the overall loss. Common choices are L1 regularization (\( \lambda ||\theta||_1 \)) and L2 regularization (\( \lambda ||\theta||_2^2 \)), where \( \lambda \) is a hyperparameter controlling the penalty strength. Regularization discourages overly complex models with excessively large parameter values, acting as a mechanism to combat overfitting and improve the model's ability to generalize. The interplay between the primary loss and regularization shapes the final learned parameters.

\subsubsection{Backprop Algorithm: Efficiently Computing Gradients}
To minimize the chosen loss function \( \mathcal{L}(\theta) \), optimization algorithms like gradient descent require the computation of the gradient of the loss with respect to each parameter in the network, \( \nabla_{\theta} \mathcal{L} \). Given the potentially vast number of parameters and the compositional nature of NNs (layers stacked upon layers), calculating these gradients efficiently is paramount. The \textit{Backpropagation algorithm} provides this efficiency. It leverages the structure of the network, often visualized as a \textit{computational graph}, where nodes represent variables (inputs, parameters, activations) or operations (matrix multiplication, activation functions, loss calculation) and edges represent the flow of data.

The process begins with a \textit{forward pass}, where an input sample (or mini-batch) \( x \) propagates through the network layer by layer, computing the pre-activations \( z^{(l)} \) and activations \( a^{(l)} \) at each layer \( l \) until the final output \( \hat{y} = a^{(L)} \) is produced. This output is then used to compute the loss \( \mathcal{L}(\hat{y}, y) \). The computational graph implicitly stores the sequence of operations performed.

Following the forward pass, the \textit{backward pass} computes the gradients. Starting from the final loss value, backpropagation applies the chain rule of calculus recursively to compute the gradient of the loss with respect to the parameters at each layer, working backward from the output layer to the input layer. For instance, the gradient with respect to the weights \( W^{(l)} \) of layer \( l \) depends on the gradient of the loss with respect to the activations \( a^{(l)} \) of that layer, which in turn depends on the gradients from the subsequent layer \( l+1 \). Symbolically, the chain rule allows us to compute \( \frac{\partial \mathcal{L}}{\partial W^{(l)}} \) by propagating gradients backward:
\[
\frac{\partial \mathcal{L}}{\partial W^{(l)}} = \frac{\partial \mathcal{L}}{\partial a^{(l)}} \frac{\partial a^{(l)}}{\partial z^{(l)}} \frac{\partial z^{(l)}}{\partial W^{(l)}}
\]
where \( \frac{\partial a^{(l)}}{\partial z^{(l)}} \) involves the derivative of the activation function and \( \frac{\partial z^{(l)}}{\partial W^{(l)}} \) depends on the activations from the previous layer \( a^{(l-1)} \). Modern deep learning frameworks automate this gradient calculation via automatic differentiation based on the constructed computational graph.

Once the gradients \( \nabla_{\theta} \mathcal{L} \) are computed for all parameters \( \theta \), an \textit{optimization algorithm} is used to update the parameters. The simplest is \textit{gradient descent}, which updates parameters in the opposite direction of the gradient:
\[
\theta \leftarrow \theta - \eta \nabla_{\theta} \mathcal{L}
\]
where \( \eta \) is the \textit{learning rate}, a hyperparameter controlling the step size. Various sophisticated optimizers exist (discussed in Section \ref{ssec:training}) that adapt the learning rate or incorporate momentum to improve convergence speed and stability.

Backpropagation, while powerful, faces challenges in very deep networks. \textit{Vanishing gradients} can occur when gradients become extremely small during the backward pass (especially with saturating activation functions like sigmoid or tanh ; cf. \ref{ssec:neuron_act}), effectively halting learning in earlier layers. Conversely, \textit{exploding gradients} involve gradients growing excessively large, leading to numerical instability. Techniques like using non-saturating activations (e.g., ReLU), careful initialization, normalization layers (e.g., Batch Normalization), residual connections, and \textit{gradient clipping} (capping the maximum magnitude of gradients) are employed to mitigate these issues. Proper tuning of the learning rate is also crucial for stable and effective training.

\subsubsection{Hyperparameter finetuning}\label{sec:finetuning}

The performance of a NN is highly sensitive not only to its learned parameters (weights and biases) but also to a set of \textit{hyperparameters} that are chosen before the training process begins. These hyperparameters define the network's architecture and the training procedure itself. Examples include the \textit{learning rate} \( \eta \), the \textit{batch size} (number of samples processed before each parameter update), the number of hidden layers, the number of units (neurons) in each layer, the choice of \textit{activation functions}, the type of optimizer used, and regularization strengths (e.g., the \( \lambda \) in L1/L2 regularization or the dropout rate).

\textit{Finetuning} or \textit{hyperparameter optimization} is the process of systematically searching for the combination of hyperparameters that yields the best performance, typically measured on the validation set. The goal is to find settings that allow the model to learn effectively from the training data while also generalizing well to unseen data. Several strategies exist for this search. \textit{Grid Search} involves defining a discrete set of values for each hyperparameter and evaluating the model performance for every possible combination. While exhaustive, this approach can be computationally very expensive, especially with many hyperparameters. \textit{Random Search}, where hyperparameter values are sampled randomly from specified distributions or ranges, is often found to be more efficient in practice, as model performance is frequently more sensitive to some hyperparameters than others. More sophisticated methods like \textit{Bayesian Optimization} build a probabilistic model of the relationship between hyperparameters and performance, using past evaluations to intelligently select the next set of hyperparameters to try, aiming to find good configurations more quickly. Evaluating each hyperparameter configuration often involves \textit{cross-validation} on the training set, where the data is split into multiple folds, and the model is trained and evaluated multiple times, rotating which fold is used for validation. This provides a more robust estimate of performance for a given hyperparameter setting compared to a single train-validation split.

\subsubsection{Overfitting}

A central challenge in training complex models like NNs is \textit{overfitting}. This occurs when the model learns the training data too well, capturing not only the underlying patterns but also the noise and specific idiosyncrasies of the training samples. An overfitted model exhibits excellent performance on the training data but fails to generalize to new, unseen data, resulting in poor performance on the validation and test sets. It essentially memorizes the training examples rather than learning the underlying concepts.

Overfitting can be detected by monitoring the model's performance on both the training and validation sets throughout the training process (as alluded to in Figure \ref{fig:toy_example}). Typically, both training loss and validation loss will decrease initially. However, if training continues for too long, the training loss might keep decreasing while the validation loss starts to increase. This divergence is a clear sign of overfitting: the model is becoming overly specialized to the training data at the expense of generalization.

Several techniques are commonly employed to combat overfitting. \textit{Regularization} methods add a penalty term to the loss function based on the magnitude of the network's weights (e.g., L1 or L2 regularization) or introduce randomness during training. \textit{Dropout} is a popular regularization technique that randomly sets a fraction of neuron activations to zero during each training update, preventing the network from becoming too reliant on any specific neurons or pathways. \textit{Early Stopping} involves monitoring the validation loss (or another validation metric) and stopping the training process when this metric ceases to improve or starts to worsen, even if the training loss is still decreasing. This prevents the model from entering the overfitting regime. \textit{Data Augmentation} artificially increases the size and diversity of the training dataset by applying random transformations (e.g., rotations, flips, color shifts for images; synonym replacement for text) to existing samples. This exposes the model to more variations, encouraging it to learn more robust features. \textit{Simplifying the Model} by reducing the number of layers or units can decrease its capacity, making it less prone to fitting noise, though this must be balanced against the risk of underfitting (where the model is too simple to capture the underlying patterns). Finally, techniques like \textit{Cross-Validation}, while primarily used for evaluation and hyperparameter tuning, also give a better indication of generalization performance than a single train-validation split, indirectly helping in the selection of models that are less overfitted.

\subsection{Sophisticated Model Architectures}\label{ssec:otharch}
There exists a model architecture for every possible application of a NN, like image classification, sentiment analysis, text generation, etc. In the following sections, we will delve into two of the most common ones: Autoencoders (Section \ref{sec:autoencoders}) and Transformers (Section \ref{sec:attention_and_transformers}). Other most common architectures will be given in Section \ref{sec:otherarch}, as they are not building blocks for the Chapter \ref{chap:mech_int}.

\section{Autoencoders}\label{sec:autoencoders}

Autoencoders (AEs) represent a class of NNs primarily utilized for unsupervised learning paradigms. In such settings, the learning objective is intrinsically defined by the structure within the data itself, rather than relying on explicit target labels \citep{Goodfellow2016}. The fundamental goal of an AE is to learn a compressed, lower-dimensional latent representation of the input data, from which the original input can be subsequently reconstructed with minimal information loss. This process effectively learns a non-linear projection onto a potentially lower-dimensional manifold capturing the principal factors of variation within the data distribution.

\subsection{The Encoder-Decoder Architecture}

An archetypal autoencoder comprises two main components: an encoder network and a decoder network.

The \textit{encoder}, denoted by a function \( f_{\theta} \), maps an input vector \( x \in \mathcal{X} \) (typically \( \mathcal{X} = \mathbb{R}^d \)) to a latent representation \( z \in \mathcal{Z} \) (typically \( \mathcal{Z} = \mathbb{R}^k \) with \( k \le d \)). This mapping is parameterized by \( \theta \):
\[ z = f_{\theta}(x) \]
The aim of this encoding process is to break down the most prominent features of the input into a compact representation, thereby reducing dimensionality or extracting features.

The \textit{decoder}, denoted by a function \( g_{\phi} \), maps the latent representation \( z \) back to the original input space \( \mathcal{X} \), producing a reconstruction \( \hat{x} \). This mapping is parameterized by \( \phi \):
\[ \hat{x} = g_{\phi}(z) \]
Thus, the autoencoder seeks to learn parameters \( \theta \) and \( \phi \) that approximate the identity function through the composition \( g_{\phi} \circ f_{\theta} \). The learning objective is typically formulated as minimizing a reconstruction loss function \( \mathcal{L}(x, \hat{x}) \) averaged over the data distribution \( p_{\text{data}}(x) \):
\[ \min_{\theta, \phi} \mathbb{E}_{x \sim p_{\text{data}}(x)} [\mathcal{L}(x, g_{\phi}(f_{\theta}(x)))] \]
For continuous data, the Mean Squared Error (MSE) is commonly employed:
\[ \mathcal{L}_{\text{MSE}}(x, \hat{x}) = \| x - \hat{x} \|_2^2 = \frac{1}{d} \sum_{i=1}^{d} (x_i - \hat{x}_i)^2 \]
where \( d \) is the dimensionality of the input vector \( x \). For binary data, the Binary Cross-Entropy (BCE) loss is often more appropriate.

\subsection{Variational Autoencoders}\label{ssec:vae}

Variational Autoencoders (VAEs) \citep{Kingma2013AutoEncodingVB, Rezende2014} extend the standard AE framework by introducing a probabilistic perspective, framing the autoencoder within the context of latent variable models and variational inference. Instead of mapping the input \( x \) to a single point \( z \) in the latent space, the VAE encoder learns a mapping to the parameters of a probability distribution over the latent space.

Specifically, the encoder, often called the \textit{inference network} \( q_{\theta}(z|x) \), typically outputs the parameters of a diagonal Gaussian distribution: the mean vector \( \mu_{\theta}(x) \) and the log-variance vector \( \log(\sigma_{\theta}^2(x)) \).
\[ \mu_{\theta}(x), \log(\sigma_{\theta}^2(x)) = f_{\theta}(x) \]
A latent vector \( z \) is then sampled from this approximate posterior distribution:
\[ z \sim q_{\theta}(z|x) = \mathcal{N}(z | \mu_{\theta}(x), \text{diag}(\sigma_{\theta}^2(x))) \]
The decoder, which we call the \textit{generative network} \( p_{\phi}(x|z) \), then reconstructs the input \( \hat{x} \) from the sampled latent vector \( z \):
\[ \hat{x} \sim p_{\phi}(x|z) \]
where \( p_{\phi}(x|z) \) is often modeled as a Gaussian \( \mathcal{N}(x | g_{\phi}(z), \sigma^2 I) \) (leading to an MSE-like reconstruction term) or a Bernoulli distribution (leading to a BCE-like term).

Training a VAE involves maximizing the Evidence Lower Bound (ELBO) on the marginal log-likelihood \( \log p(x) \), which is equivalent to minimizing the negative ELBO:
\[ \mathcal{L}_{\text{VAE}}(\theta, \phi; x) = -\mathbb{E}_{z \sim q_{\theta}(z|x)}[\log p_{\phi}(x|z)] + D_{\text{KL}}(q_{\theta}(z|x) \| p(z)) \]
The first term represents the expected negative log-likelihood, acting as the reconstruction loss. The second term is the Kullback-Leibler (KL) divergence between the approximate posterior \( q_{\theta}(z|x) \) and a prior distribution over the latent variables \( p(z) \), typically chosen as a standard multivariate Gaussian \( \mathcal{N}(0, I) \). This KL divergence acts as a regularization term, encouraging the learned latent distributions to resemble the prior. A weighting factor \( \beta \) is often introduced to balance the two terms \citep{Higgins2017betaVAE}, particularly for disentanglement purposes:
\[ \mathcal{L}_{\text{VAE}} = \mathcal{L}_{\text{reconstruction}} + \beta \mathcal{L}_{\text{KL}} \]

\subsection{Applications}

Autoencoders, through their capacity to learn meaningful data representations, find utility across diverse machine learning applications, despite not being optimized for a singular discriminative or predictive task. Key applications include:

\paragraph{Dimensionality Reduction}
The encoder \( f_{\theta} \) provides a mapping to a lower-dimensional latent space \( \mathcal{Z} \), offering a potentially non-linear alternative to methods like Principal Component Analysis (PCA) for data visualization or compact representation.

\paragraph{Feature Learning}
The learned latent representations \( z = f_{\theta}(x) \) or \( z_q(x) \) can serve as input features for subsequent supervised learning tasks, such as classification or regression, potentially improving performance by capturing essential data characteristics.

\paragraph{Anomaly Detection}
Trained predominantly on nominal data, autoencoders typically exhibit higher reconstruction errors \( \mathcal{L}(x, \hat{x}) \) when presented with inputs \( x \) that deviate significantly from the learned data manifold. This property allows for their use in detecting outliers or anomalies.

\paragraph{Generative Modeling}
Probabilistic variants like VAEs excel as generative models. New data samples resembling the training distribution can be synthesized by sampling from the latent prior distribution \( p(z) \) (for VAEs), and subsequently passing these samples through the decoder network \( g_{\phi} \).

\section{Attention Mechanism \& Transformer Architecture}\label{sec:attention_and_transformers}

Another very popular type of NN are attention-based networks, at the base of the Transformer (\ref{ssec:transformers}) architecture. Transformers are particularly useful in Natural Language Processing (NLP) tasks, especially since the emergence of Large Language Models (LLMs) in recent years. They demonstrated State of the Art capabilities in language understanding, text generation, and complex problem solving.

\subsection{Attention Mechanism}\label{ssec:attention}
Attention mechanisms have emerged as the biggest breakthrough in deep learning, particularly for processing sequential data such as natural language, speech, and time-series. Originally introduced in to enhance encoder-decoder architectures in tasks like machine translation, it was later reformalized by \citep{VaswaniSPUJGKP17}. Attention allows models to dynamically focus on relevant parts of the input sequence, addressing the limitations of fixed-size representations.
The intuition behind is derived from classical linear networks (like MLP \ref{ssec:mlp}) by \textit{dynamically} learning weight matrices $W^{(l)}$. If we think each of the weight matrix's row as a selector for the input's linear combinations, then the attention mechanism computes a learned, data-dependent, weighted selection these linear combinations of the input. The latter is first encoded into a matrix $V$, similarly to linear networks, and then a weight matrix refines it into a selection of important features. The weight matrix is constructed by first creating "queries" of the inputs, encoded into a matrix $Q$, and matching those queries against a database of keys to compute a score value. Those keys, encoded from the input into a matrix $K$, compute \textit{attention scores} of the inputs into a matrix $QK^\top$. Those scores are then mapped to probabilities using the \textit{softmax} activation (see \ref{ssec:neuron_act}), which we call the weight matrix.

This mechanism thus aim to refine the feature selection to be more data-driven and context dependent. More formally, if we define the input $X\in\R^{n\times d}$, $n$ being the sequence length and $d$ the embedding size of the data, then the self-attention (cf. section \ref{sec:differentattention} below) matrices $K$, $V$, and $Q$ are defined as follow :
\[
Q = X W_Q, \quad K = X W_K, \quad V = X W_V
\]
where \( W_Q, W_K, W_V \in \mathbb{R}^{d \times d_k} \) are learned projections. The formula for the full attention mechanism is thus given by:
\[
\text{Attention}(Q, K, V) = \text{softmax}\left(\frac{QK^T}{\sqrt{d_k}}\right) V
\]

where the scaling term $\frac{1}{\sqrt{d_k}}$ has been introduced to prevent large magnitudes, ensuring stable gradients in softmax.

\subsection{Different type of attention}\label{sec:differentattention}

\subsubsection{Self-attention}
This is the most straightforward one, the key-value matrix pair is computed from the previous hidden states, as well as the query matrix, and as described in \ref{ssec:attention}.

\subsubsection{Cross-attention}

We also differentiate \textit{Self-Attention} from \textit{Cross-attention}. Contrary to self-attention, cross-attention computes the key-value pair from the encoder's output, and use the precedent decoder's hidden layer to compute the queries. It is particularly useful in the setting where we are decoding the next datapoint from a encoded sequence, and the decoder's architecture uses the learned representation of the encoder (cf. section \ref{ssec:transformers}). The difference also lies in the dimension of matrices $K$ and $V$, which now has dimension $m\times d_k$ ($m$ being the dimension of the encoder).

\subsubsection{Multi-Head Attention}
Another architecture change that improves the capture of diverse relationships it the Multi-Head Attention (MHA). Multiple attention mechanisms run in parallel sourcing from the same hidden layer, and each capturing there own semantic relationship of the data. For each of the $h$ heads, we have 
\[
\text{head}_i = \text{Attention}(Q W_{Q_i}, K W_{K_i}, V W_{V_i})
\]

where \( W_{Q_i}, W_{K_i}, W_{V_i} \in \mathbb{R}^{d \times d_k} \). The final output is:
\[
\text{MultiHead}(Q, K, V) = \text{Concat}(\text{head}_1, \dots, \text{head}_h) W_O
\]

with \( W_O \in \mathbb{R}^{h d_k \times d} \) allowing to map back into the hidden layer's original dimension.

\subsubsection{Causal Attention/Masking}
One last building block for the Transformer architecture is the masking, which is more generally called \textit{Causal Attention}. The usual technique makes sort that no weight is put on forward lookups by elements of the sequence. For example, in a sentence, the words at its beginning has an effect on the end of the sentence, but the words at the end can't have an effect on those from the beginning. This is represented in the attention weights by setting the upper-triangle part of the $QK^\top$ matrix to $0$s. In general, other type of intervention can be done in the attention weights, in order to enforce no relations at all (setting an entry to 0) or strong relation to only one point/a set of points.

\subsection{Transformer Architecture}\label{ssec:transformers}
The transformer architecture was introduced by \citep{VaswaniSPUJGKP17} as a state-of-the-art method for machine translation. It heavily relies on the attention mechanism, combining self and cross attention, MHA and causal attention. It is structured as an encoder-decoder, each consisting of sequential attention layers: One MHA followed by a FFN. Each step also applies normalization \citep{LayerNorm} of the data and residual skip connections \citep{he2016deep}, techniques used to improve gradient computations. Cross-attention and masking is applied in the decoder part of the transformer, where we first input the targets' into a masked MHA, feeding the outputted hidden layer to the next MHA's query, while its key-value pair is computed from the encoder's output. Finally, the resulting attention is passed through a FFN then into the next attention layer. More formally, for the encoder, we have:
\[
\begin{aligned}
h_1 &= \text{LayerNorm}(x + \text{MHA}(x, x, x)) \\
h_2 &= \text{LayerNorm}(h_1 + \text{FFN}(h_1))
\end{aligned}
\]
where $x$ is the input to the encoder, and for the decoder we have:
\[
\begin{aligned}
h_1 &= \text{LayerNorm}(y + \text{MaskedMHA}(y, y, y)) \\
h_2 &= \text{LayerNorm}(h_1 + \text{MHA}(h_1, h_{\text{enc}}, h_{\text{enc}})) \\
h_3 &= \text{LayerNorm}(h_2 + \text{FFN}(h_2))
\end{aligned}
\]
where $y$ is the input to the decoder, \textit{MaskedMHA} is the masked MHA, and $h_{\text{enc}}$ is the output from the encoder. \textit{MHA} (and consequently \textit{MaskedMHA}) takes as input, in the order, the query, the key, and the value matrices.

\subsection{Large Language Models are Decoder-only transformers}
The advent of the transformer architecture have been majorly influenced by Large Language Models (LLMs), a class of deep learning models built on top of the attention mechanism and specialized in NLP tasks.

\begin{figure}[htp]
    \centering
    \includegraphics[width=\linewidth]{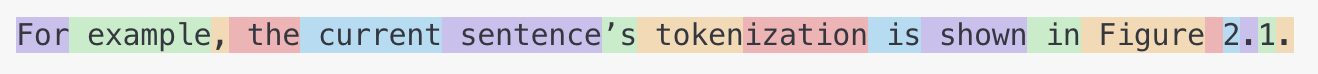}
    \caption{Tokenization of the sentence "For example, the current sentence's tokenization is shown in Figure \ref{fig:tokenization}." with OpenAI's GPT-4o tokenizer.}
    \label{fig:tokenization}
\end{figure}

LLMs architecture tweaked a bit the classical transformer's architecture, by namely skipping the encoding part. Indeed, as each text input must first be embedded into a vector -- i.e. a numerical representation allowing computations -- this sort of acts as an encoder. We call this part the \textit{"tokenizing"} of the data and each fragment of text is thus called a \textit{token}, corresponding to one element of the sequence. For example, the current sentence's \textit{tokenization} is shown in Figure \ref{fig:tokenization}. The embedded tokens are then learned to be decoded in a self-masked-multi-head-attention setting.

\section{Other Architectures \& Techniques}\label{sec:otherarch}
While FFNs (cf. Section \ref{sec:ffn}), AEs and VAEs (cf. Section \ref{sec:autoencoders}), and the Transformer architecture (cf. Section \ref{sec:attention_and_transformers}) represent foundational and highly successful NN paradigms, the field encompasses a broader range of specialized architectures. This section briefly discusses three prominent examples: Convolutional Neural Networks (CNNs), primarily used for grid-like data such as images; Recurrent Neural Networks (RNNs), designed for sequential data; and Diffusion Models, a powerful class of modern generative models. 

\subsection{Convolutional Neural Networks}\label{ssec:cnn}
\textbf{Convolutional Neural Networks (CNNs)} \citep{homma1987convolution} are a class of deep NNs particularly effective for processing data with a known grid-like topology, such as images. Their architecture is inspired by the organization of the animal visual cortex and designed to automatically and adaptively learn spatial hierarchies of features, from low-level edges and textures to high-level concepts.

\subsubsection{The Convolution Operation}
The core of CNNs is the convolutional layer, which applies a convolution operation to the input, passing the result through a non-linear activation function. Unlike fully connected layers, convolutional layers leverage parameter sharing and local connectivity.

\paragraph{Local Connectivity and Weight Sharing} Instead of connecting every input neuron to every output neuron, each output neuron in a convolutional layer is connected only to a small region of the input volume, known as the \textit{receptive field}. Furthermore, all neurons in a single output feature map share the same set of weights (the \textit{kernel} or \textit{filter}). This dramatically reduces the number of parameters compared to a fully connected network and makes the model somewhat invariant to the location of features in the input.

\paragraph{Mathematical Formulation} Consider a 2D input volume (e.g., a grayscale image or a feature map from a previous layer) denoted by $X \in \mathbb{R}^{H \times W}$. A convolutional layer applies a set of $K$ filters, where each filter $F_k \in \mathbb{R}^{h \times w}$ (with $h \ll H, w \ll W$) slides across the input. The convolution operation (*) for a single filter $F_k$ at spatial location $(i, j)$ in the output feature map $Z_k$ is defined as:
\begin{multline}
    Z_k[i, j] = (X * F_k)[i, j] + b_k \\
    = \sum_{m=1}^{h} \sum_{n=1}^{w} X[i \cdot S + m - P - 1, j \cdot S + n - P - 1] \cdot F_k[m, n] + b_k
\end{multline}
where $b_k \in \mathbb{R}$ is a bias term associated with filter $F_k$, $S$ is the \textit{stride} dictating the step size of the filter across the input, and $P$ is the amount of zero-padding applied to the input borders (often used to control the spatial dimensions of the output). The indices must be handled carefully at the boundaries, especially considering padding. For multi-channel inputs $X \in \mathbb{R}^{H \times W \times C_{in}}$, each filter becomes a volume $F_k \in \mathbb{R}^{h \times w \times C_{in}}$, and the summation extends over the input channels as well, producing an output $Z_k \in \mathbb{R}^{H' \times W'}$, where $H'$ and $W'$ depend on $H, W, h, w, S, P$. The full output volume $Z \in \mathbb{R}^{H' \times W' \times K}$ comprises the feature maps from all $K$ filters.

\subsubsection{Pooling Layers}
Another important component of many CNN architectures is the pooling layer. Its function is to progressively reduce the spatial size of the representation, reducing the amount of parameters and computation in the network, and also helping to make the learned features more robust to small translations and distortions.

Common pooling operations include max pooling and average pooling. A pooling layer operates independently on each feature map slice $A_k$ of the input activations $A$. It partitions the feature map into non-overlapping (or sometimes overlapping) rectangular regions and, for each region, outputs a single value.
For max pooling with a pooling region of size $p \times p$ and stride $S_p$, the output $P_k$ is:
\begin{equation}
    P_k[i, j] = \max_{m=1..p, n=1..p} A_k[i \cdot S_p + m - 1, j \cdot S_p + n - 1]
\end{equation}
Average pooling computes the arithmetic mean instead of the maximum over the region.

\subsection{Recurrent Neural Networks}\label{ssec:rnn}
\textbf{Recurrent Neural Networks (RNNs)} \citep{RumelhartHintonWilliams1985} are designed to operate on sequential data, such as time series or natural language. Unlike FFNs (cf. Section \ref{sec:ffn}), RNNs possess connections that form directed cycles, allowing them to maintain an internal state or "memory" that captures information about previous elements in the sequence.

\subsubsection{The Recurrence Relation}
The core idea of an RNN is to apply the same set of weights recursively over the elements of a sequence. At each time step $t$, the network receives an input $x_t \in \mathbb{R}^{d_x}$ and the hidden state $h_{t-1} \in \mathbb{R}^{d_h}$ from the previous time step. It computes the new hidden state $h_t$ and, optionally, an output $y_t \in \mathbb{R}^{d_y}$.

\paragraph{Mathematical Formulation} The basic RNN update rules are:
\begin{align}
    h_t &= \sigma_h (W_{xh} x_t + W_{hh} h_{t-1} + b_h)  \\
    y_t &= \sigma_y (W_{hy} h_t + b_y) 
\end{align}
Here, $W_{xh} \in \mathbb{R}^{d_h \times d_x}$, $W_{hh} \in \mathbb{R}^{d_h \times d_h}$, and $W_{hy} \in \mathbb{R}^{d_y \times d_h}$ are weight matrices, and $b_h \in \mathbb{R}^{d_h}$, $b_y \in \mathbb{R}^{d_y}$ are bias vectors. These parameters are shared across all time steps. $\sigma_h$ is typically a non-linear activation function like $\tanh$ or $\relu$, while $\sigma_y$ depends on the task (e.g., $\softmax$ for classification at each step). The initial hidden state $h_0$ is often initialized to a zero vector.

\subsubsection{Training and Challenges}
RNNs are typically trained using Backpropagation Through Time (BPTT), which involves unrolling the network over the sequence length and applying standard backpropagation.

\paragraph{Vanishing and Exploding Gradients} A major challenge in training simple RNNs is the vanishing or exploding gradient problem. Due to the repeated multiplication by the recurrent weight matrix $W_{hh}$ during backpropagation, gradients can either shrink exponentially towards zero (vanishing) or grow exponentially large (exploding) as they propagate back through time. This makes it difficult for standard RNNs to learn long-range dependencies in sequences.

\paragraph{Advanced Architectures} To mitigate these issues, more sophisticated RNN variants like Long Short-Term Memory (LSTM) networks and Gated Recurrent Units (GRUs) were developed. These architectures incorporate gating mechanisms that learn to control the flow of information, allowing them to selectively remember or forget information over longer time scales and significantly improving their ability to capture long-range dependencies.

\subsubsection{Long Short-Term Memory (LSTM)}\label{sssec:lstm}
To address the difficulties simple RNNs face in learning long-range dependencies, primarily due to the vanishing and exploding gradient problems, more sophisticated recurrent architectures were developed. Among the most successful and widely adopted is the Long Short-Term Memory (LSTM) network \citep{hochreiter1997long}. LSTMs introduce a more complex internal structure designed to regulate the flow of information through time via specialized gating mechanisms.

\paragraph{Cell State and Gating Mechanisms} The main innovation of LSTM is the introduction of a \textit{cell state}, $c_t \in \mathbb{R}^{d_h}$, which acts as an information superhighway, allowing information to flow through the sequence with minimal distortion. The flow of information into and out of the cell state, as well as the generation of the hidden state $h_t$, is controlled by three primary gates: the forget gate, the input gate, and the output gate. Each gate uses a sigmoid activation function, $\sigma_g(z) = (1 + e^{-z})^{-1}$, which outputs values between 0 and 1, indicating how much of each component of information should be let through.

\subparagraph{Forget Gate ($f_t$)} This gate decides what information to discard from the previous cell state $c_{t-1}$. It looks at the previous hidden state $h_{t-1}$ and the current input $x_t$:
\begin{equation}
    f_t = \sigma_g(W_f [h_{t-1}, x_t] + b_f)
\end{equation}
Here, $[h_{t-1}, x_t]$ denotes the concatenation of the vectors $h_{t-1}$ and $x_t$. $W_f$ and $b_f$ are the weight matrix and bias vector for the forget gate, respectively.

\subparagraph{Input Gate ($i_t$) and Candidate Values ($\tilde{c}_t$)} This gate determines what new information will be stored in the cell state. It consists of two parts: first, the input gate layer decides which values to update, and second, a $\tanh$ layer creates a vector of new candidate values, $\tilde{c}_t$, that could be added to the state.
\begin{align}
    i_t &= \sigma_g(W_i [h_{t-1}, x_t] + b_i)  \\
    \tilde{c}_t &= \tanh(W_C [h_{t-1}, x_t] + b_C) 
\end{align}
$W_i, b_i$ and $W_C, b_C$ are the parameters for the input gate and candidate value layers.

\subparagraph{Cell State Update} The previous cell state $c_{t-1}$ is updated to the new cell state $c_t$. The old state is multiplied element-wise ($\odot$) by the forget gate $f_t$ (forgetting things), and then the product of the input gate $i_t$ and the candidate values $\tilde{c}_t$ is added (adding new information).
\begin{equation}
    c_t = f_t \odot c_{t-1} + i_t \odot \tilde{c}_t 
\end{equation}

\subparagraph{Output Gate ($o_t$) and Hidden State ($h_t$)} Finally, the output gate decides what parts of the cell state will be output. The cell state is passed through $\tanh$ (pushing values to be between -1 and 1) and then multiplied element-wise by the output of the sigmoid gate $o_t$. This result is the new hidden state $h_t$.
\begin{align}
    o_t &= \sigma_g(W_o [h_{t-1}, x_t] + b_o)  \\
    h_t &= o_t \odot \tanh(c_t) 
\end{align}
$W_o, b_o$ are the parameters for the output gate. The hidden state $h_t$ is then used for predictions, and is also fed back into the LSTM unit at the next time step.

\paragraph{Information Flow Control} The gating mechanisms, combined with the additive interaction for the cell state update, allow LSTMs to learn long-term dependencies. By setting gates appropriately (close to 0 or 1), the network can choose to remember information over many time steps (if $f_t \approx 1, i_t \approx 0$) or forget past information and incorporate new inputs. This structure provides a more stable path for gradients during backpropagation compared to simple RNNs, significantly mitigating the vanishing gradient problem.

\subsection{Diffusion Models}\label{ssec:diffusion}
Diffusion models, specifically \textbf{Denoising Diffusion Probabilistic Models (DDPMs)} \citep{ho2020denoising}, represent a powerful class of generative models that have recently achieved state-of-the-art results in generating high-fidelity data, particularly images. They operate by systematically destroying structure in data through a forward diffusion process and then learning a reverse denoising process to generate data from noise.

\subsubsection{Forward Diffusion Process}
The forward process gradually adds Gaussian noise to a data sample $x_0$ drawn from the true data distribution $q(x_0)$ over a sequence of $T$ time steps. This is typically defined as a Markov chain where the noise added at each step $t$ is controlled by a variance schedule $\{\beta_t\}_{t=1}^T$, with $\beta_t \in (0, 1)$.
\begin{equation}
    q(x_t | x_{t-1}) = \mathcal{N}(x_t; \sqrt{1 - \beta_t} x_{t-1}, \beta_t \mathbf{I})    
\end{equation}
A useful property of this process is that we can sample $x_t$ directly from $x_0$ in closed form. Let $\alpha_t = 1 - \beta_t$ and $\bar{\alpha}_t = \prod_{i=1}^t \alpha_i$. Then:
\begin{equation}
    q(x_t | x_0) = \mathcal{N}(x_t; \sqrt{\bar{\alpha}_t} x_0, (1 - \bar{\alpha}_t) \mathbf{I})
\end{equation}
As $t \to T$, if the schedule $\{\beta_t\}$ and $T$ are chosen appropriately, $x_T$ becomes approximately distributed according to a standard Gaussian distribution $\mathcal{N}(0, \mathbf{I})$, effectively destroying all information about the original data sample $x_0$.

\subsubsection{Reverse Denoising Process}
The generative process involves reversing the diffusion. Starting from pure noise $x_T \sim \mathcal{N}(0, \mathbf{I})$, the model learns to iteratively denoise the sample to produce a data point $x_0$. This reverse process is also modeled as a Markov chain $p_\theta(x_{t-1} | x_t)$, parameterized by a NN $\theta$.
\begin{equation}
    p_\theta(x_{t-1} | x_t) = \mathcal{N}(x_{t-1}; \mu_\theta(x_t, t), \Sigma_\theta(x_t, t))
\end{equation}
If $\beta_t$ are small, the true reverse conditional $q(x_{t-1} | x_t, x_0)$ is also Gaussian. The goal is to train the network $\theta$ to approximate this true conditional. The variance $\Sigma_\theta(x_t, t)$ is often fixed (e.g., $\Sigma_\theta(x_t, t) = \beta_t \mathbf{I}$ or $\tilde{\beta}_t \mathbf{I}$ where $\tilde{\beta}_t = \frac{1-\bar{\alpha}_{t-1}}{1-\bar{\alpha}_t}\beta_t$), while the network learns the mean $\mu_\theta(x_t, t)$.

\subsubsection{Training Objective}
Instead of directly predicting $\mu_\theta(x_t, t)$, it is common practice to reparameterize the network to predict the noise $\epsilon$ that was added at step $t$, denoted $\epsilon_\theta(x_t, t)$. Using the relationship $x_t = \sqrt{\bar{\alpha}_t} x_0 + \sqrt{1 - \bar{\alpha}_t} \epsilon$, where $\epsilon \sim \mathcal{N}(0, \mathbf{I})$, the mean can be expressed in terms of the predicted noise. A simplified training objective, derived from the variational lower bound on the data log-likelihood, often involves minimizing the mean squared error between the true noise and the predicted noise:
\begin{equation}
    L_{simple}(\theta) = \mathbb{E}_{t \sim U(1, T), x_0 \sim q(x_0), \epsilon \sim \mathcal{N}(0, \mathbf{I})} \left[ || \epsilon - \epsilon_\theta(\sqrt{\bar{\alpha}_t} x_0 + \sqrt{1 - \bar{\alpha}_t} \epsilon, t) ||^2 \right]
\end{equation}
The NN $\epsilon_\theta$, often implemented using architectures like U-Nets which are adept at image-to-image tasks, takes the noisy data $x_t$ and the time step $t$ as input and outputs a prediction of the noise component. During generation, one samples $x_T \sim \mathcal{N}(0, \mathbf{I})$ and iteratively samples $x_{t-1} \sim p_\theta(x_{t-1} | x_t)$ for $t = T, T-1, \dots, 1$ to obtain the final sample $x_0$.

\chapter{Causal Estimation: From TMLE to Deep Learning}\label{chap:causality_tl}

\section{Fundamentals of Causal Inference in Bio-statistics}
\label{sec:fundamentals_causal_bio}

Bio-statistics frequently confronts questions extending beyond mere association towards establishing causal relationships. Determining whether a new drug \textit{causes} improved patient outcomes, or if an environmental exposure \textit{causes} an increased risk of disease, necessitates a framework distinct from standard correlational analysis. While randomized controlled trials (RCTs) provide a gold standard by balancing confounders through randomization, much bio-statistical data originates from observational studies where treatments or exposures are not assigned randomly. This prevalence of observational data underscores the critical need for robust causal inference methodologies capable of addressing inherent biases.

\subsection{Statistical Preliminaries for Causal Questions}
\label{subsec:stat_preliminaries}

Before formalizing causal concepts, we revisit essential statistical inference tools. Hypothesis testing provides a framework for evaluating evidence against a specific claim. We typically formulate a \textit{null hypothesis} ($H_0$), often representing no effect or no difference (e.g., the drug has no effect), and an \textit{alternative hypothesis} ($H_1$) representing the presence of an effect. A test statistic, calculated from sample data, quantifies the discrepancy between the observed data and $H_0$. The resulting \textit{p-value} is the probability of observing a test statistic at least as extreme as the one computed, assuming $H_0$ is true. If the p-value falls below a pre-specified significance level $\alpha$ (commonly 0.05), we reject $H_0$. This process carries risks: a \textit{Type I error} (probability $\alpha$) occurs when rejecting a true $H_0$, while a \textit{Type II error} (probability $\beta$) occurs when failing to reject a false $H_0$.

While hypothesis testing asks \textit{whether} an effect exists, estimation quantifies its \textit{magnitude}. A \textit{point estimate}, $\hat{\theta}$, provides a single best guess for a population parameter (e.g., the difference in mean recovery times) based on the sample. However, it lacks a measure of uncertainty. A \textit{confidence interval} (CI), typically expressed as $[L, U]$, provides a range of plausible values for the true parameter, calculated from the data, with a specified confidence level (e.g., 95\%). Narrower intervals indicate greater precision.

The ability of a study to detect a true effect, if one exists, is its \textit{statistical power}, defined as $1-\beta$. Adequate power (often targeted at 80\% or higher) is crucial for study validity. Power depends on the chosen significance level $\alpha$, the sample size $n$, the magnitude of the true effect (effect size), and the variability of the data (e.g., outcome variance $\sigma^2$). Power calculations are essential during study design to determine the necessary sample size $n$ to achieve a desired power for detecting a clinically meaningful effect.

\subsection{Defining Causality: The Potential Outcomes Framework}
\label{subsec:potential_outcomes}

The Potential Outcomes framework, also known as the Neyman-Rubin Causal Model (RCM) \citep{neyman-rubin-model}, provides a formal language for defining causal effects. Consider a simple scenario with a binary treatment $A$ (where $A=1$ denotes treatment and $A=0$ denotes control/no treatment) and an outcome $Y$. For each individual unit $i$ in the population, we define two \textit{potential outcomes}: $Y_i(1)$, the outcome that \textit{would have been observed} had individual $i$ received the treatment ($A=1$), and $Y_i(0)$, the outcome that \textit{would have been observed} had individual $i$ received the control ($A=0$).

The \textit{individual treatment effect} (ITE) for unit $i$ is the difference between their potential outcomes: $ITE_i = Y_i(1) - Y_i(0)$. This represents the true causal effect of the treatment specifically for that individual. However, the \textit{fundamental problem of causal inference} is that we can only observe one potential outcome for each individual. The observed outcome $Y_i$ can be written as $Y_i = A_i Y_i(1) + (1-A_i)Y_i(0)$. If individual $i$ received the treatment ($A_i=1$), we observe $Y_i(1)$ but $Y_i(0)$ remains unobserved (counterfactual). Conversely, if $A_i=0$, we observe $Y_i(0)$ but $Y_i(1)$ is counterfactual.

Since ITEs are typically unobservable, causal inference often focuses on estimating average causal effects at the population level. The primary target is often the \textit{Average Treatment Effect} (ATE), defined as the expected difference in potential outcomes across the entire population:
\[ \text{ATE} = E[Y(1) - Y(0)] = E[Y(1)] - E[Y(0)] \]
The ATE represents the average effect of assigning the entire population to treatment versus control. Another relevant quantity is the \textit{Average Treatment effect on the Treated} (ATT), defined as $E[Y(1) - Y(0) | A=1]$.

\subsection{Challenges in Estimating Causal Effects from Observational Data}
\label{subsec:observational_challenges}

A fundamental principle is that \textit{association is not causation}. In observational studies, a naive comparison of outcomes between treated and untreated groups, $E[Y|A=1] - E[Y|A=0]$, generally does not equal the ATE, $E[Y(1)] - E[Y(0)]$. This discrepancy arises because the groups receiving $A=1$ and $A=0$ may differ systematically in ways other than the treatment itself.

The most critical challenge is \textit{confounding}. A confounder $W$ is a variable associated with both the treatment $A$ and the outcome $Y$ (potentially through separate paths not mediated by $A$'s effect on $Y$). For example, if physicians preferentially prescribe a new drug ($A=1$) to sicker patients, and sicker patients ($W$) naturally have worse outcomes ($Y$), then $W$ is a confounder. Failing to account for $W$ leads to \textit{confounding bias}. The naive comparison captures both the true causal effect and the bias introduced by the baseline differences between groups:
\[ E[Y|A=1] - E[Y|A=0] = \text{ATE} + \underbrace{E[Y(0)|A=1] - E[Y(0)|A=0]}_{\text{Bias Term}} \]
(assuming homogeneous treatment effect for simplicity, otherwise ATE is replaced by ATT and the bias term involves $Y(1)$ as well). If $A$ is randomly assigned (as in an ideal RCT), then $A$ is independent of the potential outcomes, $Y(a) \perp A$, making the bias term zero on average.

Another significant challenge is \textit{selection bias}. This occurs when the selection of individuals into the study or analysis sample is related to both the exposure/treatment and the outcome, leading to a distorted estimate. Examples include differential loss-to-follow-up, where patient dropout is related to both treatment assignment and prognosis, or the healthy worker effect, where occupational cohorts are inherently healthier than the general population, biasing mortality comparisons. Conditioning on a \textit{collider} (a variable caused by both $A$ and $Y$, or factors related to them) can also induce spurious associations and bias. Addressing confounding and selection bias is paramount for drawing valid causal conclusions from observational data.

\subsection{Representing Causal Assumptions: SCMs and DAGs}
\label{ssec:scms_dags}

Structural Causal Models (SCMs), as described in \citep{Pearl2009}, provide a formal framework for encoding causal assumptions. An SCM consists of a set of endogenous variables $V$, a set of exogenous (unmeasured) variables $U$, and a set of functions $f = \{f_j\}$ such that each $V_j \in V$ is determined by its function and corresponding exogenous variable: $V_j = f_j(\text{PA}_j, U_j)$, where $\text{PA}_j \subseteq V \setminus \{V_j\}$ are the endogenous parents (direct causes) of $V_j$. The exogenous variables $U_j$ represent stochasticity or unmodelled factors.

Directed Acyclic Graphs (DAGs) $\mathcal{G} = (V, E)$ are often used to visualize the causal relationships assumed in an SCM, omitting the exogenous variables for clarity (assuming they are independent). The nodes $V$ represent the variables (e.g., $A, Y, W$), and the directed edges $E$ represent direct causal effects ($X \to Y$ implies $X$ is a direct cause of $Y$ relative to the other variables in $V$). \textit{Acyclicity} means there are no directed paths starting and ending at the same node, reflecting the assumption that a variable cannot cause itself.

DAGs are powerful tools for making causal assumptions explicit. For instance, confounding between $A$ and $Y$ by $W$ is represented as $A \leftarrow W \to Y$. They allow researchers to reason about conditional independencies implied by the causal structure. Furthermore, graphical criteria, such as the \textit{back-door criterion} \citep{Pearl2009}, can be applied to a DAG to determine if a causal effect (like the ATE) is \textit{identifiable} from the observed data distribution. Identifiability means the causal quantity can be expressed purely in terms of the observational distribution $P(V)$. For example, if a set of covariates $W$ satisfies the back-door criterion relative to $(A, Y)$, it implies conditional ignorability ($Y(a) \perp A | W$), allowing the ATE to be estimated via adjustment: $E[Y(a)] = \sum_w E[Y|A=a, W=w]P(W=w)$. SCMs and DAGs thus provide the formal language and graphical tools necessary to define causal questions and assess the possibility of answering them from available data.

\section{Targeted Maximum Likelihood Estimation (TMLE)}
\label{sec:tmle}

The challenges inherent in estimating causal effects from observational data, particularly confounding and selection bias (Section \ref{sec:fundamentals_causal_bio}), motivate the development of advanced statistical methodologies. Traditional parametric models (e.g., linear or logistic regression) often impose strong, potentially incorrect assumptions about functional forms, leading to biased estimates if misspecified. Conversely, while modern ML excels at flexible prediction, standard ML algorithms are typically optimized for predictive accuracy ($E[(Y - \hat{Y})^2]$) rather than for estimating a specific causal parameter $\psi$. They may lack mechanisms for bias reduction specific to the causal target and often do not readily provide valid statistical inference (e.g., confidence intervals). The Targeted Learning (TL) framework provides a principled approach to bridge this gap, leveraging ML's flexibility while ensuring robust and efficient estimation of causal parameters.

\subsection{Motivation and Principles of Targeted Learning}
\label{ssec:motivation_tl}

The core idea of Targeted Learning \citep{van_der_laan_rose_2018} is to formulate causal estimation as a problem of estimating a parameter $\psi_0 = \Psi(P_0)$ of the true data-generating distribution $P_0 \in \mathcal{M}$, where $\mathcal{M}$ is a statistical model representing our assumptions about $P_0$. TL advocates for using large, realistic statistical models $\mathcal{M}$ (often nonparametric or semiparametric) that minimize untestable assumptions. The TL roadmap involves a two-stage procedure.

First, \textit{initial estimation of nuisance parameters}: Relevant features of $P_0$ that are necessary for estimating $\psi_0$, but not the target itself, are estimated. For the ATE, $\psi_0 = E[Y(1)] - E[Y(0)]$, key nuisance parameters are the conditional outcome expectation $\bar{Q}_0(A,W) = E_0[Y|A,W]$ and the treatment mechanism (propensity score) $g_0(A|W) = P_0(A|W)$ (or just $g_0(W) = P_0(A=1|W)$ for binary $A$). TL encourages using flexible, data-adaptive methods for this stage. The \textit{Super Learner} algorithm \citep{vanderLaan2007} is often employed, which uses cross-validation to find the optimal convex combination of predictions from a diverse library of candidate ML algorithms (e.g., GLMs, random forests, gradient boosting), yielding an asymptotically optimal estimator for the nuisance function under certain conditions. Let $\bar{Q}^0_n$ and $g^0_n$ denote these initial estimates based on a sample of size $n$.

Second, \textit{targeting the initial estimates}: The initial estimate, typically $\bar{Q}^0_n$, is updated to a new estimate $\bar{Q}^*_n$ in a way that specifically targets the parameter $\psi_0$. This "targeting" or "fluctuation" step aims to reduce bias for $\psi_0$ while minimally impacting the fit to the data elsewhere. The update is constructed to ensure the final estimator $\hat{\psi}_n = \Psi(P^*_n)$, where $P^*_n$ is the distribution implied by the targeted nuisance estimates, solves the \textit{efficient influence curve (EIC)} estimating equation. This step is crucial for achieving the desirable properties of double robustness and efficiency.

Two foundational principles guide the TL framework. Firstly, the \textit{target parameter} $\psi_0 = \Psi(P_0)$ must be clearly and formally defined based on the scientific question. This definition dictates the entire estimation procedure. Secondly, the estimation occurs within a chosen \textit{statistical model} $\mathcal{M}$, which codifies assumptions about $P_0$. The goal is to find an efficient and robust estimator for $\Psi(P)$ within $\mathcal{M}$, respecting these assumptions while leveraging data as much as possible.

\subsection{The TMLE Algorithm}
\label{ssec:tmle_algorithm}

Targeted Maximum Likelihood Estimation (TMLE) provides a general template for implementing the TL roadmap. We illustrate it for estimating the ATE, $\psi_0 = E_W[E[Y|A=1,W]-E[Y|A=0,W]]$, assuming observed data $O_1, \dots, O_n$ where $O_i = (W_i, A_i, Y_i)$.

\textit{1. Initial Nuisance Parameter Estimation}: Obtain initial estimates $\bar{Q}^0_n(A, W)$ for $E_0[Y|A,W]$ and $g^0_n(W)$ for $P_0(A=1|W)$. These can be derived using any suitable method, preferably Super Learner for flexibility \citep{vanderLaan2007, van_der_laan_rose_2018}. For simplicity, we denote $g^0_n(1|W) = g^0_n(W)$ and $g^0_n(0|W) = 1 - g^0_n(W)$.

\textit{2. Define Clever Covariate and Fluctuation Model}: Construct the "clever covariate", which is related to the part of the EIC involving the treatment mechanism:
\[ H_n(A,W) = \frac{I(A=1)}{g^0_n(W)} - \frac{I(A=0)}{1-g^0_n(W)} \]
Define a parametric working model (fluctuation model) that fluctuates the initial estimate $\bar{Q}^0_n$ along a path indexed by a parameter $\epsilon$. This is often done using a logistic regression model for binary $Y$ or a linear regression model for continuous $Y$, treating $\text{logit}(\bar{Q}^0_n)$ or $\bar{Q}^0_n$ as an offset:
\[ \text{logit}(P_\epsilon(Y=1|A,W)) = \text{logit}(\bar{Q}^0_n(A,W)) + \epsilon H_n(A,W) \quad \text{(for binary Y)} \]
\[ E_\epsilon[Y|A,W] = \bar{Q}^0_n(A,W) + \epsilon H_n(A,W) \quad \text{(for continuous Y)} \]

\textit{3. Estimate Fluctuation Parameter}: Estimate $\epsilon$ by fitting this parametric model to the observed data $(A_i, W_i, Y_i)$ using Maximum Likelihood Estimation (MLE). This yields $\hat{\epsilon}_n$. This MLE step originally gave TMLE its name\footnote{Although the literature sometimes refer to TMLE as \textit{Targeted Minimum Loss-based estimation}, it was initially introduced in reference to the MLE in \citep{vanderLaanRubin2006}. The name \textit{Minimum Loss-based} comes from \citep{vanDerLaan2013TargetedLearning}, where the fluctuation alongside the clever covariate is redefined as an optimization problem, thus loss-based.}.

\textit{4. Update Outcome Regression Estimate}: Update the initial estimate using the fitted fluctuation parameter:
\[ \text{logit}(\bar{Q}^*_n(A,W)) = \text{logit}(\bar{Q}^0_n(A,W)) + \hat{\epsilon}_n H_n(A,W) \quad \text{(for binary Y)} \]
\[ \bar{Q}^*_n(A,W) = \bar{Q}^0_n(A,W) + \hat{\epsilon}_n H_n(A,W) \quad \text{(for continuous Y)} \]
This $\bar{Q}^*_n$ is the targeted estimate of the conditional outcome expectation.

\textit{5. Compute Final Parameter Estimate}: Calculate the plug-in estimate of the ATE using the targeted outcome regression estimate, averaging over the empirical distribution $P_n$ of $W$:
\[ \hat{\psi}_{TMLE} = \frac{1}{n} \sum_{i=1}^n [\bar{Q}^*_n(1, W_i) - \bar{Q}^*_n(0, W_i)] \]

\textit{6. Statistical Inference}: Inference is based on the estimated EIC. For the ATE, the EIC at $P_0$ is:
\[ D^*(O; \bar{Q}_0, g_0) = H_0(A,W)(Y - \bar{Q}_0(A,W)) + \bar{Q}_0(1,W) - \bar{Q}_0(0,W) - \psi_0 \]
where $H_0(A,W)$ uses the true $g_0$. TMLE is constructed such that the empirical mean of the estimated EIC is approximately zero: $P_n D^*(O; \bar{Q}^*_n, g^0_n) = \frac{1}{n}\sum_i D^*(O_i; \bar{Q}^*_n, g^0_n) \approx 0$. The variance of the TMLE estimator can be estimated as the sample variance of the estimated EIC values, divided by $n$:
\[ \hat{\sigma}^2_{TMLE} = \frac{1}{n} \widehat{\text{Var}}(D^*(O; \bar{Q}^*_n, g^0_n)) = \frac{1}{n^2} \sum_{i=1}^n [D^*(O_i; \bar{Q}^*_n, g^0_n)]^2 \]
(assuming the EIC mean is zero). A $100(1-\alpha)\%$ Wald-type confidence interval is then constructed as $\hat{\psi}_{TMLE} \pm z_{1-\alpha/2} \hat{\sigma}_{TMLE} / \sqrt{n}$.

\subsection{Key Properties of TMLE}
\label{subsec:tmle_properties}

The TMLE construction yields estimators with highly desirable statistical properties.

\textit{Double Robustness}: The TMLE estimator $\hat{\psi}_{TMLE}$ is consistent for $\psi_0$ if \textit{either} the initial outcome regression estimate $\bar{Q}^0_n$ is consistent for $\bar{Q}_0$, \textit{or} the initial propensity score estimate $g^0_n$ is consistent for $g_0$ (but not necessarily both). This property arises because the targeting step ensures the estimator solves the EIC estimating equation ($P_n D^* \approx 0$). If $g^0_n$ is consistent, the bias of $\hat{\psi}_{TMLE}$ depends on how well $\bar{Q}^*_n$ approximates $\bar{Q}_0$ in a specific sense related to the fluctuation. If $\bar{Q}^0_n$ (and thus $\bar{Q}^*_n$) is consistent, the term $(Y - \bar{Q}^*_n(A,W))$ in the EIC averages to zero, removing the dependence on the consistency of $g^0_n$. Using Super Learner for both nuisance functions increases the likelihood that at least one is well-estimated, enhancing practical robustness.

\textit{Asymptotic Efficiency}: If \textit{both} $\bar{Q}^0_n$ and $g^0_n$ are consistently estimated, then $\hat{\psi}_{TMLE}$ is asymptotically efficient among regular estimators within the statistical model $\mathcal{M}$. This means it achieves the lowest possible asymptotic variance, known as the semiparametric efficiency bound, which is equal to the variance of the EIC under $P_0$. It makes the "best" use of the data for estimating $\psi_0$ under the model assumptions.

\subsection{Extensions and Variants of TMLE}
\label{subsec:tmle_extensions}

The TMLE framework's flexibility allows adaptation to various complex data structures and target parameters beyond the static ATE.

\textit{Longitudinal TMLE (LTMLE)} \citep{vanderLaanRobins2003, bang_robins_2005, vanderLaanPetersenHubbard2008}: Addresses time-varying treatments $A_t$, confounders $W_t$, and outcomes $Y_t$. A key challenge is time-varying confounding affected by past treatment, where a variable $W_t$ can be both a confounder for $A_t$'s effect on $Y_{t+k}$ and affected by past treatments $A_{<t}$. LTMLE typically involves sequentially estimating time-varying nuisance parameters (e.g., $Q_t(\bar{A}_t, \bar{W}_t) = E[Y_{final}|\bar{A}_t, \bar{W}_t]$ and $g_t(A_t|\bar{A}_{t-1}, \bar{W}_t)$) often backward in time, followed by a sequential targeting step designed to solve the longitudinal EIC equation. This provides doubly robust estimates of the effects of dynamic treatment regimes.

\textit{Conditional Average Treatment Effects (C-TMLE)} \citep{vanderLaan2018CTMLE}: Estimates the ATE within specific subgroups defined by baseline covariates $V \subset W$, i.e., $\psi_{0,V} = E[Y(1) - Y(0) | V]$. The TMLE procedure is adapted to target this conditional parameter, often involving modifications to the clever covariate or the final estimation step.

\textit{Mediation Analysis TMLE} \citep{Zheng2018Mediation}: Used to decompose the total effect of a treatment into a direct effect (not acting through a specific mediator $M$) and an indirect effect (acting through $M$). This requires estimating models for the outcome conditional on $A, M, W$ and for the mediator conditional on $A, W$. TMLE is adapted to estimate these path-specific effects.

It is crucial to emphasize that while these extensions modify the target parameter or the structure of the nuisance functions and targeting steps, the underlying principle of utilizing flexible estimators for the initial nuisance functions remains. Therefore, the outcome regressions ($\bar{Q}$) and treatment mechanisms ($g$), whether static or time-varying as required by LTMLE, C-TMLE, or other variants, can themselves be estimated using deep learning models. Utilizing neural networks, Transformers, or other deep architectures for these initial estimation steps, as discussed further in Section \ref{sec:deep_causal}, can be particularly advantageous when dealing with high-dimensional covariates or complex dependencies, aiming to improve the accuracy of the nuisance estimates upon which the subsequent targeting step operates.

\subsection{Comparison with Other Classical Causal Inference Methods}
\label{subsec:tmle_comparison}

TMLE stands alongside other established methods for causal effect estimation from observational data.

\textit{Inverse Probability Weighting (IPW)}: IPW creates a pseudo-population where treatment assignment is independent of measured confounders $W$. Each individual $i$ is weighted by the inverse probability of receiving the treatment they actually received, conditional on $W_i$: weight $h_i = A_i/g^0_n(W_i) - (1-A_i)/(1-g^0_n(W_i))$. The ATE is estimated by means: $\hat{\psi}_{IPW} = \frac{1}{n}\sum_i h_i Y_i$. IPW is \textit{singly robust}, relying crucially on the correct specification of the propensity score model $g_0(W)$. It can suffer from high variance and instability if estimated propensity scores are very close to 0 or 1 (extreme weights).

\textit{Standardization / G-computation}: This method relies on estimating the outcome regression model $\bar{Q}_0(A,W) = E_0[Y|A,W]$. It then predicts the potential outcomes for each individual under both treatment ($A=1$) and control ($A=0$) using their observed covariates $W_i$, based on the fitted model $\bar{Q}^0_n$. The ATE is estimated by averaging the difference between these predicted potential outcomes: $\hat{\psi}_{Gcomp} = \frac{1}{n}\sum_i [\bar{Q}^0_n(1, W_i) - \bar{Q}^0_n(0, W_i)]$. G-computation is also \textit{singly robust}, relying crucially on the correct specification of the outcome model $\bar{Q}_0(A,W)$. Misspecification leads to bias.

\textit{Propensity Score Matching (PSM)}: PSM first estimates the propensity score $g^0_n(W_i)$ for each unit. It then attempts to match treated units ($A=1$) with control units ($A=0$) that have similar propensity scores. The treatment effect is estimated by comparing outcomes within matched pairs or sets. PSM primarily relies on the correct specification of $g_0(W)$ for creating balanced groups. It often discards unmatched units, potentially reducing efficiency and changing the target estimand if matching is incomplete. Deriving standard errors can be complex.

In contrast to these singly robust methods, TMLE utilizes estimates of \textit{both} the outcome regression $\bar{Q}_0$ and the propensity score $g_0$. The targeting step integrates information from $g^0_n$ to update $\bar{Q}^0_n$ specifically to reduce bias for the target parameter $\psi_0$, achieving double robustness. Furthermore, by solving the EIC equation, TMLE aims for semiparametric efficiency, often leading to lower variance and narrower confidence intervals than IPW or G-computation when correctly specified, while providing robustness against misspecification of one nuisance function.

\section{Deep Causal Learning}\label{sec:deep_causal}

\subsection{Integrating Deep Learning into Existing Frameworks}\label{ssec:dl_integration}
As established in Section~\ref{sec:tmle}, causal inference frameworks like TMLE rely on estimating nuisance parameters, namely the propensity score $g(x) = P(A=1|W)$ and the conditional outcome models $\bar{Q}(x) = \mathbb{E}[Y|A=a, W]$. Deep learning models offer a powerful alternative to traditional statistical methods for estimating these functions \citep{shi21a, schulte2025adjustment}. NNs, autoencoders, or other deep architectures can be employed as flexible function approximators, $\hat{g}(w; \theta)$ and $\hat{Q}(w; \phi)$, capable of capturing complex non-linear relationships and high-order interactions directly from high-dimensional covariate data $W$, without extensive manual feature engineering \citep{weberpals2021deep}. These DL-based estimators can then serve as inputs to the subsequent steps of the chosen causal framework.

In longitudinal settings with time-varying treatments $A_t$ and confounders $L_t$, LTMLE provides a doubly robust approach by sequentially estimating nuisance parameters, such as the conditional expectations $Q_k(H_k, A_k) = \mathbb{E}[Y | H_k, A_k]$ and treatment mechanisms $g_k(A_k | H_k)$, where $H_k = (\bar{L}_k, \bar{A}_{k-1})$ represents the history up to time $k$. Traditionally, this sequential estimation might use separate regressions or a longitudinal Super Learner at each time step. Shirakawa et al. proposed DeepLTMLE, which leverages the capabilities of Transformer networks (Section~\ref{ssec:transformers}) to perform this initial nuisance function estimation \citep{shirakawa_deepltmle_2024}. The Transformer's self-attention mechanism is adept at modeling \textit{long-range dependencies} within the sequential patient history $H_k$. In DeepLTMLE, a single Transformer model is trained to simultaneously estimate the sequences of relevant conditional expectations or densities required by the LTMLE algorithm over time. These DL-based initial estimates are then plugged into the standard LTMLE targeting step, which adjusts the initial outcome model estimates to solve the EIC estimating equation, thereby retaining the desirable statistical properties of LTMLE.

\subsection{Neural Network Architectures for Causal Effect Estimation}\label{subsec:nn_architectures_causal}

Beyond using DL for nuisance parameter estimation within existing frameworks, specialized NN architectures have been developed to directly estimate causal effects, particularly heterogeneous treatment effects (HTE), defined as $\tau(x) = \mathbb{E}[Y(1) - Y(0) | X=x]$. A common strategy involves multi-head architectures with shared layers for learning a representation $\Phi(X)$ from covariates, followed by separate heads for specific estimation tasks.

\subsubsection{TARNet}
The Treatment-Agnostic Representation Network (TARNet) \citep{Shalit2017} employs shared layers to learn a representation $\Phi(X) = f(X; \theta_\Phi)$, followed by separate outcome prediction heads $h_1$ and $h_0$ to estimate the conditional outcomes for the treated and control groups, respectively: $\hat{\mu}_1(x) = h_1(\Phi(x); \theta_1)$ and $\hat{\mu}_0(x) = h_0(\Phi(x); \theta_0)$. The HTE is then estimated as $\hat{\tau}(x) = \hat{\mu}_1(x) - \hat{\mu}_0(x)$. The network is typically trained by minimizing the prediction error on the observed factual outcomes.

\subsubsection{Dragonnet}
Dragonnet \citep{shi2019adapting} extends TARNet by adding a third head, $h_e$, which uses the same shared representation $\Phi(X)$ to predict the propensity score: $\hat{e}(x) = \sigma(h_e(\Phi(x); \theta_e))$, where $\sigma$ is the sigmoid function \citep{ShiBleiVeitch2019}. Explicitly modeling the propensity score alongside the outcome models can potentially improve the quality of the learned representation $\Phi(X)$ and provide diagnostics. The loss function typically combines the factual outcome prediction loss (e.g. MSE) with the propensity score prediction loss (e.g. BCE):
\begin{multline*}
 L(\theta_\Phi, \theta_0, \theta_1, \theta_e) = \frac{1}{n} \sum_{i=1}^n \left( (1-T_i) \ell(Y_i, \hat{\mu}_0(X_i)) + T_i \ell(Y_i, \hat{\mu}_1(X_i)) \right)\\ + \alpha \cdot BCE(T_i, \hat{e}(X_i))
\end{multline*}

where $\ell$ is the outcome loss and $\alpha$ is a hyperparameter balancing the terms. Variations may include additional regularization terms, such as those enforcing balance between the treatment groups in the representation space \citep{Farajtabar2020Balance}.

\subsection{Representation Learning for Confounding Adjustment}\label{subsec:representation_confounding}
A central challenge in observational causal inference is adjusting for confounding variables $W$ that affect both treatment $A$ and outcome $Y$. Simply adjusting for all measured covariates $X$ can be problematic if $X$ is high-dimensional or contains "bad controls" like mediators or colliders, which can induce bias \citep{shi21a}. Representation learning \citep{pmlr-v48-johansson16} aims to use DL to learn a transformation $\Phi: \mathcal{X} \rightarrow \mathcal{Z}$ that maps high-dimensional covariates $X$ to a lower-dimensional representation $\Phi(X)$ in a latent space $\mathcal{Z}$. The goal is for $\Phi(X)$ to be a sufficient adjustment set, meaning $Y(a) \perp A | \Phi(X)$ for $a \in \{0, 1\}$, while satisfying overlap, $0 < P(A=1 | \Phi(X) = z) < 1$. This learned representation ideally captures confounding information while potentially discarding irrelevant or harmful variables.

\subsubsection{Invariant Representation Learning}
This approach seeks representations $\Phi(X)$ that are stable or invariant across different environments or domains (e.g., data from different hospitals or time periods). Techniques like Invariant Risk Minimization (IRM) \citep{arjovsky2019invariant} aim to learn $\Phi$ such that predictors built upon it (e.g., $w \circ \Phi$) generalize well across environments. The IRM objective often involves minimizing the sum of risks across environments plus a penalty on the gradient norm measuring predictor invariance: $\min_{\Phi} \sum_{e} R_e(w \circ \Phi) + \lambda \|\nabla_{w|w=1.0} R_e(w \circ \Phi)\|^2$. Under certain assumptions, this can help isolate stable causal relationships and remove spurious correlations or bad controls, as explored in methods like Nearly Invariant Causal Estimation (NICE) \citep{shi21a}. This typically requires access to data from multiple environments.

\subsubsection{Distribution Balancing}
These methods aim to learn a representation $\Phi(X)$ such that the distribution of the representation is similar for the treated and control groups, i.e., minimizing a distance metric $D[\, P(\Phi(X)|A=1) \, , \, P(\Phi(X)|A=0) \,]$. Common choices for $D$ include Maximum Mean Discrepancy (MMD) \citep{Iyer2014} or Wasserstein distance. Enforcing balance in the representation space is closely related to propensity score methods and can be incorporated via regularization terms in the loss function when training architectures like TARNet or Dragonnet.

\subsubsection{Other Approaches}
Other strategies include using unsupervised methods like autoencoders to learn a compressed representation $\Phi(X)$, which is then used for downstream causal estimation (e.g., estimating $P(A=1|\Phi(X))$) \citep{weberpals2021deep}. Representations learned by large models pre-trained on related data (e.g., image or language models) can sometimes be leveraged as features for causal adjustment \citep{schulte2025adjustment}. Once a suitable representation $\Phi(X)$ is obtained, causal effects are estimated by adjusting for $\Phi(X)$, for example, via outcome modeling: $ATE = \mathbb{E}_{\Phi(X)}[\mathbb{E}[Y|A=1, \Phi(X)] - \mathbb{E}[Y|A=0, \Phi(X)]]$. The success of these methods hinges on the extent to which the learned representation accurately captures the true confounding structure \citep{Schlkopf2021TowardCR}.

\chapter{Mechanistic Interpretability: Understanding Neural Networks}\label{chap:mech_int}


Deep learning models offer powerful tools for prediction and function approximation, finding increasing application within bio-statistics and causal inference frameworks like TL. However, their complex, often non-linear internal structure typically renders them "black boxes," making it difficult to understand how they arrive at their predictions based solely on input-output relationships. This opacity poses significant challenges in high-stakes domains like healthcare, where understanding the reasoning behind a prediction is crucial for trust, validation, and scientific discovery.

Mechanistic Interpretability (MI) provides a collection of techniques aimed at moving beyond this input-output view to understand the internal \textit{mechanisms} by which NNs compute their functions. The goal is to \textit{reverse-engineer} the algorithms learned by the network from data, mapping internal computations (represented by activations, weights, and their interactions) to human-understandable concepts and causal processes.

Within the specific context of applying NNs for causal estimation in bio-statistics, MI offers several critical aims. \textit{Validation} is essential: MI techniques can help verify if NN components, such as those used to estimate nuisance functions (cf. Section~\ref{ssec:dl_integration}), learn representations consistent with existing domain knowledge, correctly identifying known risk factors or confounders. Beyond confirmation, MI enables \textit{Discovery}, potentially uncovering novel biological or causal relationships, or identifying distinct patient subgroups implicitly learned by the network from complex data patterns.

Furthermore, MI serves crucial roles in \textit{Debugging \& Robustness} analysis. It allows investigation into whether a network relies on genuine causal features or exploits spurious correlations and data artifacts, which is vital for assessing the reliability of causal effect estimates derived from NN components. MI also facilitates \textit{Comparison}, enabling researchers to understand how the mechanisms learned by NNs differ from or align with the assumptions embedded in traditional statistical models commonly used in causal inference (e.g., linear models, logistic regression). Ultimately, achieving these aims contributes to enhancing \textit{Trustworthiness}. By providing insights into \textit{why} an NN makes a specific prediction or estimates a certain effect, MI helps build confidence in the application of these powerful but often opaque models in bio-statistical research and practice. This chapter explores various MI techniques and their application towards achieving these goals.

\section{Observational Techniques: What Information is Encoded?}\label{sec:observational_techniques}
Observational techniques aim to understand the information represented within a NN's internal states, i.e. activations, without altering the network's computation during inference. The core question these methods address is: \textit{Given a specific input, what kind of information is encoded in the activation patterns at different layers of the network?} These techniques passively observe the network's function to map its internal representations to human-understandable concepts.

\subsection{Probing}
\textbf{Probing}, often referred to as linear probing, is a technique used to determine if specific properties or concepts are encoded in a NN's intermediate representations \citep{niven2019probing}. The fundamental idea is to train a \textit{simple} (thus often linear) supervised model, the \textbf{probe}, on the activation vectors extracted from a specific layer of a pre-trained network to predict a target property associated with the original input.

Let $f$ be a pre-trained NN. We are interested in understanding the representations at a specific layer $l$. For a dataset of inputs $\{x_i\}$, we first compute the corresponding activation vectors $\{h^{(l)}_i\}$ at layer $l$ by performing a forward pass: $h^{(l)}_i = f^{(l)}(x_i)$, where $f^{(l)}$ denotes the computation up to layer $l$.
Alongside the inputs $\{x_i\}$, we have corresponding labels $\{y_i\}$ representing the property we want to probe for (e.g., a biological marker, a demographic attribute, a linguistic feature). These labels $y_i$ are \textit{not} necessarily the original labels used to train the main network $f$.

We then train a separate, simple classifier $g$ (the probe) that takes the activations $h^{(l)}_i$ as input and attempts to predict the property labels $y_i$:
$$ \hat{y}_i = g(h^{(l)}_i) $$
The probe $g$ is deliberately kept simple (e.g., logistic regression, a linear Support Vector Machine (SVM), or a shallow (MLP)) for a crucial reason. The goal is not to achieve the absolute best possible prediction accuracy for $y_i$, but rather to assess how \textit{easily} (e.g., linearly) the information about $y_i$ can be extracted from the activations $h^{(l)}_i$. If a simple probe $g$ achieves high performance (measured on a held-out test set using metrics like accuracy, F1-score, etc.) in predicting $y_i$ from $h^{(l)}_i$, it suggests that the information related to the property $y$ is explicitly represented or readily available in the activations at layer $l$. Conversely, poor performance suggests the information is either absent, obfuscated, or encoded in a highly non-linear way that the simple probe cannot capture.

\paragraph{Correlation, not Causation}
A critical limitation of probing, inherent to its observational nature, is that it can only reveal \textbf{correlation}, not \textbf{causation}. A successful probe demonstrates that information about property $y$ is \textit{present} in the activations $h^{(l)}$, meaning the activations correlate with $y$. However, it \textit{does not} prove that the main network $f$ actually \textit{uses} this information causally to arrive at its final prediction. The network might encode the information incidentally, perhaps because the probed property $y$ is correlated with other features the network \textit{does} use. As highlighted by works like \citep{ProbingClassifiers2022}, interpreting probe results requires caution; high accuracy indicates representation, but not necessarily utilization. To assess causal reliance, interventional techniques, discussed in Section~\ref{sec:interventional_techniques}, are necessary.

\subsection{Feature Analysis}\label{ssec:feat_anal}
\subsubsection{Sparse Autoencoders} 
\textbf{Sparse Autoencoders (SAEs)} \citep{bricken2023sparse} are an unsupervised machine learning technique specifically adapted to analyze internal network activations. They are structured as AEs (cf Section~\ref{sec:autoencoders}), comprising an encoder function $f_{enc}$ and a decoder function $f_{dec}$, trained not on the original input data, but on the activation vectors $h^{(l)}$ extracted from an intermediate layer $l$ of a pre-trained base model $f$. The primary objective is to \textbf{reconstruct these activations sparsely}. That is, the SAE learns a mapping such that $h^{(l)} \approx f_{dec}(f_{enc}(h^{(l)}))$, but with a crucial constraint imposed on the latent representation $z = f_{enc}(h^{(l)})$: the vector $z$ must be sparse, meaning most of its components are zero for any given activation $h^{(l)}$.

The core goal of employing SAEs is to find potentially \textbf{monosemantic features} within the network's representations, thereby helping to overcome the common issue of \textbf{polysemanticity} \citep{scherlis2022polysemanticity}. Polysemanticity occurs when individual neurons or units in the base model $f$ respond to multiple, seemingly unrelated concepts or features in the input data. An SAE attempts to address this by decomposing the potentially polysemantic activation vector $h^{(l)} \in \mathbb{R}^d$ into a typically higher-dimensional latent code $z \in \mathbb{R}^k$ (where $k \gg d$) that is constrained to be sparse. The hypothesis is that this sparsity encourages each non-zero component $z_j$ of the latent code (often called a "feature activation") to correspond to a single, more interpretable underlying concept or property represented within $h^{(l)}$.

The standard SAE approach trains the encoder and decoder parameters ($\theta_{enc}, \theta_{dec}$) to minimize a loss function balancing reconstruction accuracy with sparsity. A common formulation uses L1 regularization:
$$ L_{SAE}(\theta_{enc}, \theta_{dec}) = \mathbb{E}_{h^{(l)}} \left[ ||h^{(l)} - f_{dec}(f_{enc}(h^{(l)}))||^2_2 + \lambda ||f_{enc}(h^{(l)})||_1 \right] $$
Here, the first term is the reconstruction loss (e.g., MSE), and the second term penalizes the L1 norm of the latent code $z = f_{enc}(h^{(l)})$, weighted by $\lambda$, encouraging sparsity. However, several distinct variants of SAEs have emerged. An alternative method is the \textbf{TopK sparse autoencoder} \citep{gao2024scaling}, which directly controls sparsity by keeping only the $k$ largest activations in $z$ and zeroing out the rest. This approach can simplify hyperparameter tuning compared to tuning $\lambda$ and potentially achieve a better trade-off on the reconstruction-sparsity frontier. The choice of activation function itself significantly impacts feature stability; for instance, \textbf{ReLU-based SAEs} have shown greater consistency in learned features across different random initializations compared to some state-of-the-art TopK variants \citep{Paulo2025}. Other strong alternatives include \textbf{JumpReLU} SAEs \citep{Rajamanoharan2024} employing the following specific activation function within the encoder:
$$\text{JumpReLU}_\theta(z) \coloneqq z H(z-\theta)$$
where $H$ is the  Heaviside step function (cf. Section~\ref{ssec:mlp}) and $\theta>0$ is the JumpReLU’s threshold.

\subsubsection{Transcoders} 
\textbf{Transcoders} \citep{dunefsky2024transcoders} are another mechanistic interpretability technique aiming for \textbf{sparse and interpretable mappings}, but specifically designed to analyze NN \textbf{computation flows across consecutive layers}, particularly within FFN sublayers. Unlike Sparse Autoencoders which typically decompose activation vectors $h^{(l)}$ at a single layer $l$, transcoders focus on approximating the transformation performed by a network component mapping from the pre-component activations to the post-component activations. Structurally, a transcoder utilizes an encoder $f_{enc}$ and a decoder $f_{dec}$ to mimic the function $f_{FFN}$ of a target FFN layer, often using an overparameterized, wider latent space $z = f_{enc}(h^{(l)})$ that is constrained to be sparse.

The primary objective is to \textbf{decompose complex neural computations into interpretable circuits}. This addresses the challenge that neurons in standard FFNs often activate densely, making direct circuit analysis difficult. By enforcing sparsity in the latent space that represents the transformation itself, transcoders encourage the discovery of factorized computational pathways, potentially separating input-dependent and input-invariant aspects of the layer's function and yielding more interpretable circuits.

Transcoders are trained to minimize a loss function that balances the fidelity of approximating the FFN's transformation with the sparsity of the intermediate latent representation $z$. This loss typically takes the form:
$$ L_{transcoder}(\theta_{enc}, \theta_{dec}) = \mathbb{E}_{h^{(l)}} \left[ ||f_{FFN}(h^{(l)}) - f_{dec}(f_{enc}(h^{(l)}))||^2_2 + \lambda_1 ||f_{enc}(h^{(l)})||_1 \right] $$
Here, $h^{(l)}$ represents the input activation to the target FFN layer, $f_{FFN}(h^{(l)})$ is the true output activation of that layer, and the first term measures the reconstruction error between the true output and the transcoder's approximation. The second term is an L1 sparsity penalty on the latent code $z = f_{enc}(h^{(l)})$, controlled by the hyperparameter $\lambda_1$.

By focusing on the transformation $h^{(l)} \rightarrow f_{MLP}(h^{(l)})$ rather than just the representation $h^{(l)}$, transcoders enable a different kind of analysis compared to SAEs. They facilitate weights-based circuit analysis \textit{through} a layer, offering insights into how information is processed and transformed, potentially providing a more robust way to elicit and reverse-engineer latent computational circuits within larger models.

\section{Interventional Techniques: What Mechanisms Does the Model Causally Rely On?}\label{sec:interventional_techniques}

While observational techniques like probing (Section~\ref{sec:observational_techniques}) reveal correlations between NN activations and specific concepts, they cannot establish \textit{causal} links. They tell us what information is \textit{present}, but not necessarily what information the model \textit{uses} to make its predictions. \textbf{Interventional techniques} address this gap by actively manipulating the network's internal state during computation and observing the effect on the output. By changing specific activations or component functions, we can test hypotheses about whether those components causally contribute to the model's behavior.

\subsection{Activation Patching / Interchange Interventions}\label{ssec:act_patch_ii}
\textbf{Activation Patching} \citep{nanda2023attributionpatching} and \textbf{Interchange Interventions} \citep{geiger2021causal} are powerful interventional techniques designed to isolate the causal effect of specific internal activations on the model's output.

The core idea is to perform \textbf{swapping activations between runs on different inputs}. It involves running the model on two inputs:
\begin{enumerate}
    \item A \textbf{base input} ($x_{base}$), representing the default or clean state for which we want to understand the computation.
    \item A \textbf{source input} ($x_{source}$), representing a state containing a specific feature or condition whose effect we want to isolate.
\end{enumerate}
The model is run forward on both inputs. At a specific, pre-determined point in the network (e.g., the output of a specific neuron, a set of neurons, an attention head output, or an entire layer activation vector $h^{(l)}$), the activation value computed during the forward pass on the base input ($h^{(l)}_{base}$) is discarded. Instead, the corresponding activation value computed from the source input ($h^{(l)}_{source}$) is "patched" into the computation for the base input. The forward pass for the base input then resumes from this point using the patched activation $h^{(l)}_{source}$.

Mathematically, if $f$ is the network and $f_{>l}$ denotes the computation from layer $l+1$ onwards, the original output for the base input is $y_{base} = f(x_{base}) = f_{>l}(f^{(l)}(x_{base}))$. The patched output is $y_{patched} = f_{>l}(h^{(l)}_{source})$. The difference between $y_{patched}$ and $y_{base}$ measures the causal effect of the information encoded in the activation at layer $l$ specifically related to the difference between $x_{source}$ and $x_{base}$, on the final output for $x_{base}$.

\subsection{Ablation}
\textbf{Ablation} \citep{meyes2019ablation} is a conceptually simpler, yet fundamental, interventional technique. It involves \textbf{zeroing/neutralizing components} of the NN to understand their contribution. This typically means setting the output of specific components (e.g., individual neurons, attention heads, residual stream components, or even entire layers) to zero during the forward pass. Essentially, the component is "removed" from the computation for that specific run. 

The primary goal of ablation is to identify \textbf{critical neurons/layers} or other components necessary for the model's performance or a specific behavior. By observing the degradation in performance (e.g., drop in accuracy, increase in loss) or the change in output when a component is ablated, we can infer its importance. If ablating a neuron significantly harms performance on a task, that neuron is considered \textit{causally important} for that task. 

Ablation is often a \textit{first step} in understanding model mechanisms due to its simplicity. While less nuanced than activation patching (which replaces information rather than just removing it), ablation effectively answers the question: "\textit{Is this component necessary for the observed behavior?}"

\subsection{Causal Tracing}
\textbf{Causal Tracing} \citep{meng2022locating} is an interventional technique, often built upon activation patching, that aims to \textbf{trace the flow of effects from interventions} through the network's layers. While a single activation patch measures the effect of a source activation at a specific layer $l$ on the final output, causal tracing extends this to understand how that effect propagates through intermediate layers.

The core idea is to investigate not just the final output change, but also how activations at intermediate layers $k > l$ are affected when an activation at layer $l$ is patched from a source input $x_{source}$ into the computation of a base input $x_{base}$. Specifically, one might perform an activation patch at layer $l$ (replacing $h^{(l)}_{base}$ with $h^{(l)}_{source}$) and then measure the change in activations at a subsequent layer $k$: $\Delta h^{(k)} = h^{(k)}_{patched} - h^{(k)}_{base}$. A significant change $\Delta h^{(k)}$ suggests that the causal pathway from the information difference introduced at layer $l$ passes through layer $k$.

Causal tracing helps pinpoint \textit{where} an effect originating earlier in the network (represented by the patched activation) has its impact on later representations, and \textit{how} different components (e.g., specific neurons or attention heads in subsequent layers) respond to the patched information, revealing the computational pathway.

\subsection{Steering with SAEs and Transcoder}

Once we have learned the sparse and interpretable mappings of the features within an SAE or a transcoder (\ref{ssec:feat_anal}), we can intervene on these mappings, a method called \textbf{steering}. The core idea is to perform \textit{Interchange Intervention} (\ref{ssec:act_patch_ii}) within the SAE/transcoder's latent. We identify, from a group of source inputs with a desired feature (e.g. obesity), the few latent's corresponding activations encoding this feature, meaning when active they indicate its presence. Once identified, we can run our modified model with the base input, where we replaced the activations at the desired layer with the reconstruction from the SAE/transcoder plus the shift in the sparse latent space to activate the feature. 

\section{Explanation Synthesis \& Validation: How Can Model Understanding Be Structured and Tested?}
The techniques discussed so far (probing, patching, ablation, SAEs, transcoders) provide tools to investigate specific components or representations. Frameworks for understanding computations aim to synthesize these findings into a more holistic view of how NNs perform tasks, often by identifying interconnected pathways or relating network operations to higher-level causal models.

\subsection{Circuit Analysis}
\textbf{Circuit Analysis} \citep{elhage2021mathematical} refers to the process of \textbf{understanding computational pathways involving multiple interconnected components} within a NN. The goal is often to identify a subgraph of the full network responsible for a specific capability or behavior. This typically involves \textbf{combining insights from multiple techniques}: observational methods like probing or SAEs might identify candidate components or features, while interventional techniques like activation patching, ablation, or causal tracing are used to confirm the causal roles of these components and map out their functional connections within the identified circuit. Circuit analysis moves beyond single-neuron or single-layer analysis to explain how information flows and is transformed through sequences or groups of network elements to achieve a particular outcome.

\subsection{Causal Abstraction}
\textbf{Causal Abstraction} \citep{geiger2021causal} provides a formal framework for relating the complex, low-level computations within a NN to simpler, higher-level causal models. Its core idea is the \textbf{mapping of low-level NN computations to higher-level causal graphs}, such as SCMs or DAGs often used in fields like bio-statistics (cf. Section~\ref{ssec:scms_dags}). This framework aims to determine if the network's internal processing can be faithfully represented by a simpler causal model involving higher-level variables.

Let $f$ represent the NN (the low-level system) operating on inputs $x$, producing internal states $h$ and outputs $y$. A causal abstraction proposes a high-level causal model $M_{HL}$ (e.g., an SCM) with variables $U, V, W, \dots$ and a mapping (abstraction function) $\alpha$ that connects the states of the low-level system to the states of the high-level variables (e.g., $\alpha(h)$ might determine the value of $V$). The key question is whether interventions performed on the low-level system $f$ produce effects consistent with the causal relationships defined in $M_{HL}$.

\textbf{Interventions, particularly activation patching, are crucial within this framework for testing the faithfulness of the abstraction}. An intervention $do(V=v)$ in the high-level model $M_{HL}$ corresponds to interventions on the low-level network $f$. For example, using activation patching, we can force the network's internal state $h$ into a configuration $\tilde{h}$ such that $\alpha(\tilde{h})$ corresponds to the high-level state $V=v$. Let $f_{patched}$ denote the network under this intervention. The abstraction is considered faithful if the subsequent behavior of $f_{patched}$ aligns with the predictions of $M_{HL}$ under the intervention $do(V=v)$. For instance, if $M_{HL}$ posits that $V$ causes $Z$ but is independent of $X$ given its parents, then patching the network components corresponding to $V$ should affect the network outputs corresponding to $Z$, but ideally not those corresponding to $X$ (in a way not mediated by the parents). By systematically performing such interventions mirroring the causal structure of $M_{HL}$ and checking for consistency in the network's behavior, causal abstraction aims to validate whether the high-level causal model is a meaningful and accurate description of the network's underlying computational mechanisms.

\subsection{Causal Scrubbing}
\textbf{Causal Scrubbing} \citep{Chan2022CausalScrubbing} provides a methodology to evaluate whether a hypothesized computational structure, represented as an abstraction of the NN $f$, is sufficient to explain a target behavior, often measured via a loss function $L$.

This evaluation begins by formalizing the hypothesis. This typically involves defining a \textbf{high-level interpretation graph}, $G_{HL} = (V_{HL}, E_{HL})$, which captures the hypothesized critical computational components $V_{HL}$ and their causal dependencies $E_{HL}$. This graph is often conceptualized as a SCM (cf. Section~\ref{ssec:scms_dags}). A \textbf{translation function}, $\tau: V_{HL} \rightarrow \mathcal{P}(V_{LL})$, maps these high-level nodes to their corresponding elements (sets of neurons, attention heads, etc.) $V_{LL}$ within the actual network's computational graph (the \textbf{low-level graph}).

The methodology then proceeds by performing targeted interventions on the network's internal activations during a forward pass for a given input $x_{base}$. Specifically, it uses \textbf{behavior-preserving resampling ablations}. For certain nodes $v \in V_{LL}$ in the low-level graph, their original activation $h_v(x_{base})$ is replaced (``scrubbed'') with an activation $h'_v$. This replacement activation $h'_v$ is typically generated by running the network on a different input, $x_{resample}$, i.e., $h'_v = h_v(x_{resample})$. Critically, the choice of which nodes $v$ to scrub and how to select $x_{resample}$ is constrained by the dependencies specified in the high-level hypothesis graph $G_{HL}$. If a node $v$ is part of a high-level component $u \in V_{HL}$ (i.e., $v \in \tau(u)$), its resampling might depend on ensuring consistency with the parents of $u$ in $G_{HL}$, $Pa_{HL}(u)$. Nodes $v$ corresponding to high-level components hypothesized to be irrelevant, or nodes considered conditionally independent based on $G_{HL}$'s structure, can be resampled more freely (e.g., using a randomly chosen $x_{resample}$ or one matching only necessary parent conditions). Let $f_{scrubbed}(x_{base})$ denote the output of the network after performing all specified scrubbing interventions based on $G_{HL}$.

The validity of the hypothesis $G_{HL}$ is then quantitatively assessed by measuring how well the model's original behavior is preserved under these interventions, typically evaluated over a dataset $D$. This is often quantified by comparing the expected loss after scrubbing, $\mathbb{E}_{x \sim D}[L(f_{scrubbed}(x))]$, to the original expected loss, $\mathbb{E}_{x \sim D}[L(f(x))]$. If the loss remains largely unchanged, meaning the difference $\Delta L = \mathbb{E}_{x \sim D}[L(f_{scrubbed}(x))] - \mathbb{E}_{x \sim D}[L(f(x))]$ is close to zero (high recovery), it provides strong evidence that the hypothesized structure $G_{HL}$ accurately captures the necessary mechanisms for the behavior. This is because altering parts of the network outside this hypothesized structure (or altering parts within it according to the hypothesized conditional independencies) did not significantly disrupt performance. Conversely, a significant increase in loss ($\Delta L \gg 0$, low recovery) indicates the hypothesis $G_{HL}$ is incomplete or incorrect, as the scrubbing interventions disrupted computations essential to the behavior that were not accounted for in the hypothesis.

\section{Building and Training Explainable Models: How Can Causality Be Integrated?}

\subsection{Causal Proxy Models}
\textbf{Causal Proxy Models (CPMs)} \citep{wu2022causal} represent a distinct approach to explainability, focusing on \textit{training} simpler, inherently interpretable models that mimic a complex black-box model $f$ not only in its factual predictions but also in its causal, counterfactual behavior. Let $f(x)$ be the prediction of the black-box model for input $x$. A CPM, denoted $f_{CPM}$, is trained to satisfy two objectives.

First, it minimizes a distillation-style loss to match the factual predictions of the original model: $\mathcal{L}_{factual} = \mathbb{E}_{x \sim D} [d(f_{CPM}(x), f(x))]$, where $d$ is a distance metric. Second, it aims to mimic the counterfactual behavior of $f$. Since obtaining true counterfactual inputs $x'$ and corresponding $f(x')$ is often infeasible, CPMs leverage \textit{approximate counterfactuals} $(\tilde{x}', f(\tilde{x}'))$. These might be generated via heuristics, domain knowledge, or sampling guided by metadata associated with inputs. The CPM is trained to match these approximate counterfactual predictions, often by incorporating interventions directly on its learned representations $z_{CPM} = \phi_{CPM}(x)$. For an intervention $do(\text{concept}=c)$ intended to simulate a change in input concept, the CPM's prediction $f_{CPM}(\psi_{CPM}(z'_{CPM}))$ (where $z'_{CPM}$ is the intervened representation and $\psi_{CPM}$ is the CPM's decoder/predictor head) should align with the target counterfactual outcome, minimizing a loss $\mathcal{L}_{counterfactual}$.

The overall objective combines these: $\mathcal{L}_{CPM} = \mathcal{L}_{factual} + \lambda \mathcal{L}_{counterfactual}$. Different CPM variants exist based on how interventions are implemented, such as appending intervention tokens to the input (\textbf{CPMIN}) or manipulating hidden representations (\textbf{CPMHI}). By construction, CPMs offer a potential pathway to directly interpretable models that retain high predictive performance while providing causal explanations grounded in their ability to answer counterfactual questions.

\subsection{Interchange Intervention Training}
\textbf{Interchange Intervention Training (IIT)} \citep{geiger2022inducing} is a \textit{training methodology} designed to explicitly enforce desired causal structures or invariances within a NN $f$ during its optimization process. It requires a predefined causal graph $G_{causal}$ specifying assumed causal relationships between input variables $X = \{X_1, ..., X_p\}$ and potentially internal model states or the final output $Y$. The core principle is to train the network to be invariant to interventions on variables deemed causally irrelevant according to $G_{causal}$.

During training, IIT performs interchange interventions (\ref{ssec:act_patch_ii}) directly on the network's hidden activations. For a given input $x$, consider an internal activation $h = f_l(x)$ at layer $l$. Suppose $G_{causal}$ implies that a subset of input variables $X_{irr} \subset X$ should not influence $h$ (or the final output $Y$) given another set $X_{rel}$. IIT constructs a 'source' input $x_{source}$ where $X_{irr}$ differs from $x$ but $X_{rel}$ is kept the same. It computes the corresponding activation $h_{source} = f_l(x_{source})$. The intervention involves patching $h_{source}$ into the forward pass for input $x$, resulting in a counterfactual output $\tilde{y} = f_{>l}(h_{source})$.

An additional loss term, $\mathcal{L}_{IIT}$, penalizes deviations between the original output $y = f(x)$ and the counterfactual output $\tilde{y}$ resulting from swapping activations corresponding to causally irrelevant variables: $\mathcal{L}_{IIT} = \mathbb{E}_{x, x_{source}} [d(y, \tilde{y})]$. The total training loss becomes $\mathcal{L}_{total} = \mathcal{L}_{task} + \lambda \mathcal{L}_{IIT}$, where $\mathcal{L}_{task}$ is the standard task loss (e.g., cross-entropy). By minimizing $\mathcal{L}_{total}$, IIT encourages the network $f$ to learn representations $h$ that are inherently invariant to changes in $X_{irr}$ when $X_{rel}$ is fixed, thus embedding the assumptions of $G_{causal}$ directly into the learned function and potentially improving robustness against spurious correlations.

\section{Challenges and Limitations}

The application of MI techniques to NNs within the demanding context of causal bio-statistics and the Targeted Learning framework holds significant promise for enhancing validation, understanding, and trust. However, realizing this potential requires acknowledging and addressing several substantial challenges and limitations, which in turn point towards important avenues for future research.

\textbf{The Interpretation of Interpretability Results} itself presents a significant challenge. MI techniques often generate complex outputs – high-dimensional SAE feature activations, intricate circuit diagrams derived from tracing, quantitative results from scrubbing experiments. Translating these outputs into clear insights requires considerable expertise and effort. There is a risk that MI results might be as opaque as the original model or misinterpreted. Defining \textbf{clear metrics} for the quality and faithfulness of MI explanations in this specific domain is an ongoing task.

\textbf{Complexity Mismatch and Representation} poses another difficulty. NNs learn internal representations by optimizing their parameters based on the training data, implicitly aiming to capture aspects of the true underlying data-generating distribution $P_0$. MI techniques then attempt to understand these learned representations, often by mapping them to simpler, localized components or causal variables, such as those depicted in bio-statistical DAGs. However, the representations learned by NNs may be highly distributed, non-sparse, or entangled, reflecting the complex correlations and non-linearities inherent in $P_0$, rather than neatly aligning with the modular structure assumed by many MI analysis tools or external causal models. Biological systems themselves exhibit intricate feedback loops and dependencies, further complicating any simple mapping. Moreover, interventional MI techniques, while powerful for probing the model's internal causal logic, necessarily involve creating internal states (e.g., via activation patching or parameter modification) that may deviate significantly from those encountered under the original distribution $P_0$. Interpreting the network's behavior under such interventions requires acknowledging that we are observing the model's response in potentially "off-distribution" internal states. While this reveals the network's learned functional pathways, the extent to which these pathways accurately reflect counterfactuals under shifts in the true $P_0$ can be difficult to ascertain, especially if the learned representations are highly optimized for the specific input manifold. Current MI techniques might struggle to fully disentangle or map these complex, distribution-dependent learned mechanisms to simple, robust causal interpretations.

\textbf{Integrating MI with Causal Inference Workflows}, such as the TL roadmap, is still in its nascent stages. While MI can provide qualitative insights for validation or hypothesis generation, translating these findings into quantitative adjustments for causal estimates, confidence intervals, or formal bias analyses within frameworks like TMLE requires significant theoretical and methodological development. 

\textbf{The Need for Domain Expertise} cannot be overstated. Meaningful application of MI in bio-statistics necessitates a deep synergy between MI/ML experts and domain specialists (biologists, epidemiologists, clinicians). Formulating relevant hypotheses to test with MI, interpreting the biological plausibility of identified circuits or features, and assessing the clinical relevance of findings requires close collaboration throughout the research process.

\chapter{Experiments of Mechanistic Interpretability for Bio-statistics}
\label{chap:experiments}

Building upon the limitations identified in the previous chapter \ref{chap:mi_intro}, this chapter presents experiments\footnote{All experiments are accessible through the code repository \url{https://github.com/jbccc/mech-interp-biostats}.} applying MI techniques within bio-statistical causal settings. These investigations aim to explore practical challenges, develop or apply relevant evaluation metrics, examine pathways for integrating MI within the TL framework, and analyze how network representations of data respond to causal interventions.

All data used in these experiments are synthetic. Their exact generation details, as well as the training implementation details of the models are provided in Appendix \ref{app:datasets_models}.

\section{Validating Nuisance Function Estimators in TMLE}
\label{sec:exp1_tmle_interp}

We investigate whether a multi-task NN ($f_{Q,g}$), trained to simultaneously estimate the outcome mechanism (Q-head: $\hat{Y}_Q \approx E[Y|A, W]$) and the propensity score (g-head: $\hat{P}_g \approx P(A = 1|W)$), adequately learns and utilizes internal representations corresponding to a known critical confounder ($W_1 = W_{crit}$). Specifically, we test if $W_1$ is clearly represented in the shared layers and if ablating neurons associated with this representation impacts the performance of each head and the final ATE estimate derived by targeting the network's outputs. We also compare different drop of performance from ablating different set of neurons from the network.

\subsection{Methodology}
The multi-task model $f_{Q,g}$ is trained on DS1. Input $W \in \mathbb{R}^{10}$ is processed by shared layers to produce $h_{\text{shared}}(W)$. The Q-head computes $\hat{Y}_Q = f_Q(h_{\text{shared}}(W), A)$, and the g-head computes $\hat{P}_g = \sigma(f_g(h_{\text{shared}}(W)))$. Training minimizes a combined loss $L = w_q \mathcal{L}_{MSE}(Y, \hat{Y}_Q) + w_g \mathcal{L}_{BCE}(A, \hat{P}_g)$.
We train linear regression probes $f_{\text{probe}}(h^{(l)}_{\text{shared}}) \approx W_1$ on activations from each shared layer $l$. Probe accuracy ($R^2$ in this case) quantifies linear representability.
Neurons $N_{W1}$ in a chosen shared layer, identified via probe weights as predictive of $W_1$, are selected. Ablation is performed by setting the output of $N_{W1}$ to zero during the forward pass. The impact on Q-head loss ($\Delta \text{MSE}_Q$) and g-head loss ($\Delta \text{BCE}_g$) is measured on a test set.
Finally, the TMLE ATE estimator, $\hat{\psi}_{TMLE}$, is computed using the original Q and g predictions from $f_{Q,g}$, and again using predictions from the ablated network. We compare the point estimates and confidence intervals to the known true ATE. Consult Appendix \ref{app:exp1_details} for more details.

\subsection{Results}
\textit{Probe Analysis:} Linear probes trained to predict the critical confounder $W_1$ from shared layer activations achieved high $R^2$ scores, indicating that $W_1$ is strongly and linearly represented within these layers. The $R^2$ values (Table \ref{tab:exp1_probe_r2}) suggested a clear encoding of the confounder, with slightly decreasing linear representability in deeper layers.

\begin{table}[h!]
\centering
\caption{Probe Accuracy for $W_1$ ($W_{crit}$) Prediction.}
\label{tab:exp1_probe_r2}
\begin{tabular}{l c c}
\hline
Shared Layer & Probe Metric ($R^2$) \\
\hline
h1 & 0.9479\\
h2 & 0.9246\\
h3 & 0.8913\\
h4 & 0.8781\\
h5 & 0.8591\\
h6 & 0.8273\\
h7 & 0.8043\\
h8 & 0.7845\\
h\_shared & 0.6886\\
\hline
\end{tabular}
\end{table}

\begin{figure}[h!]
    \centering
    \includegraphics[width=0.9\textwidth]{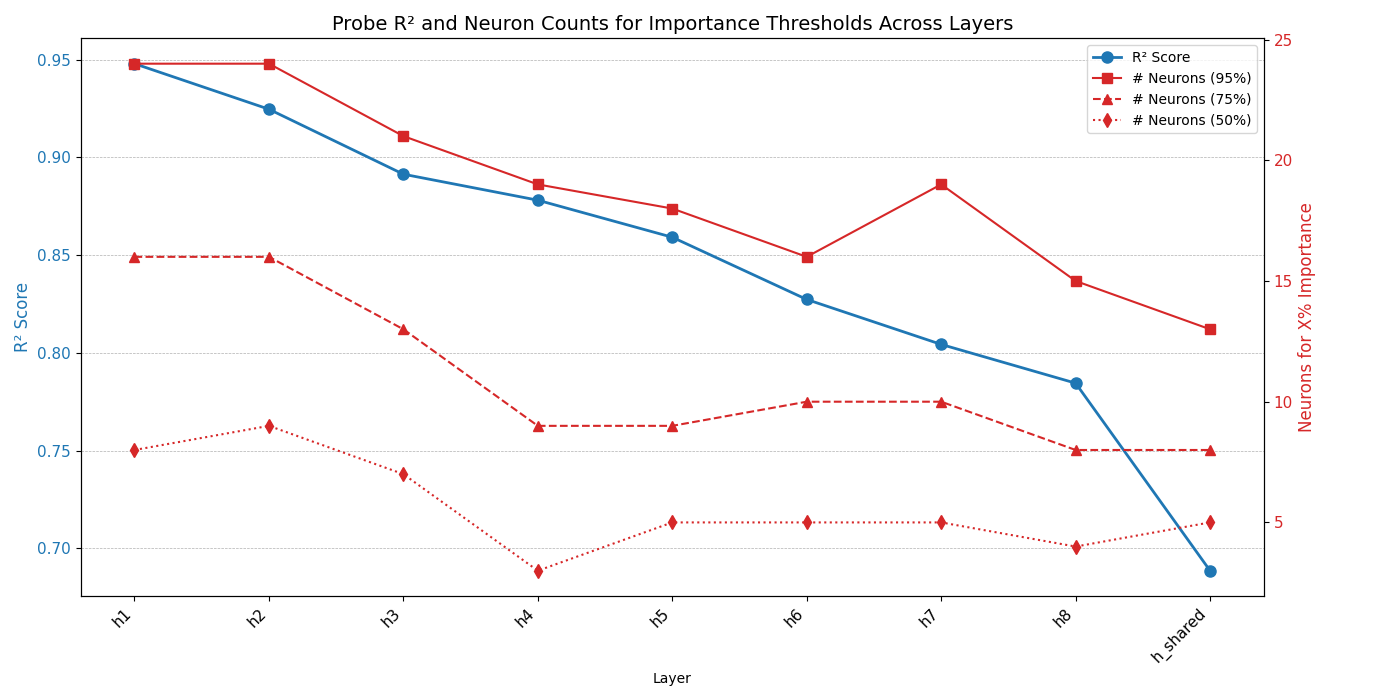} 
    \caption{Probe R² and Neuron Counts for Importance Thresholds Across Layers. This plot shows the R² score (blue line, left axis) of linear probes predicting $W_1$ from each shared layer's activations. The red lines (right axis) show the number of neurons (sorted by probe coefficient magnitude) needed to capture 95\% (solid), 75\% (dashed), and 50\% (dotted) of the total importance.}
    \label{fig:exp1_r2_neurons}
\end{figure}

\begin{figure}[h!]
    \centering
    \includegraphics[width=0.8\textwidth]{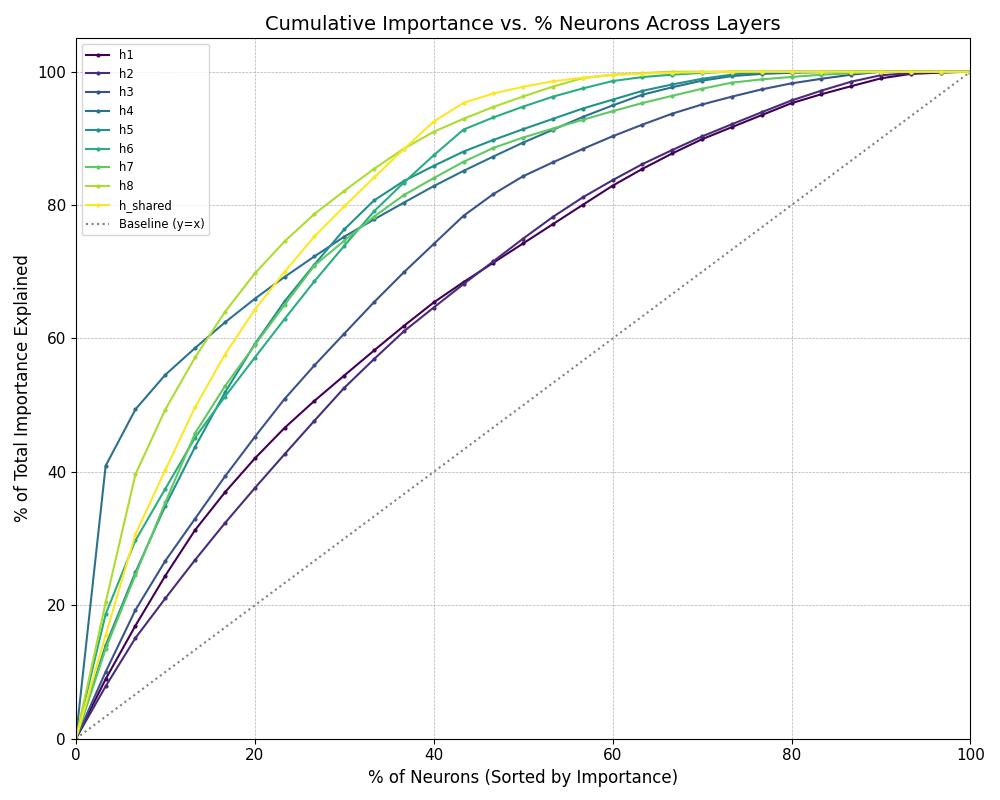} 
    \caption{Cumulative Importance vs. \% Neurons Across Layers. Each curve represents a shared layer, showing the percentage of total $W_1$ importance (sum of absolute probe coefficients) captured as more neurons (sorted by importance) are included. Steeper curves indicate higher concentration of importance in top neurons.}
    \label{fig:exp1_cumulative_importance}
\end{figure}

Figure \ref{fig:exp1_r2_neurons} illustrates the probe $R^2$ scores alongside the number of neurons required to capture 95\%, 75\%, and 50\% of the total importance (derived from probe coefficients) across layers. It visually confirms the high representability and shows that a relatively small fraction of neurons accounts for a large portion of the linear importance related to $W_1$. Figure \ref{fig:exp1_cumulative_importance} further emphasizes this concentration of importance, plotting the cumulative importance explained against the percentage of neurons sorted by importance. The curves rise steeply, particularly for later layers, indicating that the most important neurons capture the majority of the linear signal for $W_1$.

\begin{figure}[h!]
    \centering
    \includegraphics[width=\textwidth]{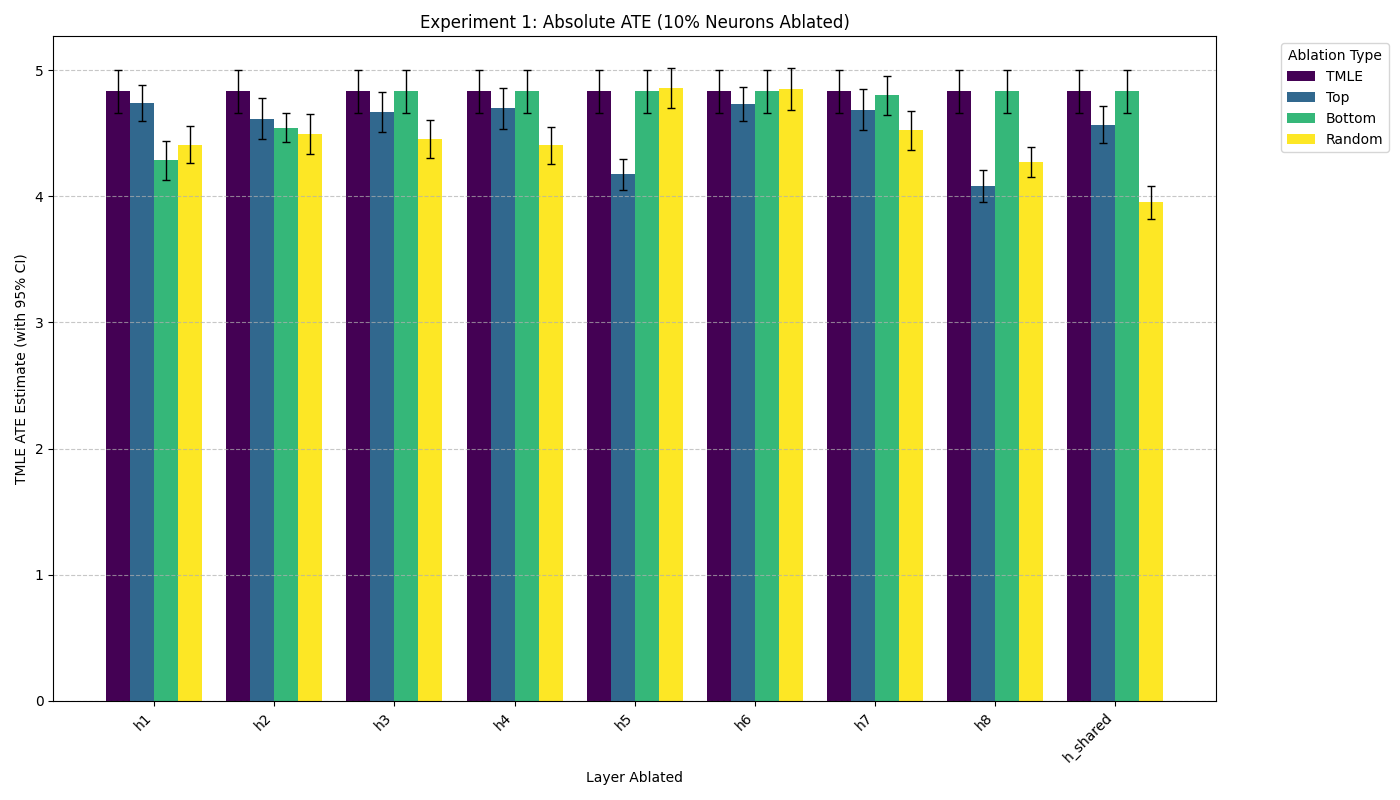} 
    \caption{Experiment 1: Absolute ATE ($10\%$ Neurons Ablated). This bar chart compares the TMLE ATE estimate (with 95\% CI) obtained using the original network outputs (Baseline 'TMLE') versus outputs after ablating the Top 10\%, Bottom 10\%, or Random 10\% most important neurons (according to $W_1$ probe) in each shared layer.}
    \label{fig:exp1_ablation_exp1}
\end{figure}

\textit{Ablation Analysis:} Ablating neurons identified as most important for representing $W_1$ had a noticeable impact on the model's performance and the final TMLE estimate, while ablating the least important or randomly selected neurons had considerably less effect. Figure \ref{fig:exp1_ablation_exp1} shows the absolute TMLE ATE estimate (with 95\% CI) under different ablation conditions (Baseline/TMLE, Top 10\%, Bottom 10\%, Random 10\% neurons ablated) for each layer. Ablating the 'Top' neurons consistently leads to a significant deviation in the ATE estimate compared to the baseline and other ablation types, suggesting these neurons are causally relevant to how the model processes the confounder $W_1$.

\begin{figure}[h!]
    \centering
    \includegraphics[width=\textwidth]{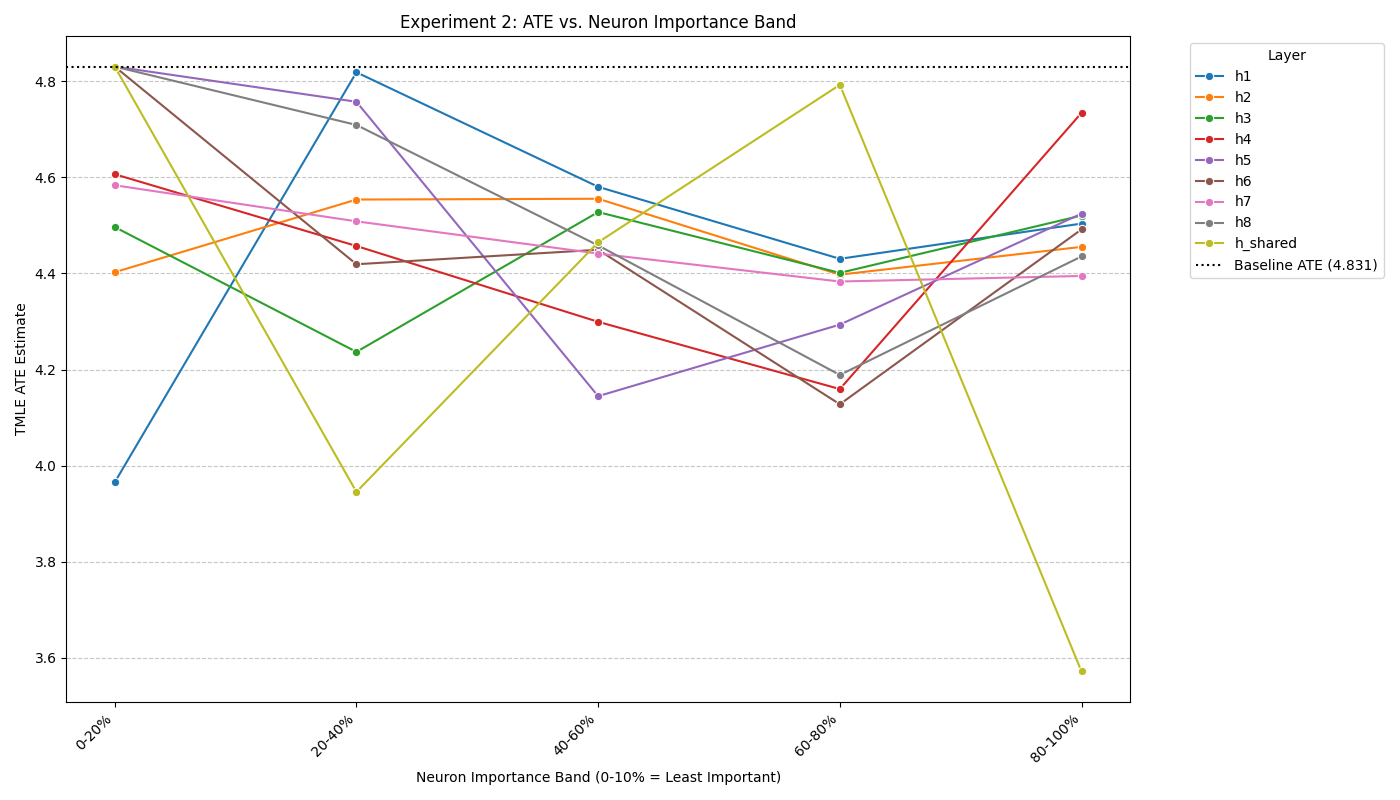}
    \caption{Experiment 2: ATE vs. Neuron Importance Band. This line plot shows the TMLE ATE estimate resulting from ablating neurons within specific 5\% importance bands (sorted from least to most important based on $W_1$ probe coefficients) for each shared layer. The baseline ATE (no ablation) is shown as a dotted line.}
    \label{fig:exp1_ablation_exp2}
\end{figure}

Figure \ref{fig:exp1_ablation_exp2} presents the results from ablating neurons in 20\% importance bands, ranging from least important (0-20\%) to most important (80-100\%). The plot shows the resulting ATE estimate for each band ablation across different layers. A clear trend emerges where ablating bands containing the most important neurons (towards the right side of the plot) causes the largest deviation from the baseline ATE (dotted line), reinforcing the finding that neuron importance, as identified by the linear probe for $W_1$, correlates with causal impact on the TMLE estimate. We can also see that the later layers are more sensible to \textit{important neurons} rather than non-important ones, whereas than the former layers has a more uniform effect of the ablation, hinting at computation being done with every neuron available. This also can be traced back to the results displayed in Figures \ref{fig:exp1_r2_neurons} and \ref{fig:exp1_cumulative_importance}, where we can see later layer carry \textbf{much more information on fewer neurons}.

These ablation results support the hypothesis that the network learns representations that capture the confounder $W_1$ and that this representation is being refined, abstracted throughout the layers. These specific representations causally affects the downstream task performance and the final causal estimate.

\section{Discovering and Visualizing Treatment vs. Confounder Pathways I}
\label{sec:exp2_cfs}

Using Causal Tracing initiated from different \textit{input neurons} of a pre-trained $f_{Q,g}$ model, we aim to identify, visualize, and compare the computational pathways through which information associated with different input covariates propagates and influences subsequent layers. We analyze pathway differences and overlaps using pathway analysis techniques to understand how the network processes distinct input signals.

\subsection{Methodology}
An $f_{Q,g}$ model is trained on DS2 (No Effect Confounders). For each input neuron $i$, a causal trace is generated by initiating an activation patch at the input layer -- changing the activation of only input neuron $i$ -- and propagating the effects forward using the \texttt{trace\_causal\_pathways} logic: we successively patch the activations of subsequent neurons to identity the pathways which are the most impacted by the current neuron. The quality of the resulting pathways for each input neuron is assessed using the two follwi metrics:
\textit{Sparsity Score:} Measures the average inverse width of activated layers within the pathway (higher score means narrower, more focused pathways).
\textit{Success Score:} Measures the fraction of intermediate pathway nodes from which the trace successfully propagated further (did not "fail" based on the activation change threshold).
The overlap between pathways originating from different input neurons $i$ and $j$ is calculated using the Jaccard index on the sets of activated (non-input, non-failed) nodes identified in their respective traces.

\subsection{Results}
\begin{figure}[h!]
    \centering
    \includegraphics[width=0.8\textwidth]{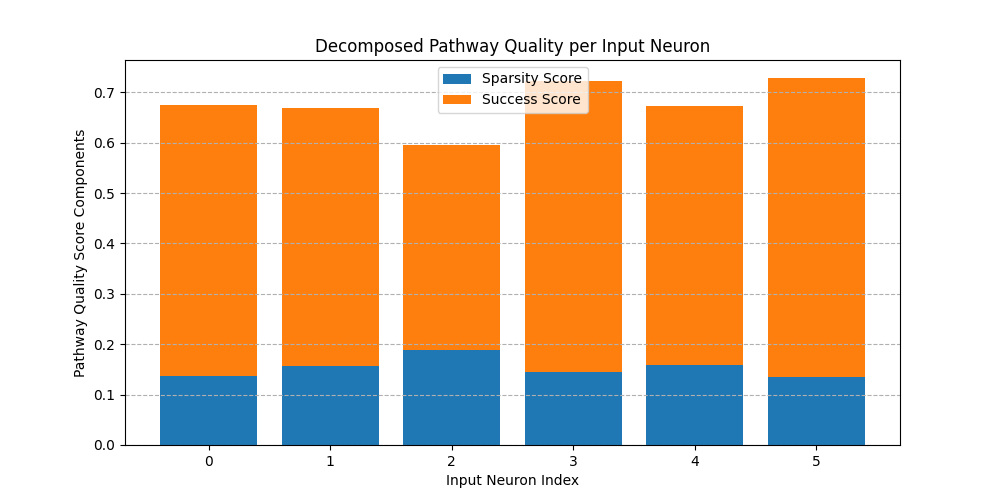}
    \caption{Decomposed Pathway Quality per Input Neuron (No Effect Data). Stacked bar chart showing sparsity score (blue) and success score (orange) for causal traces originating from each input neuron of the $f_{Q,g}$ model trained on data with no treatment effect.}
    \label{fig:exp2_pathway_quality}
\end{figure}

The analysis of pathway characteristics provides insights into how the network internally processes information from different inputs, even in the absence of a true treatment effect. Figure \ref{fig:exp2_pathway_quality} shows the decomposed quality scores for pathways traced from each input neuron of the $f_{Q,g}$ model trained on the "no effect" data. This reveals variability in how focused (sparsity) and how strongly propagating (success) pathways are depending on the input source.

\begin{figure}[h!]
    \centering
    \includegraphics[width=0.7\textwidth]{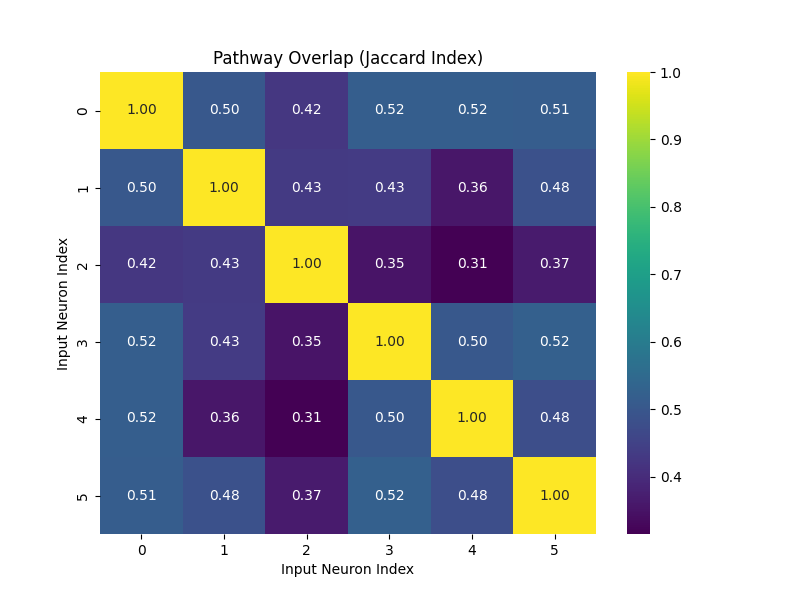}
    \caption{Pathway Overlap (Jaccard Index) Heatmap (No Effect Data). Shows the overlap between pathways originating from pairs of input neurons for the $f_{Q,g}$ model trained on data with no treatment effect.}
    \label{fig:exp2_overlap_matrix}
\end{figure}

Figure \ref{fig:exp2_overlap_matrix} visualizes the pathway overlap using a heatmap of Jaccard indices. This matrix quantifies the similarity between the sets of downstream neurons activated by interventions on pairs of different input neurons. It identifies inputs that tend to activate shared internal pathways (high overlap, yellow) versus those processed more independently (low overlap, purple). This information can be valuable for understanding potential functional grouping or separation of input processing within the network.

\section{Discovering and Visualizing Treatment vs. Confounder Pathways II}
\label{sec:exp3_visualize}

Using Causal Tracing initiated from different \textit{input neurons} of the pre-trained $f_{Q,g}$ model (from Exp \ref{sec:exp2_cfs}), we aim to identify, visualize, and compare the computational pathways through which information associated with different input covariates propagates and influences subsequent layers. We analyze pathway differences and overlaps using activation patching (\ref{ssec:act_patch_ii}) to understand how the network processes distinct input signals.

\subsection{Methodology}
This experiment utilizes the pre-trained multi-task model $f_{Q,g}$ developed in Experiment \ref{sec:exp1_tmle_interp}. The core technique employed is \textit{Causal Tracing}, as described in the last section \ref{sec:exp2_cfs}. The graph are generated to represent the causal interaction. Edges indicate significant causal influence propagating forward, while node colors represent layer progression and gray nodes indicate "failed sources" where the trace did not propagate further (the computation stopped there). More graphs are shown in the Appendix \ref{app:exp3_graphs}.

\subsection{Results}
\begin{figure}[h!]
    \centering
    \includegraphics[width=0.9\textwidth]{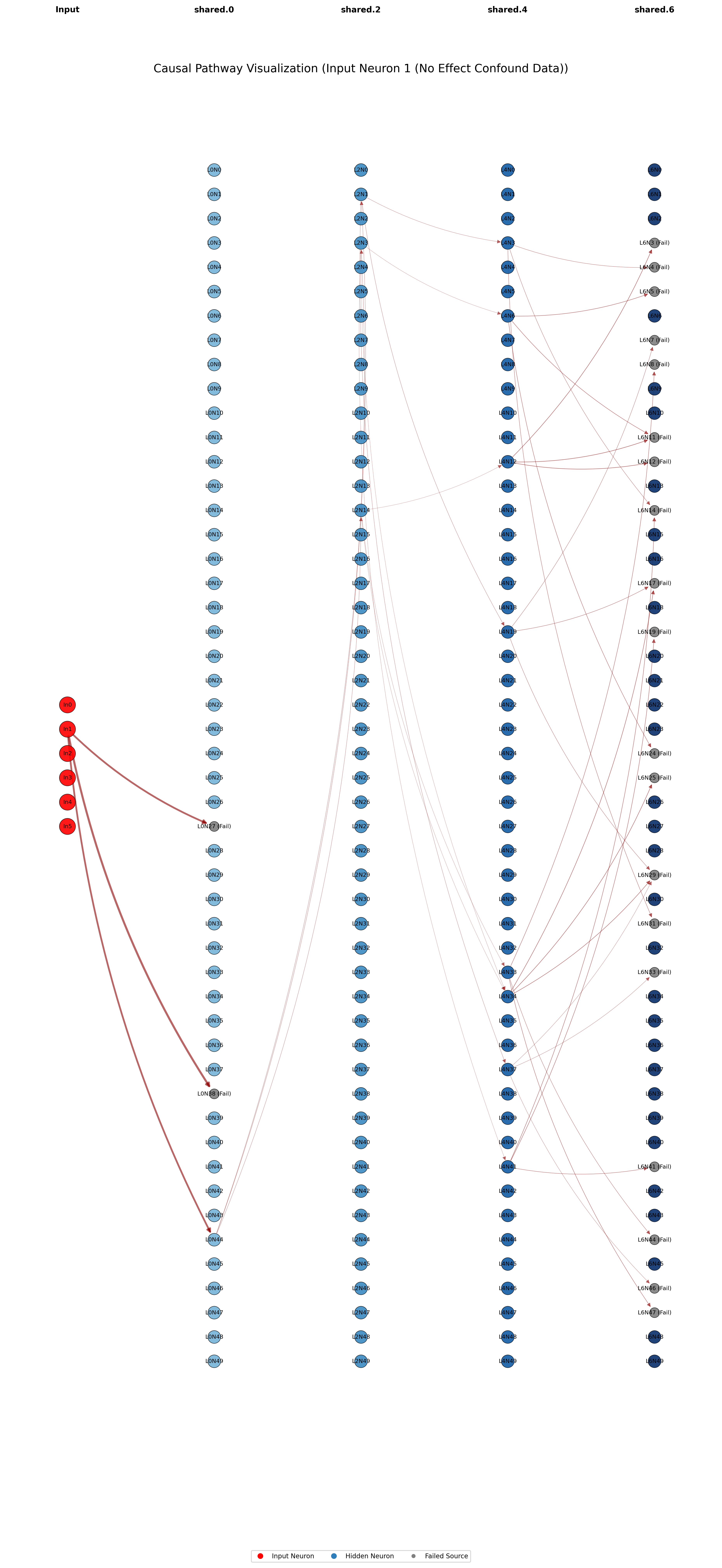}
    \caption{Causal Pathway Visualization (Example: Input Neuron 1). Shows the propagation of effects originating from activating input neuron 1 in the $f_{Q,g}$ model. Edges represent significant activation changes induced by patching.}
    \label{fig:exp3_viz_input_1}
\end{figure}

Causal tracing reveals distinct propagation patterns through the network's shared layers and into the task-specific heads. Figure \ref{fig:exp3_viz_input_1} provides an example visualization showing the pathways activated when tracing starts from input neuron 1. 

Comparing traces from different inputs highlights variations in network processing. Figure \ref{fig:exp2_overlap_matrix} from Experience \ref{sec:exp2_cfs} showed the pathway overlap matrix (Jaccard index) comparing the set of activated nodes for traces originating from each pair of input neurons in the $f_{Q,g}$ model trained on DS1 (strong confounder). This reveals which inputs tend to utilize similar downstream computational pathways.

\begin{figure}[h!]
    \centering
    \includegraphics[width=0.9\textwidth]{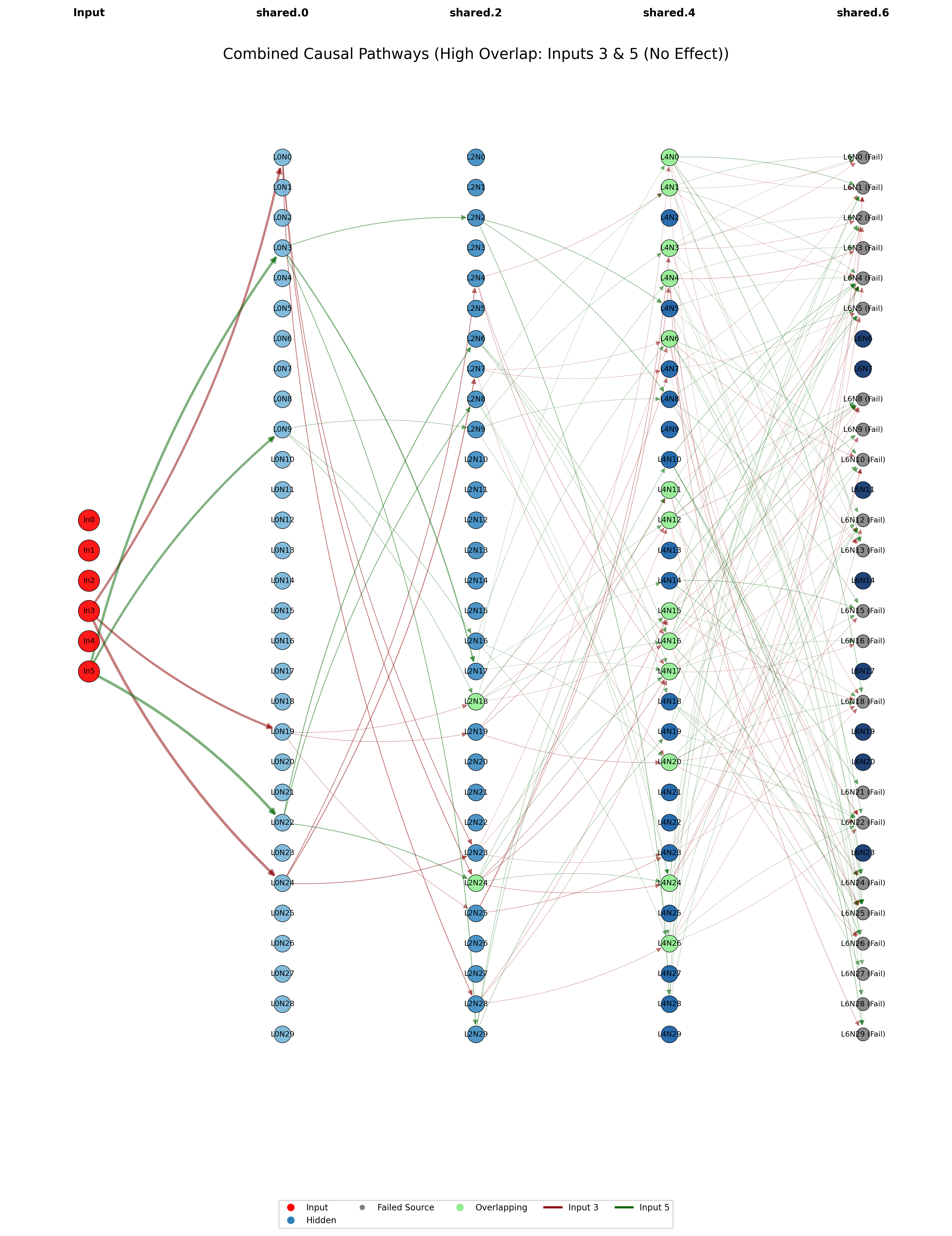} 
    \caption{Combined Causal Pathways (High Overlap Example). Overlays pathways from traces initiated from two input neurons with high overlap in the $f_{Q,g}$ model. Green nodes are activated in both traces.}
    \label{fig:exp3_viz_high_overlap_strong}
\end{figure}

\begin{figure}[h!]
    \centering
    \includegraphics[width=0.9\textwidth]{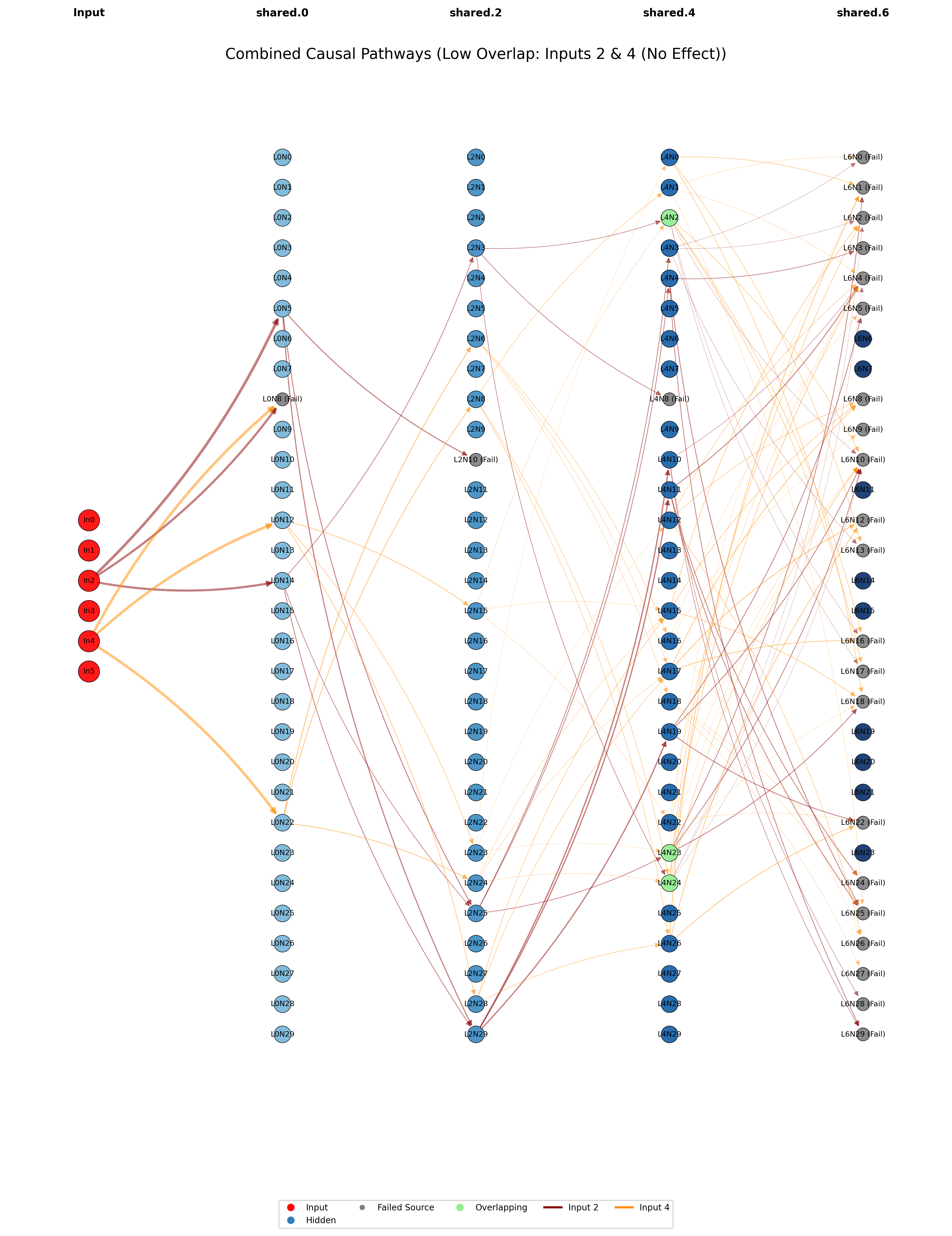} 
    \caption{Combined Causal Pathways (Low Overlap Example). Overlays pathways from traces initiated from two input neurons with low overlap in the $f_{Q,g}$ model.}
    \label{fig:exp3_viz_low_overlap_strong}
\end{figure}

Combined visualizations overlay pathways for specific input pairs. Figure \ref{fig:exp3_viz_high_overlap_strong} overlays traces for two input neurons identified as having high overlap (e.g., based on Figure \ref{fig:exp2_overlap_matrix}). Green nodes represent neurons activated in both pathways, indicating shared processing resources. Conversely, Figure \ref{fig:exp3_viz_low_overlap_strong} overlays traces for a low-overlap pair, illustrating how the network routes information from these inputs through largely separate internal pathways. These visualizations provide insights into how the network potentially segregates or integrates information from different input sources, confounders versus other.

\chapter{Conclusion}

This thesis explored the application of MI techniques, traditionally developed for large AI models, to NNs utilized within causal inference frameworks, and in particularly TMLE, in bio-statistics. Bridging the gap between the predictive power of NNs and the critical need for transparency in bio-statistical analysis, we investigated whether MI tools could elucidate the internal workings of these models in causal settings.

Our experiments demonstrated that MI techniques offer valuable insights. We successfully employed methods like probing and ablation to validate NN-based nuisance function estimators, effectively identifying internal representations of confounders and confirming their causal relevance to model predictions and the final ATE estimate. Furthermore, causal tracing proved effective in discovering and visualizing the distinct computational pathways within NNs, revealing how information from different input covariates, such as confounders versus treatment variables, is processed and segregated or integrated by the network.

These findings indicate that MI is not only feasible but also beneficial for enhancing the trustworthiness and understanding of NNs in bio-statistical causality. It provides methods for validating learned representations against domain knowledge, debugging model behavior, and potentially uncovering how models handle complex variable relationships.

However, significant challenges and opportunities for future research remain. The inherent complexity of biological systems and the representations learned by NNs necessitate further development in MI techniques, particularly in robustly mapping low-level activations to high-level causal concepts (causal abstraction). Formalizing the integration of MI findings, such as insights from pathway analysis or causal scrubbing, directly into statistical inference procedures like refining TMLE estimates or quantifying model robustness, presents a crucial next step. Continued research, fostering collaboration between MI experts and bio-statisticians, is essential to develop tailored interpretability methods that advance both fields and promote the responsible use of sophisticated AI in high-stakes scientific discovery.

\backmatter

\bibliographystyle{plainnat}
\bibliography{refs}

\backmatter

\appendix
\chapter{Appendix}

\chapter{Datasets and Models Details}
\label{app:datasets_models}

This annex provides detailed information on the synthetic datasets, neural network architectures, and training configurations used in the experiments presented in Chapter \ref{chap:experiments}.

\section{Experiment 1: Validating Nuisance Function Estimators in TMLE (Section \ref{sec:exp1_tmle_interp})}
\label{app:exp1_details}

\subsection{Dataset: DS1 - Strong Confounder}
\begin{itemize}
    \item \textbf{Generation Script:} \texttt{data\_generation/tmle\_strong\_cov\_data.py}
    \item \textbf{Description:} This dataset is designed to simulate a scenario with a strong confounder ($W_1$) that significantly influences both the treatment assignment ($A$) and the outcome ($Y$). The data includes multiple covariates ($W \in \mathbb{R}^{10}$), a binary treatment ($A \in \{0, 1\}$), and a continuous outcome ($Y$). The true ATE is known to be 2.0.
    \item \textbf{Parameters:} $N = 10000$ samples, SEED = 42 (for training script, data generation used SEED = 888 in \texttt{tmle\_estimate\_interp.py}).
\end{itemize}

\subsection{Model: $f_{Q,g}$ - ToyTMLE}
\begin{itemize}
    \item \textbf{Definition Script:} \texttt{models/toy\_tmle.py} (using \texttt{ToyTMLEConfig})
    \item \textbf{Architecture:} A multi-task feed-forward neural network.
        \begin{itemize}
            \item \textit{Shared Layers:} Processes input covariates $W$ through a sequence of hidden layers (5 layers, 100 units each, ReLU activation) defined in \texttt{ToyTMLEConfig['model\_conf']}. Input dimension is determined by the data ($d=10$).
            \item \textit{Q-Head:} Predicts the outcome $Y$. Takes the shared representation $h_{\text{shared}}$ concatenated with the treatment $A$ as input, followed by a linear output layer. $f_Q(h_{\text{shared}}, A) \rightarrow \hat{Y}_Q$.
            \item \textit{g-Head:} Predicts the propensity score $P(A=1|W)$. Takes the shared representation $h_{\text{shared}}$ as input, followed by a linear layer and a Sigmoid activation. $f_g(h_{\text{shared}}) \rightarrow \hat{P}_g$.
        \end{itemize}
    \item \textbf{Auxiliary Models:}
        \begin{itemize}
            \item \texttt{StandardScaler} (scikit-learn): Used for scaling input covariates $W$.
            \item \texttt{LinearRegression} (scikit-learn): Used as probes to predict $W_1$ from hidden layer activations (\texttt{h1}, \texttt{h2}, \texttt{h3}, \texttt{h\_shared}).
        \end{itemize}
\end{itemize}

\subsection{Training}
\begin{figure}[htpb!]
    \centering
    \includegraphics[width=\textwidth]{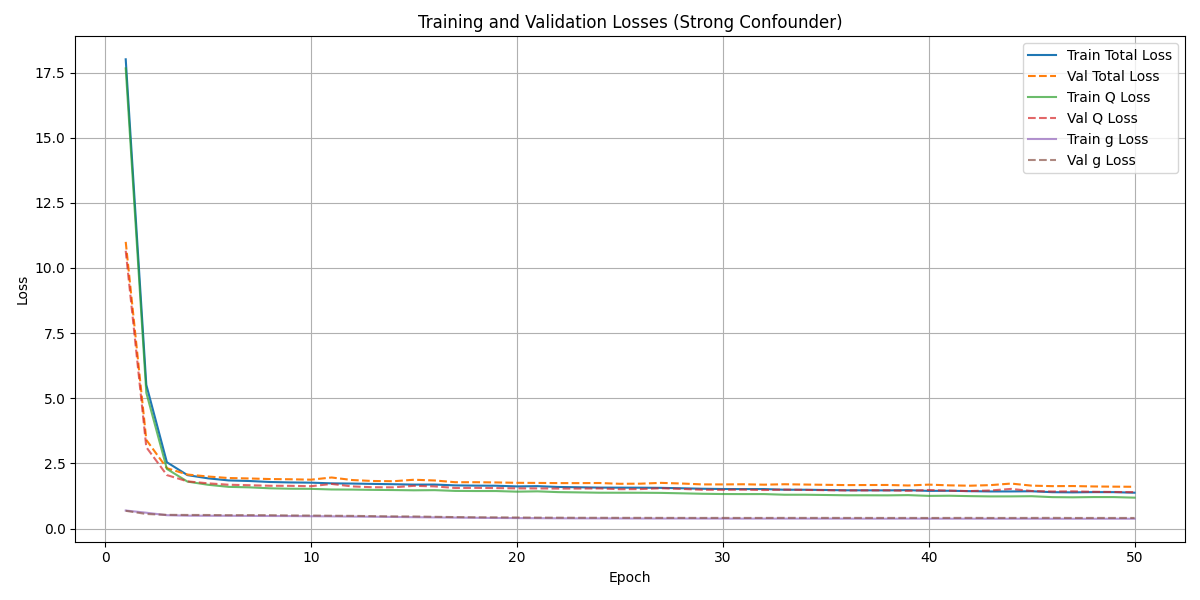}
    \caption{Training and Validation Losses for ToyTMLE on DS1 (Strong Confounder).}
    \label{fig:loss_strong_confounder}
\end{figure}

\begin{itemize}
    \item \textbf{Script:} \texttt{training/train\_tmle\_strong\_confounder.py}
    \item \textbf{Loss Function:} \texttt{ToyTMLELoss}, which is MSE loss for $f_Q$ and BCE for $f_g$. Parameter $\alpha=0.5$ balances the weight between two losses.
    \item \textbf{Optimizer:} Adam with learning rate = 0.0003.
    \item \textbf{Hyperparameters:} EPOCHS = 50, BATCH\_SIZE = 128, TEST\_SIZE = 0.2. Through experiments, we found HIDDEN\_LAYER = 10 and HIDDEN\_SIZE = 30 yielded the best results.
    \item \textbf{Loss Plot:} See Figure \ref{fig:loss_strong_confounder}.
\end{itemize}

\section{Experiment 2: Analyzing Causal Pathway Characteristics (Section \ref{sec:exp2_cfs})}
\label{app:exp2_details}

\subsection{Dataset: DS2 - No Effect Confounders}
\begin{itemize}
    \item \textbf{Generation Script:} \texttt{data\_generation/no\_effect\_confounders\_data.py}
    \item \textbf{Description:} This dataset simulates a scenario where confounders ($W$) influence treatment ($A$) and outcome ($Y$), but the treatment itself has \textit{no causal effect} on the outcome (True ATE = 0). This allows testing if pathway analysis techniques can correctly identify the lack of a direct $A\to Y$ pathway within the model, despite the presence of confounding paths ($W \to A$, $W \to Y$). The structure is similar to DS1 otherwise ($W \in \mathbb{R}^{6}$, $A \in \{0, 1\}$, $Y$ continuous).
    \item \textbf{Parameters:} $N = 10000$ samples (for training), SEED = 42. Pathway analysis often uses $N=1000$ for speed.
\end{itemize}

\subsection{Model: $NN_{HTE}$ - ToyTMLE (Pathway Analysis Version)}
\begin{itemize}
    \item \textbf{Definition Script:} \texttt{models/toy\_tmle.py} (using \texttt{ToyTMLENoEffectConfig} or \texttt{ToyTMLEConfigPathway})
    \item \textbf{Architecture:} Identical structure to the model in Experiment 1, but trained on DS2 (No Effect Confounders). Specific configuration used for pathway analysis experiments (\texttt{ToyTMLEConfigPathway}) uses 5 hidden layers of size 30.
        \begin{itemize}
            \item \textit{Shared Layers:} 5 hidden layers, 30 units each, ReLU activation. Input dimension $d=6$.
            \item \textit{Q-Head} and \textit{g-Head}: Same structure as Exp 1.
        \end{itemize}
    \item \textbf{Auxiliary Models:} Linear probes for abstraction function $\alpha$.
\end{itemize}

\subsection{Training}
\begin{figure}[htpb!]
    \centering
    \includegraphics[width=\textwidth]{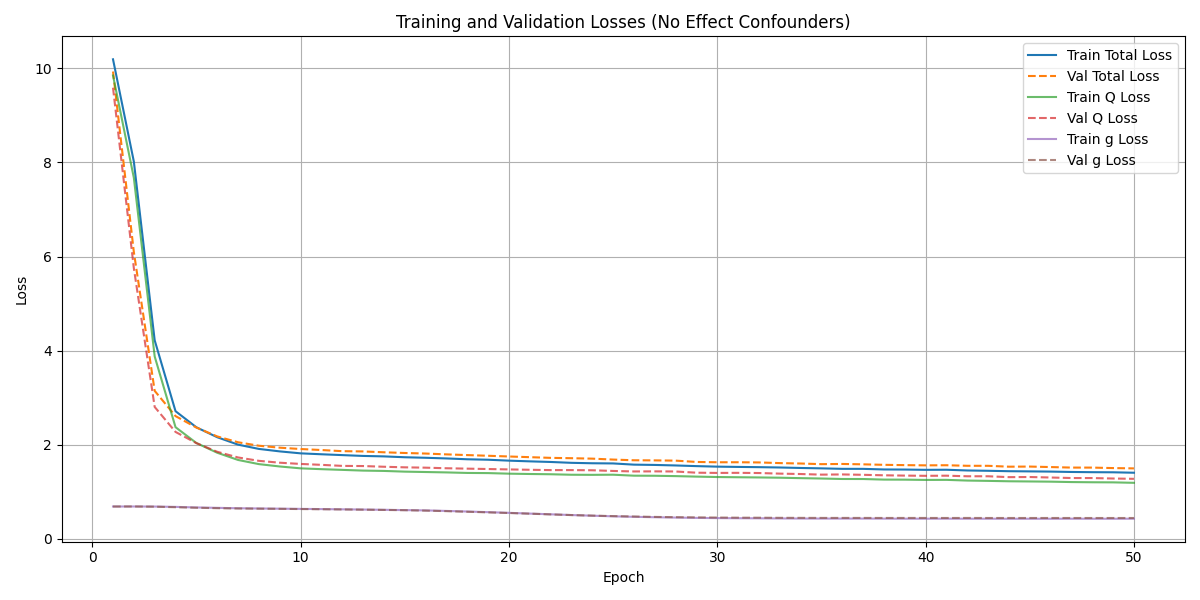}
    \caption{Training and Validation Losses for ToyTMLE on DS2 (No Effect Confounders).}
    \label{fig:loss_no_effect}
\end{figure}

\begin{itemize}
    \item \textbf{Script:} \texttt{training/train\_tmle\_no\_effect\_confounder.py}
    \item \textbf{Loss Function:} \texttt{ToyTMLELoss}, which is MSE loss for $f_Q$ and BCE for $f_g$. Parameter $\alpha=0.5$ balances the weight between two losses.
    \item \textbf{Optimizer:} Adam with learning rate = 0.0003.
    \item \textbf{Hyperparameters:} EPOCHS = 50, BATCH\_SIZE = 128, TEST\_SIZE = 0.2. Through experiments, we found HIDDEN\_LAYER = 5 and HIDDEN\_SIZE = 30 yielded the best results.
    \item \textbf{Loss Plot:} See Figure \ref{fig:loss_no_effect}.
\end{itemize}

\section{Experiment 3: Discovering and Visualizing Treatment vs. Confounder Pathways (Section \ref{sec:exp3_visualize})}
\label{app:exp3_details}

\subsection{Dataset: DS1 - Strong Confounder}
\begin{itemize}
    \item \textbf{Details:} Same as in Experiment 1 (Section \ref{app:exp1_details}).
\end{itemize}

\subsection{Model: $f_{Q,g}$ - Pre-trained ToyTMLE}
\begin{itemize}
    \item \textbf{Base Model ($f_{Q,g}$):} The \texttt{ToyTMLE} model trained in Experiment 1 (Section \ref{app:exp1_details}), loaded from \texttt{./models/toy\_tmle.pth}. Architecture details as in Section \ref{app:exp1_details}.
    \item \textbf{Interpretability Model:} No separate interpretability model (like an SAE) is trained or used in this experiment. Analysis relies on Causal Tracing applied directly to the base model $f_{Q,g}$.
\end{itemize}

\subsection{Training}
\begin{itemize}
    \item \textbf{Base Model Training:} As described in Section \ref{app:exp1_details}. No further training specific to this experiment is performed.
    \item \textbf{Loss Plot (Base Model):} See Figure \ref{fig:loss_strong_confounder}.
\end{itemize}

\section{All graphs from Experience 3 (\ref{sec:exp2_cfs})}\label{app:exp3_graphs}

\begin{figure}[htpb!]
    \centering
    \includegraphics[width=.5\textwidth]{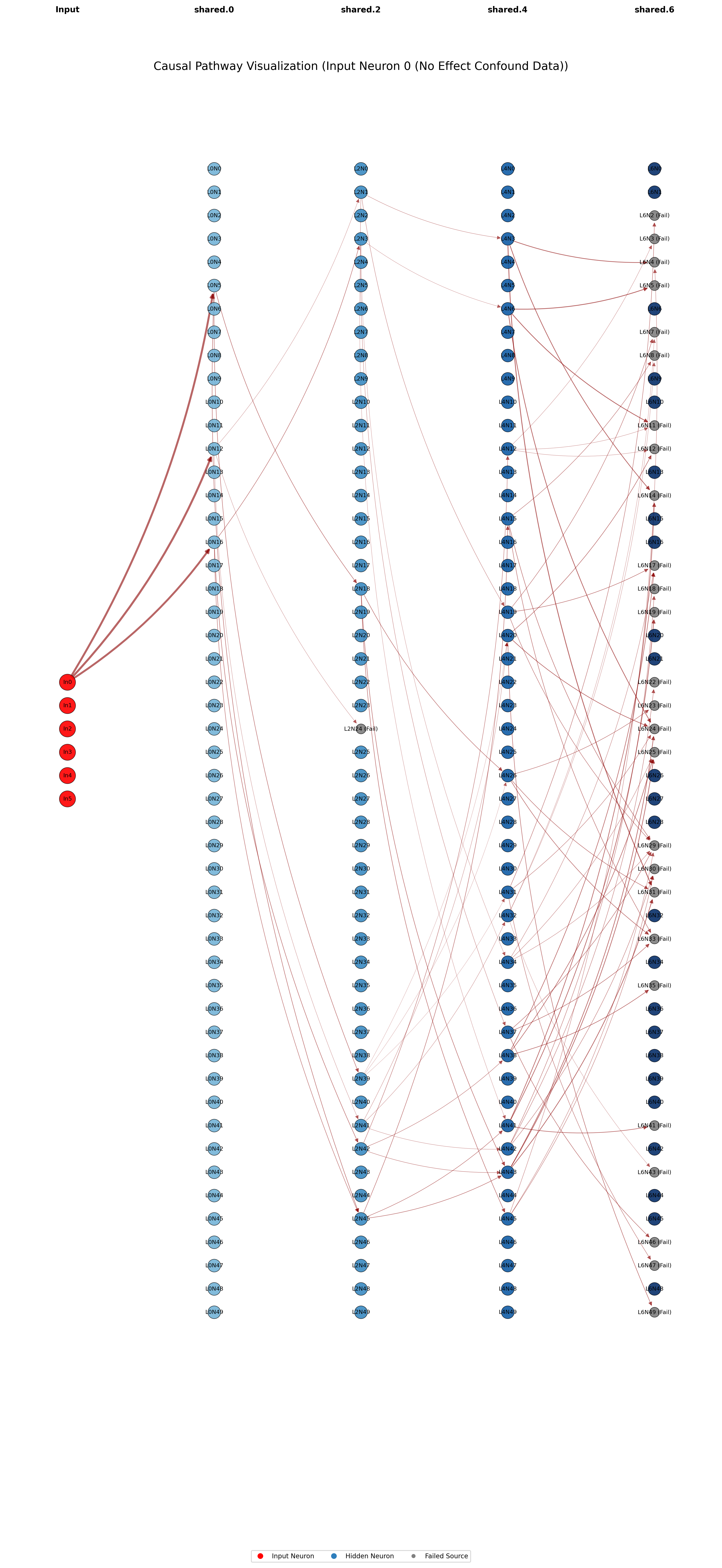}
    \caption{Pathways for confounder input 0}
    \label{fig:pathwyay0}
\end{figure}

\begin{figure}[htpb!]
    \centering
    \includegraphics[width=.5\textwidth]{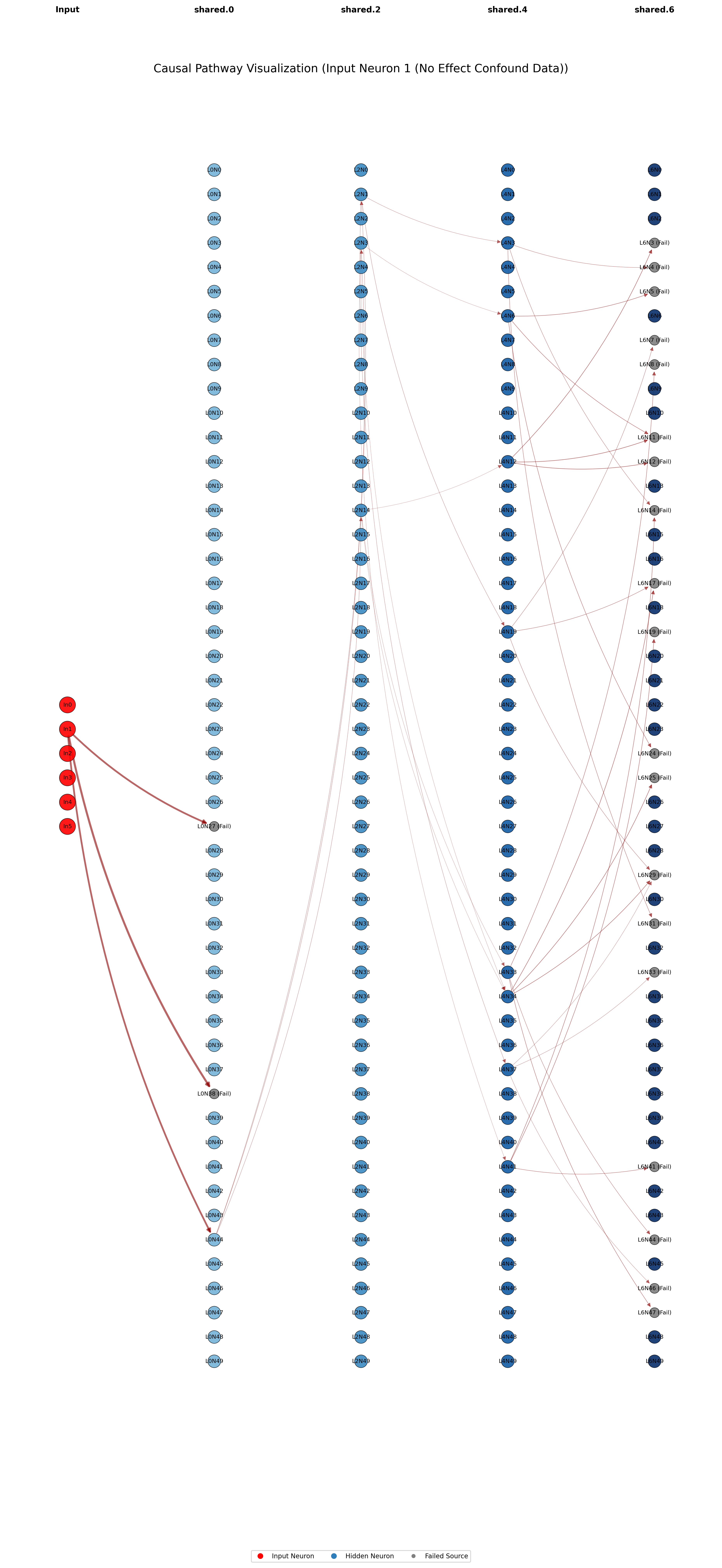}
    \caption{Pathways for confounder input 1}
    \label{fig:pathwyay1}
\end{figure}

\begin{figure}[htpb!]
    \centering
    \includegraphics[width=.5\textwidth]{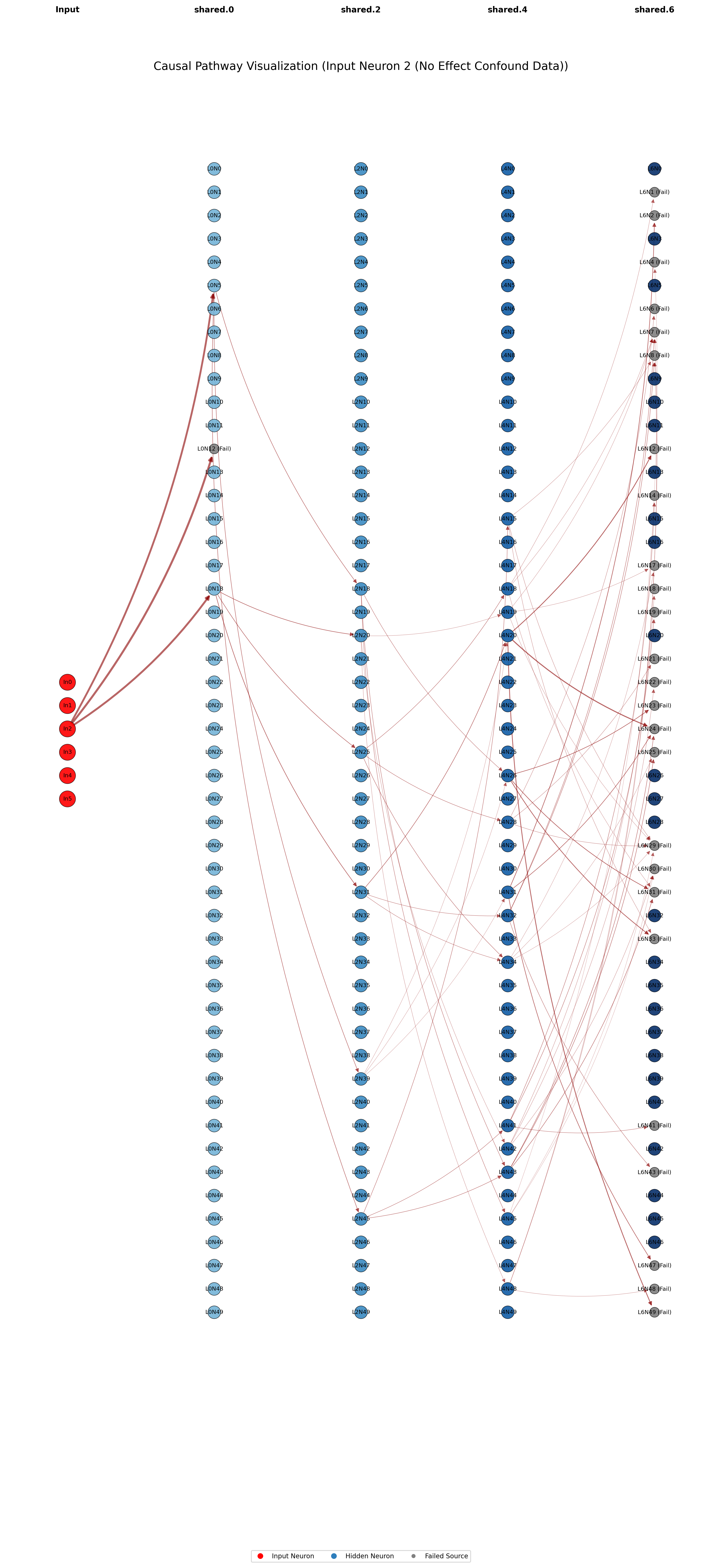}
    \caption{Pathways for confounder input 2}
    \label{fig:pathwyay2}
\end{figure}

\begin{figure}[htpb!]
    \centering
    \includegraphics[width=.5\textwidth]{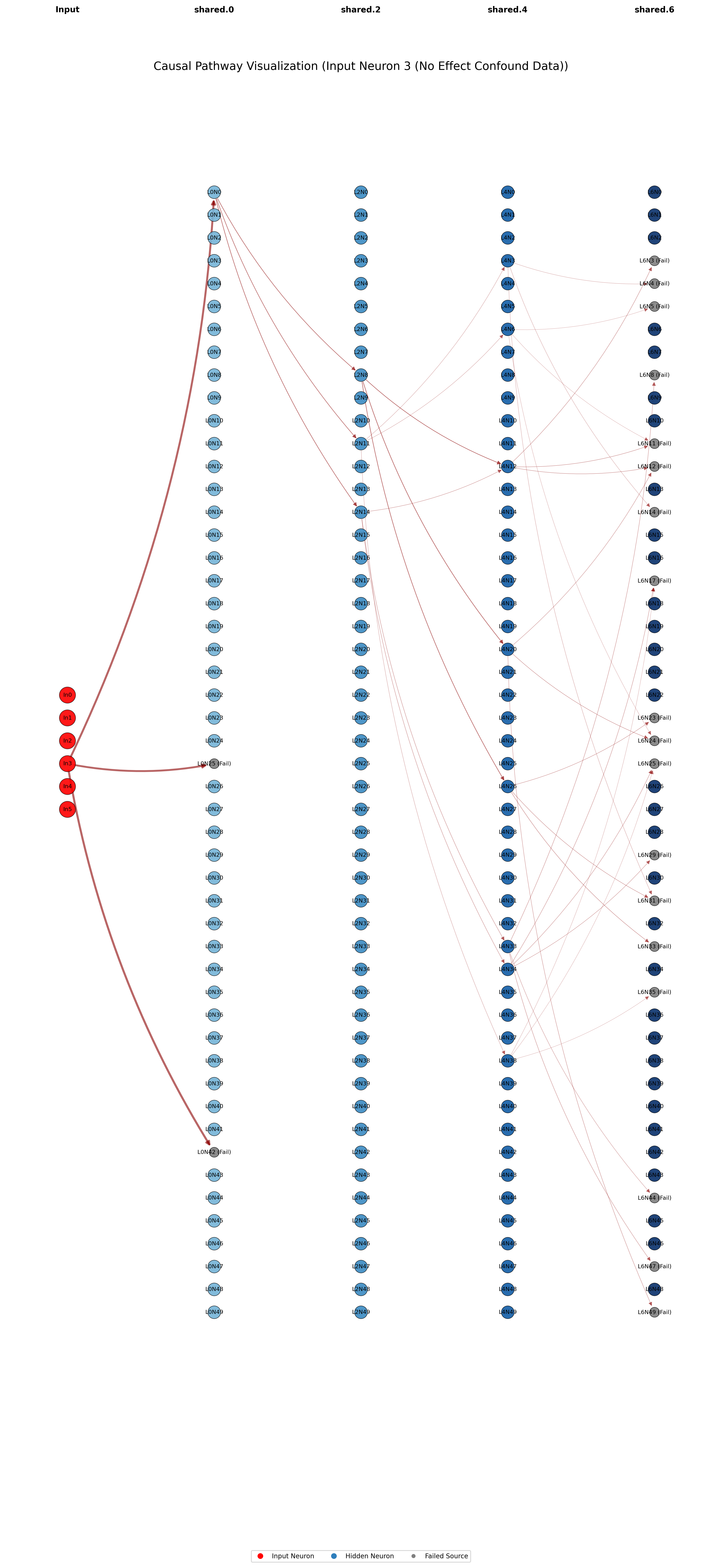}
    \caption{Pathways for confounder input 3}
    \label{fig:pathwyay3}
\end{figure}

\begin{figure}[htpb!]
    \centering
    \includegraphics[width=.5\textwidth]{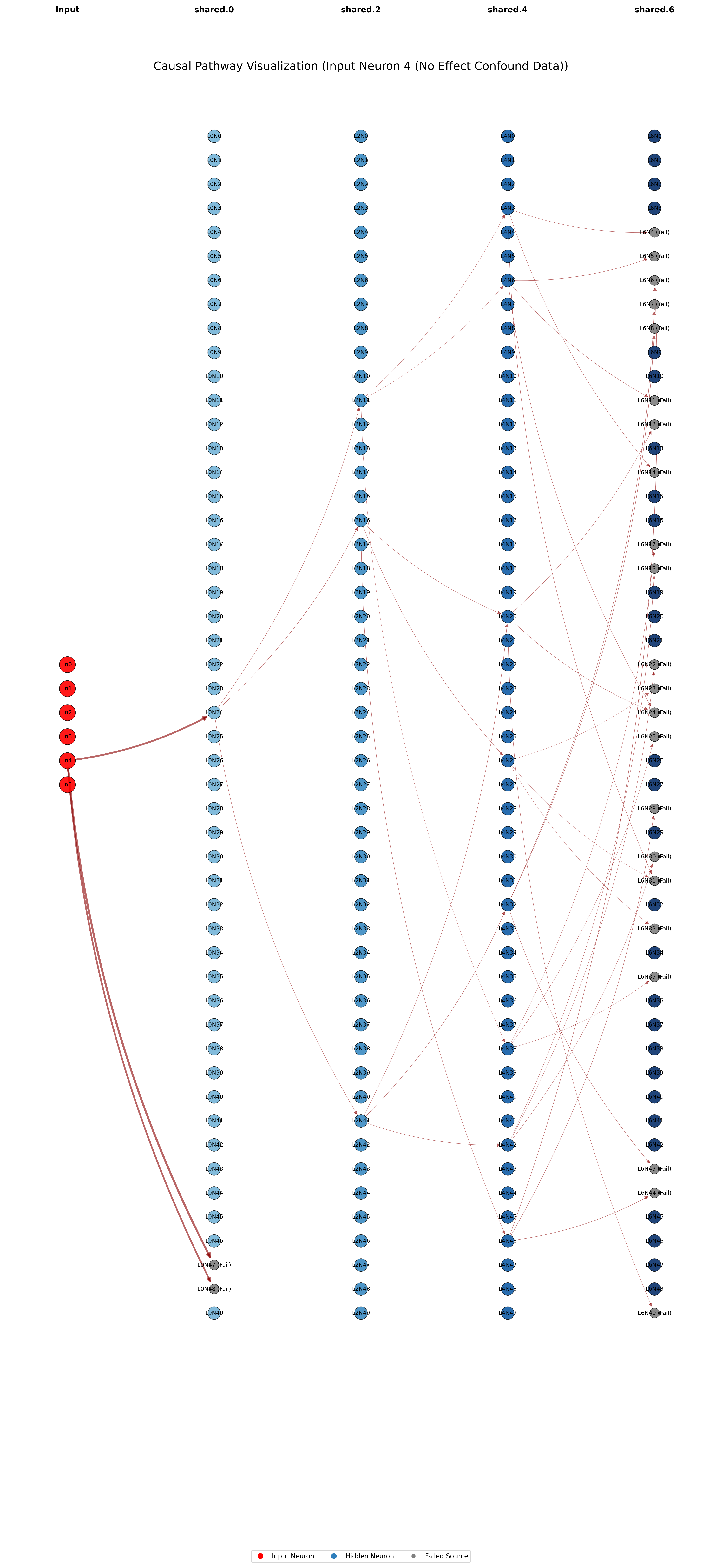}
    \caption{Pathways for confounder input 4}
    \label{fig:pathwyay4}
\end{figure}
\begin{figure}[htpb!]
    \centering
    \includegraphics[width=.5\textwidth]{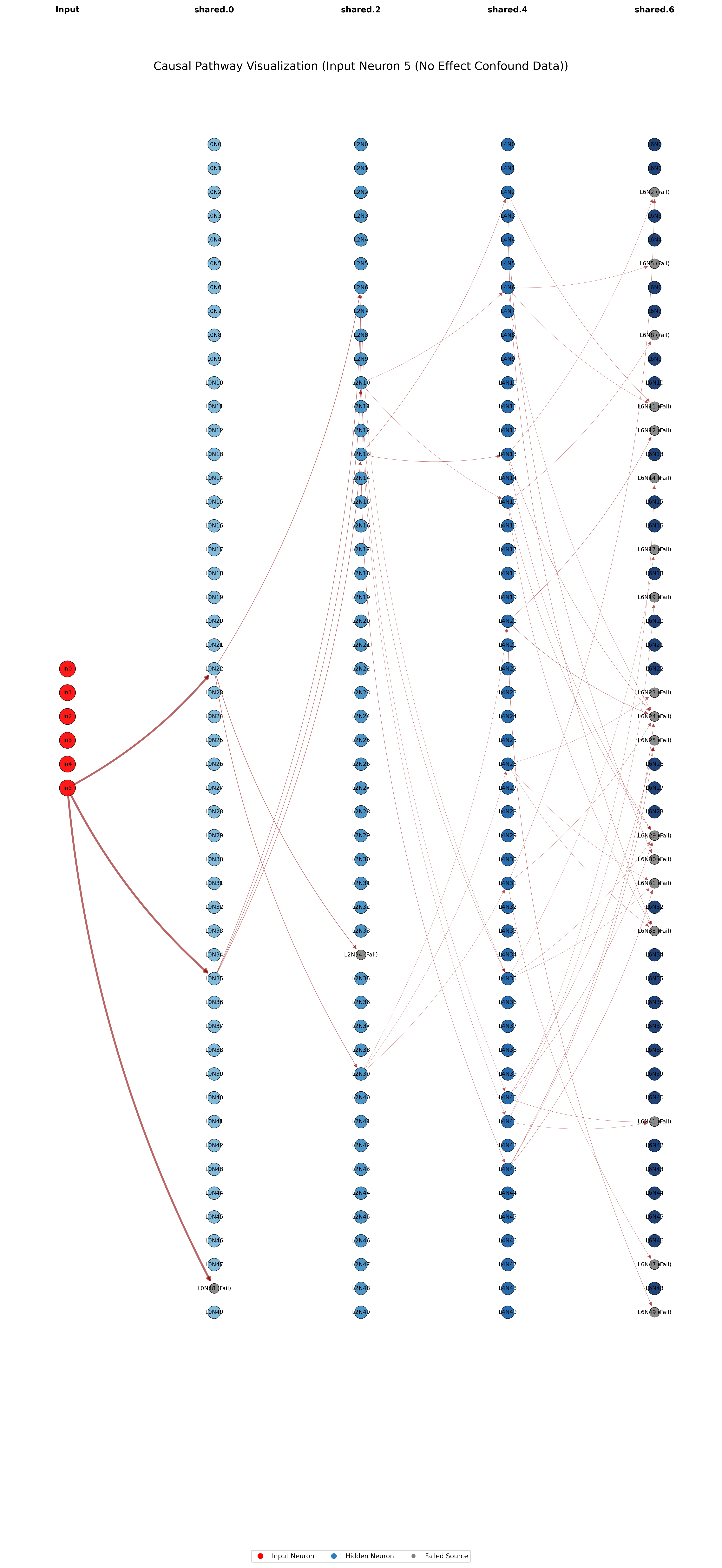}
    \caption{Pathways for confounder input 5}
    \label{fig:pathwyay5}
\end{figure}

\begin{figure}[htpb!]
    \centering
    \includegraphics[width=.5\textwidth]{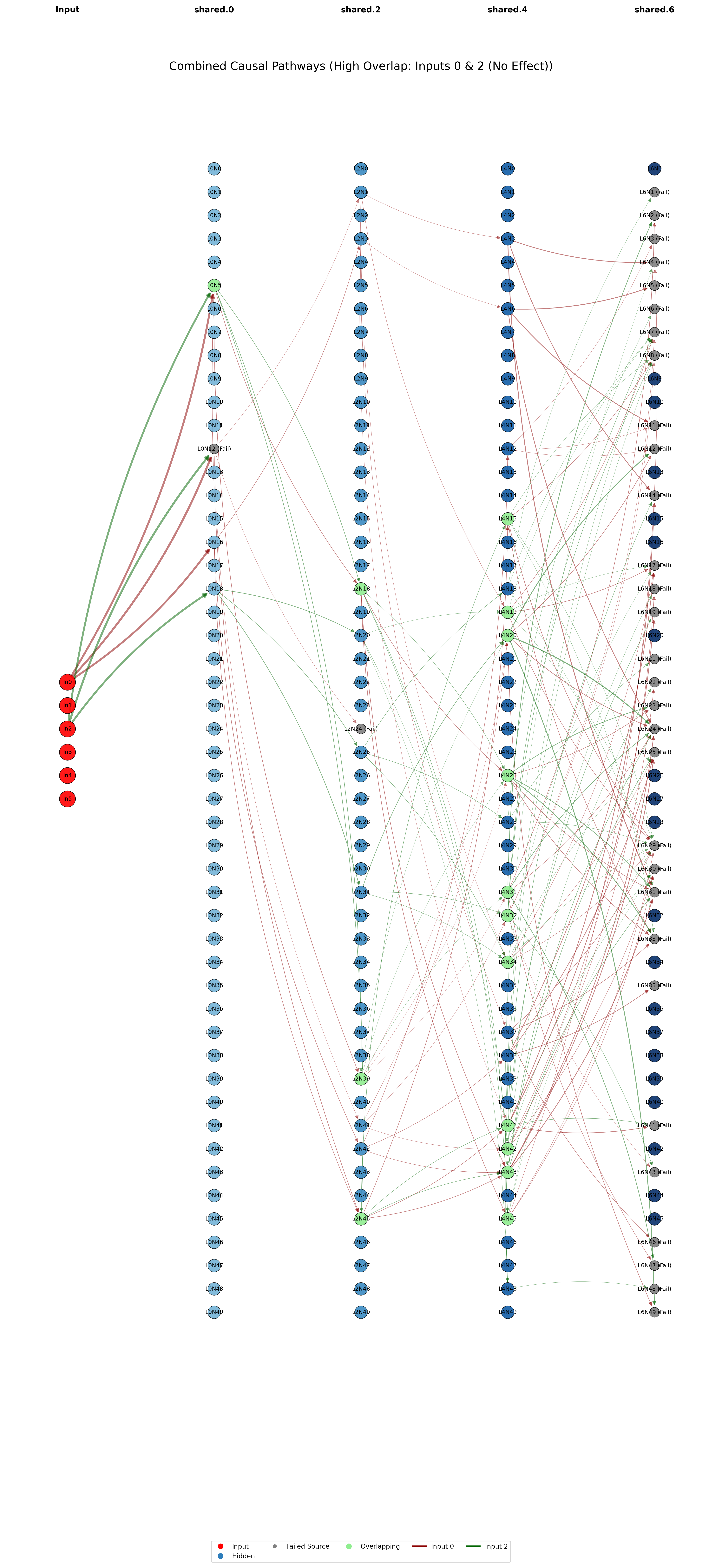}
    \caption{Combined overlap - high interaction between inputs 0 and 2}
    \label{fig:highint}
\end{figure}
\begin{figure}[htpb!]
    \centering
    \includegraphics[width=.5\textwidth]{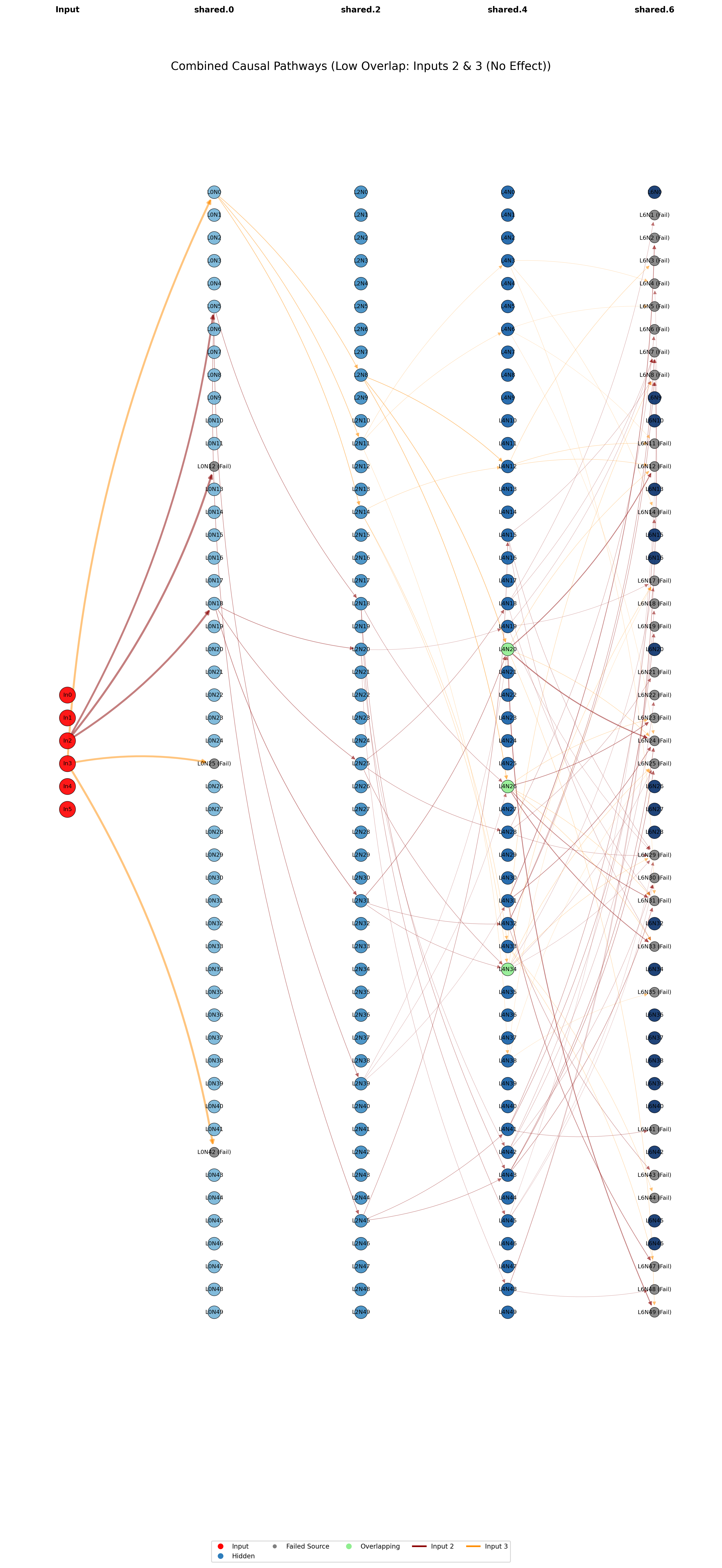}
    \caption{Combined overlap - low interaction between inputs 2 and 3}
    \label{fig:lowint}
\end{figure}

\newpage
\includepdf[pages={-}]{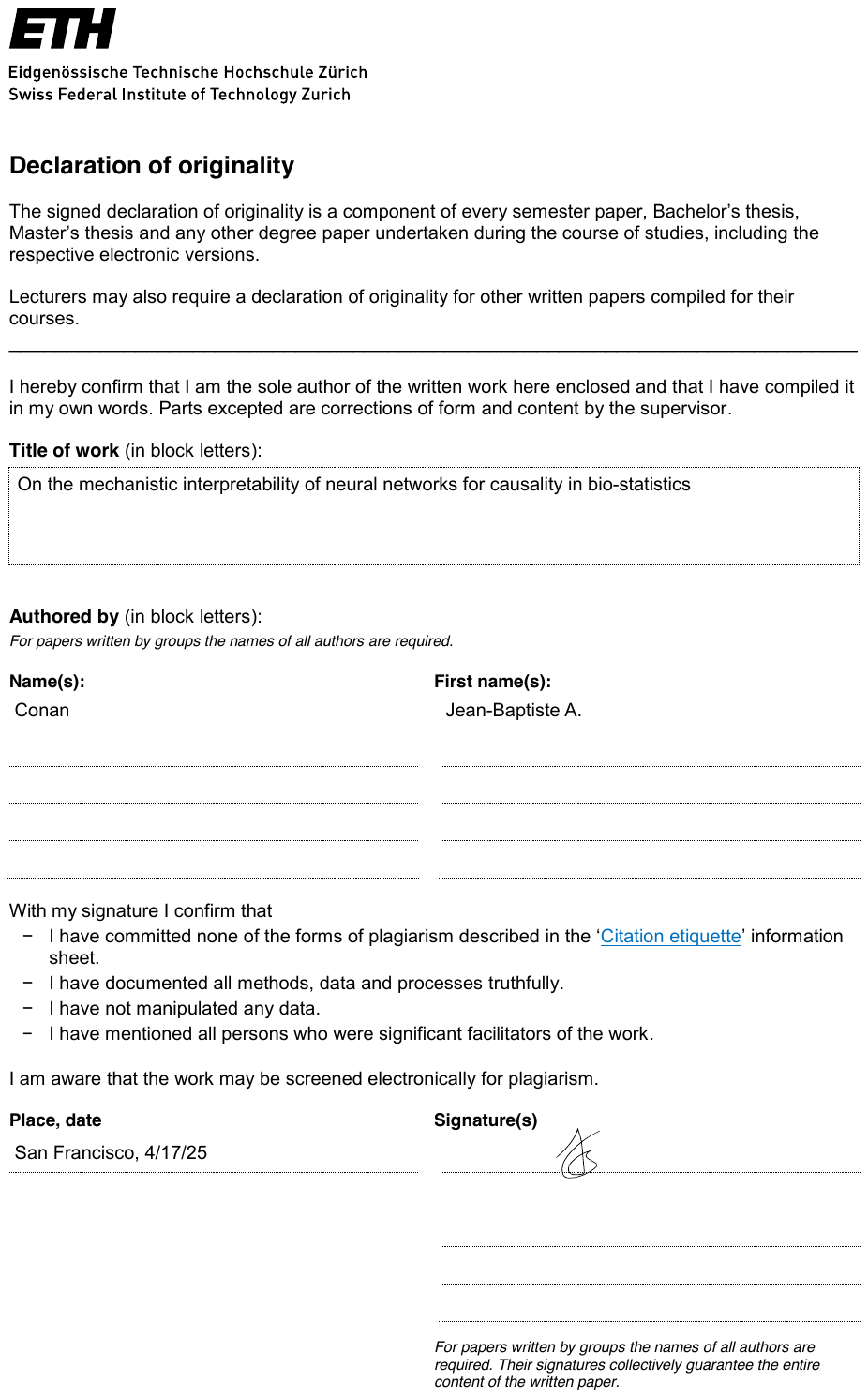}

\end{document}